# Galois Orbits of TQFTs: Symmetries and Unitarity

Matthew Buican[⊠] and Rajath Radhakrishnan[θ]

*CTP and Department of Physics and Astronomy*

*Queen Mary University of London, London E1 4NS, UK*

We study Galois actions on $2+1$D topological quantum field theories (TQFTs), characterizing their interplay with theory factorization, gauging, the structure of gapped boundaries and dualities, 0-form symmetries, 1-form symmetries, and 2-groups. In order to gain a better physical understanding of Galois actions, we prove sufficient conditions for the preservation of unitarity. We then map out the Galois orbits of various classes of unitary TQFTs. The simplest such orbits are trivial (e.g., as in various theories of physical interest like the Toric Code, Double Semion, and 3-Fermion Model), and we refer to such theories as unitary "Galois fixed point TQFTs." Starting from these fixed point theories, we study conditions for preservation of Galois invariance under gauging 0-form and 1-form symmetries (as well as under more general anyon condensation). Assuming a conjecture in the literature, we prove that all unitary Galois fixed point TQFTs can be engineered by gauging 0-form symmetries of theories built from Deligne products of certain abelian TQFTs.

# Contents

## 1. Introduction

Symmetries provide a non-perturbative way to constrain the dynamics of a quantum field theory (QFT). Depending on the spacetime dimension and the symmetry under consideration, one maybe able to, in principle, solve many or all consistent QFTs with that symmetry. A well-known example uses infinite conformal symmetry to bootstrap certain $1+1$D conformal field theories (CFTs) [1]. While conformal symmetry tightly constrains the space of CFTs, solving the conformal boostrap equations is, in general, highly non-trivial. This situation contrasts with $1+1$D topological quantum field theories (TQFTs), where topological symmetry yields certain more general statements [2,3].

Such a Lagrangian-independent approach to QFT is more universal, and it can sometimes make relations between different descriptions of a QFT more apparent. The idea is to extract the operator content and correlation functions of a QFT and encode them in natural mathematical structures. Vertex operator algebras (and their representations) are one such structure for $1+1$D CFTs, while, for $1+1$D TQFTs, Frobenius algebras play a similarly important part. For TQFTs in general spacetime dimensions, $n$-categories play a central role [3].

In this work, we focus on $2+1$D TQFTs,[1] and the corresponding algebraic structure we primarily study is a Modular Tensor Category (MTC). MTCs encode essential physical data of a TQFT without additional redundancies like the choice of a gauge group in a Lagrangian description.[2] Indeed, the fact that the same MTC can be realized by Lagrangians based on different gauge groups makes it clear that gauge groups are, as is well known, not duality invariant.

---
[1]Throughout, we will study non-spin TQFTs (i.e., TQFTs that do not depend on a choice of spin structure).

[2]Although MTCs also have redundancies related to points where particle worldlines fuse (see appendix

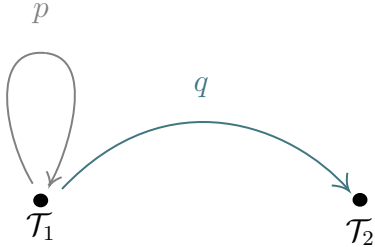

**Fig. 1:** Galois conjugation of TQFT $\mathcal{T}_1$ by elements $p, q$ in the Galois group. $\mathcal{T}_1$ is invariant under Galois action by $p$, but it transforms non-trivially to $\mathcal{T}_2$ under Galois action by $q$.

Topological symmetry is powerful enough to give us a set of constraints known as the Pentagon and Hexagon equations whose solutions give us all possible consistent MTCs. Upon making a further discrete choice, one obtains a corresponding 2+1D TQFT with a fully specified set of line operators. Therefore, finding a consistent $2 + 1$D TQFT in this sense essentially follows from finding the zeros of some multivariable polynomials.

Though these constraints are often too complicated to be solved exactly, a general mathematical result (the Ocneanu rigidity theorem) states that there are only a finite number of inequivalent solutions to the Pentagon and Hexagon equations for a given set of fusion rules [4]. This fact allows us to define a Galois group which permutes the solutions to these polynomials. In other words, Galois conjugation is a systematic way to move around the space of TQFTs. Moreover, Galois conjugation is useful in practice: it has played a role in the classification of low-rank TQFTs [5], proving the rank finiteness theorem [6], finding modular isotopes [7], studying low-dimensional lattice models [8], in connections between TQFT and other types of QFTs [9], and in the study of gapped boundaries [10].

Galois conjugate TQFTs share many important properties. In particular, they have the same fusion rules. However, other observables, like the expectation values of Wilson loop operators, can change under Galois action. Therefore Galois conjugate TQFTs are typically not dual theories. Still, one can define quantities like multi-boundary entanglement entropy, which are invariant under Galois conjugation in abelian TQFTs, and in an infinite set of links in non-abelian TQFTs [11].

In this work, we will show that Galois conjugate TQFTs share a lot more structure. More precisely, we will argue that Galois conjugate TQFTs have isomorphic 0-form, 1-form, and 2-group symmetry structure (up to a mild assumption, this result also holds for anti-

---

B).

unitary symmetries). Moreover, there is a well-defined map between the gapped boundaries of Galois conjugate TQFTs. These results show that, compared to other procedures relating distinct TQFTs like gauging, condensation, etc., there is a sense in which Galois conjugation is a particularly mild change to the TQFT.

On the other hand, unlike gauging, Galois actions can map a unitary TQFT to a non-unitary one. While non-unitary TQFTs (and more general non-unitary QFTs) are interesting in their own right, one of the motivations for our work is to better understand when unitary TQFTs are related by a Galois action. In other words, we would like to ask: Given a unitary TQFT, when is a Galois conjugation guaranteed to land on another unitary TQFT? One common way in which this can happen is if we consider a unitary theory without a time-reversal symmetry. In this case, applying time reversal takes us to a different theory that should also be unitary (examples of such phenomena include $SU(2)_1 \leftrightarrow (E_7)_1$ and $SU(3)_1 \leftrightarrow (E_6)_1$ in Chern-Simons theory). This procedure gives a simple example of a Galois action that preserves unitarity, but we will see that the story is more complex and interesting.

Another motivation for our work comes from the observation that several important low-rank unitary TQFTs like the Toric Code, Double Semion, and the 3-Fermion Model are Galois invariant.[3] These examples illustrate that, while most TQFTs transform under a Galois action, a potentially important subset are Galois invariant. This discussion begs the question of what this more general set of unitary "Galois fixed point TQFTs" looks like. As we will see, this set is substantially simpler than its non-unitary counterpart.[4] It also leads to questions of whether this Galois invariance is preserved under other operations like gauging and anyon condensation. We will see that, while Galois invariance is generally preserved under anyon condensation (which includes 1-form symmetry gauging as a special case), it can be violated when 0-form symmetries are gauged. We will prove some general statements about when such anomalous violation is allowed.

The plan of the paper is as follows. In the next section, we define Galois conjugation of a TQFT and study unitary Galois orbits. We continue with an analysis of Galois actions on various classes of unitary theories: abelian TQFTs, discrete gauge theories, and certain weakly integral MTCs. In section 3 we study theories with gapped boundaries and explain how Galois conjugation relates gapped boundaries of Galois conjugate TQFTs. In section 4 we discuss the relationship between symmetries of Galois conjugate TQFTs. Following this, we look at how Galois conjugation interacts with gauging 0-form symmetries and anyon condensation. We use these results to characterize Galois invariant TQFTs. Section 5 contains several additional examples of Galois conjugation of TQFTs which concretely illustrate our ideas and compliment our discussion. Finally, we conclude with some comments and a discussion of future directions.

---

[3]Note that by Galois invariant, we do not mean that all the data of the TQFT is invariant. For example, in the Double Semion, the anyon, $s$, has its twist $\theta_s = i$ Galois conjugated to $g(\theta_s) = -i$. However, the anyon, $\tilde{s}$, has its twist $\theta_{\tilde{s}} = -i$ Galois conjugated to $g(\theta_{\tilde{s}}) = i$. Therefore, this Galois action can be compensated by the time reversal symmetry that exchanges $s \leftrightarrow \tilde{s}$.

[4]Perhaps this relative simplicity hints at even deeper simplifications in the space of unitary TQFTs.

**Note added:** While completing our paper, the beautiful results in [10] appeared. There is partial overlap of this work with our section 3.

## 2. Galois Conjugation of TQFTs

Let us consider a $2+1$D TQFT, $\mathcal{T}$, corresponding to an MTC, $C$ (see appendix B for a further review of MTCs and [12] for an in-depth discussion of the construction of $2+1$D TQFTs from MTCs).[5] The MTC is built out of "simple" objects that correspond to anyonic line operators (e.g., Wilson lines) in the TQFT from which all other lines can be built. The simple objects are denoted as $\{a, b, \cdots\}$, and they satisfy fusion rules

$$a \otimes b = \sum_c N_{ab}^c c , \quad N_{ab}^c \in \mathbb{Z}_{\geq 0} .\qquad(2.1)$$

The fusion rules capture the position-independent operator product expansion of the line operators in the TQFT. The quantities, $N_{ab}^c$, are dimensions of vector spaces, denoted $V_{ab}^c$, known as fusion spaces. The associativity and commutativity of fusion defines isomorphisms of these fusion spaces.

$$\vcenter{\hbox{$a\ \ b\ \ c$}}\ \Y_f^d\ =\ \sum_e \left(F_{abc}^d\right)_f^e\ \vcenter{\hbox{$a\ \ b\ \ c$}}\ \Y_e^d$$

**Fig. 2:** Pictorial definition of the F-matrix

---

[5] Given the close relationship between $\mathcal{T}$ and $C$, we will sometimes drop the distinction.

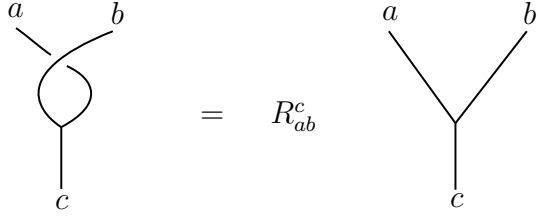

**Fig. 3:** Pictorial definition of the R-matrix

The matrices corresponding to these isomorphisms are the fusion matrices, $F$ (see Fig. 2), and the braiding matrices, $R$ (see Fig. 3). These isomorphisms have to satisfy consistency conditions known as the Pentagon and Hexagon equations.

Moreover, every MTC realizes a (projective) unitary representation of $SL(2,\mathbb{Z})$

$$T_{aa} = d_a^{-1} \sum_c d_c R_{aa}^c = \theta_a \ , \quad S_{ab} = \frac{1}{\sqrt{\sum_a d_a}} \sum_c d_c \text{Tr}(R_{ab}^c R_{ba}^c) = \frac{1}{\sqrt{\sum_a d_a}} \tilde{S}_{ab} \ , \quad (2.2)$$

where $d_a$ is the $S^3$ link invariant of an unknot labelled by the simple object $a$, and $\theta_a$ gives the self-statistics of $a$. Indeed, one can immediately verify that the above matrices form a projective representation of the modular group

$$(ST)^3 = \Theta C \ , \quad S^2 = C \ , \quad C^2 = I \ , \quad (2.3)$$

where $\Theta := \frac{1}{\sqrt{\sum_c d_c^2}} \sum_a d_a^2 T_{aa}$, and $C$ is the charge conjugation matrix. The central charge of a TQFT, $c$, is given in terms of $\Theta$ through the relation

$$e^{\frac{2\pi i c}{8}} = \Theta \ . \quad (2.4)$$

We will think of $C$ as being determined through the action of the $F$ symbols, $R$ symbols, and pivotal coefficients, $\epsilon_a \in \{\pm 1\}$, on the trivalent fusion vertices of the simple objects / anyons [13]. Note that explicitly writing down the $F$ and $R$ symbols requires a choice of gauge.[6] As another subtlety, we remark that this data only determines the total quantum dimension, $\mathcal{D} := \pm\sqrt{\sum_a d_a^2}$, and hence the normalized $S$ in (2.2) up to an overall sign. On the other hand, $\mathcal{D}$ is an important quantity in $\mathcal{T}$. For example, $\mathcal{D} > 0$ is a necessary condition for a unitary TQFT (as follows from positivity of the TQFT inner product [12]). In particular, for a given $C$, there are two TQFTs, $\mathcal{T}_\pm$, that differ by $\mathcal{D} \to -\mathcal{D}$, $S \to -S$, and $c \to c + 4 \pmod 8$ (at least one of these TQFTs must be non-unitary). When the distinction between the two TQFTs is clear from the context or does not matter, we will simply write $\mathcal{T}$.

---

[6]This gauge choice amounts to picking bases for the $V_{ab}^c$ fusion spaces.

From this discussion, we can describe the number fields that enter our analysis and set the stage for the appearance of Galois groups. To that end, first construct a field extension, $K'_C = \mathbb{Q}(F, R)$, from the adjunction of the elements of $F$ and $R$ to the field of rational numbers [14] [15]. Using the gauge freedom alluded to above, the authors of [14] showed that there is a gauge in which $K'_C$ is particularly simple: it is a finite field extension.

To understand how this finite field arises, let us consider the case of a system of multivariable polynomial equations over the rational numbers, $p_1(x_1, \cdots, x_n) = \cdots = p_k(x_1, \cdots, x_n) = 0$, with a finite number of solutions. Any solution of this system belongs to a finite extension of $\mathbb{Q}$.[7] On the other hand, the Pentagon and Hexagon equations are multivariable polynomials over $\mathbb{Q}$ with an infinite number of solutions (because of the gauge freedom). Therefore, in this case, we have an algebraic variety, $V$, in which some points do not belong to a finite field extension of $\mathbb{Q}$. However, algebraic points of a complex affine algebraic variety defined over $\overline{\mathbb{Q}}$ are dense in the Zariski topology [17] ($\bar{\mathbb{Q}}$ is the algebraic closure of $\mathbb{Q}$).[8] The upshot is that, given a set of multivariable polynomials with coefficients in $\bar{\mathbb{Q}}$, we can always find solutions that are algebraic. Therefore, there are solutions to the Pentagon and Hexagon equations that are algebraic. To show that all MTCs allow a gauge in which $F$ and $R$ are algebraic, the authors of [14] showed that the gauge freedom acts on $V$ as an algebraic group and that each orbit of this action has an algebraic point.

Next let us discuss how Galois groups enter our story. Recall that, given a number field, we can study its automorphisms. As a simple example, consider the polynomial equation $x^2 = 2$. This is a polynomial over $\mathbb{Q}$, but its solutions are $\pm\sqrt{2} \notin \mathbb{Q}$. To describe these solutions, we can construct the field extension $\mathbb{Q}(\sqrt{2})$, which consists of elements of the form $a + b\sqrt{2}$ where $a, b \in \mathbb{Q}$. Note that the field $\mathbb{Q}(\sqrt{2})$ has an automorphism given by $\sqrt{2} \to -\sqrt{2}$. In particular, any algebraic equation involving the elements of $\mathbb{Q}(\sqrt{2})$ does not change under the exchange $\sqrt{2} \to -\sqrt{2}$. Note that this action permutes the two roots of the polynomial $x^2 = 2$ we started with. In this simple case, this is the only non-trivial permutation of the roots. However, in more general cases, the automorphisms of the number field obtained from the roots of a polynomial may not exhaust all possible permutations of the roots.

---

[7]One way to show this statement involves proving that the ideal, $I$, generated by $p_1, \cdots, p_k$ in the polynomial ring $\mathbb{Q}[x_1, \cdots, x_n]$ is zero dimensional. Given a solution $a_1, \cdots, a_n$ to the set of polynomial equations $p_1 = 0, \cdots, p_k = 0$, one then shows that there exists some polynomial $r_i(x_i) \in \mathbb{Q}[x_i]$, $1 \le i \le n$, such that $r_i(a_i) = 0$. For more details, see [16].

[8]This result will play an important role in our analysis.

Throughout this paper, we will work in a gauge in which $F$ and $R$ belong to a number field. Now, any finite field extension over $\mathbb{Q}$ is separable. However, it need not be normal. Since normal closures have useful algebraic properties, let us consider the normal closure of $K'_C$, and call it $K_C$. Because $K_C$ is normal and separable, it is a Galois field, and we will refer to it as *the defining number field of $C$*. Note that $K_C$ need not contain the total quantum dimension, $\mathcal{D}$, and therefore need not contain the normalization of the $S$ matrix in (2.2) (it does contain $\mathcal{D}^2$ and $\tilde{S}$).[9]

The authors of [14] conjectured that there is a gauge in which the defining number field of an MTC is cyclotomic (in other words, the number field can be obtained by appending a primitive $n^{\text{th}}$ root of unity to $\mathbb{Q}$).[10] Note that this claim does not hold for general fusion categories. For example, the fusion category obtained from the principal even part of the Haagerup subfactor does not admit a gauge in which the defining number field is cyclotomic [18].

Given the above construction, we can act on $K_C$ with some element, $q$, of the Galois group, $\text{Gal}(K_C)$. Since $F$ and $R$ are elements of $K_C$, they get acted on by $q$; we denote the result as $q(F)$ and $q(R)$ respectively. Recall that the automorphisms of the field, $K_C$, preserve all algebraic equations involving the elements of $K_C$. The Pentagon and Hexagon equations are algebraic equations saitsified by some elements of $K_C$. Therefore, they are preserved under a Galois action. That is, if $F$ and $R$ satisfies the Pentagon and Hexagon equations, so do $q(F)$ and $q(R)$! Therefore, $q(F)$ and $q(R)$ defines an MTC, which we denote as $q(C)$.

**Definition:** Note that we define the Galois action on TQFTs through the Galois action on the defining MTC data, $F$ and $R$. In particular, we choose not to act to reverse the sign of the total quantum dimension, $\mathcal{D}$, or, equivalently, the sign of the normalization of $S$ (this choice amounts to working with the $(\tilde{S}, T)$ modular pair in (2.2)). We lose no generality since, after performing such a Galois action, we can, in principle, consider TQFTs with either sign of $\mathcal{D}$ and normalization of $S$.

Various authors have established that the modular data of a TQFT is always contained in a cyclotomic field extension [19–21]. In the language of these references, the modular data is given by the pair, $(S, \varphi \cdot T)$, where $\varphi := \exp(-\pi i c/24)$ (here $c$ can essentially be thought of as the central charge of the associated 2D RCFT[11]). Let this cyclotomic extension be $\mathbb{Q}(\xi_{N'})$, where $\xi_{N'}$ is a primitive $N'^{\text{th}}$ root of unity. Since our MTC is insensitive to the sign of $\mathcal{D}$, and since we do not consider Galois actions that take $(\mathcal{D}, S) \to (-\mathcal{D}, -S)$, it is more natural to work with the modular data field $\mathbb{Q}(\xi_N)$ (with $N \leq N'$) for $(\tilde{S}, T)$ in (2.2).[12]

---

[9]For example, in abelian TQFTs with $\mathbb{Z}_3$ fusion rules (see Table (1) for the explicit MTC data), $F$ and $R$ can be chosen to belong to the cyclotomic field $\mathbb{Q}(\xi_3)$, while $\mathcal{D} = \sqrt{3} \notin \mathbb{Q}(\xi_3)$ is only an element of $\mathbb{Q}(\xi_{12})$. Here, $\xi_n$ is a primitive $n^{\text{th}}$ root of unity.

[10]To the best of our knowledge, there are no known counterexamples to this conjecture.

Let us now connect this discussion with the defining number field. To that end, note that $\mathbb{Q}(\xi_N) \subset K_C$ as a subfield. Now, every element of $q \in \mathbb{Z}_N^{\times} = \text{Gal}(\mathbb{Q}(\xi_N))$ acts on the modular data to give potentially new modular data, $q(\tilde{S}), q(T)$. Then, for every $q \in \mathbb{Z}_N^{\times}$ acting on the modular data, we have some $\sigma \in \text{Gal}(K_C)$ such that $\sigma|_{\mathbb{Q}(\xi_N)} = q$. This statement holds because $K_C$ is normal.[13]

As we have seen from the above discussion, Galois conjugation permutes the solutions of the pentagon and hexagon equations. Hence, it relates distinct TQFTs with the same fusion rules. However, as the following example illustrates, there may not be a Galois conjugation relating any two solutions of the Pentagon and Hexagon equations for particular fixed fusion rules:

**Example:** Consider the Toric Code (a.k.a. $\mathbb{Z}_2$ discrete gauge theory), the 3-Fermion Model (a.k.a. $\text{Spin}(8)_1$ Chern-Simons theory), and Double Semion (a.k.a. $SU(2)_1 \boxtimes (E_6)_1$ Chern-Simons theory or twisted $\mathbb{Z}_2$ discrete gauge theory). All these theories have $\mathbb{Z}_2 \times \mathbb{Z}_2$ fusion rules. For abelian theories (i.e., theories whose fusion rules are abelian groups), it turns out that all the defining data discussed above—the $F$ symbols, $R$ symbols, and pivotal coefficients—can be determined in terms of the twists of the anyons, $\theta_i$. Moreover, for abelian theories, we can choose a gauge in which $K_C$ is the number field determined by the twists.[14] For Toric Code, we have anyons $1, e, \mu, f$ (where $f = e \times \mu$) with twists

$$\theta_1 = \theta_e = \theta_\mu = 1 \,, \quad \theta_f = -1 \,, \tag{2.5}$$

while the 3-Fermion Model has anyons $1, f_1, f_2, f_3$ (where $f_3 = f_1 \times f_2$) with twists

$$\theta_1 = 1 \,, \quad \theta_{f_1} = \theta_{f_2} = \theta_{f_3} = -1 \,, \tag{2.6}$$

---

[11]Although, see [21] for a more RCFT-independent discussion.

[12]We have $\mathcal{D}^2 \in \mathbb{Q}(\xi_N)$ since the quantum dimensions are in $\tilde{S}$ but, in general, $\mathcal{D}, \varphi \notin \mathbb{Q}(\xi_N)$ (this last fact follows from the observation in [20, 21] that the elements of $\varphi \cdot T$ determine the cyclotomic extension of the modular data).

[13]This discussion explains why it is better to work with the normal field $K_C$ instead of $K'_C$ itself.

[14]This statement follows from (2.17) and (2.18) of [22] along with the fact that the twists are valued in a cyclotomic (and hence Galois) field.

and Double Semion has anyons $1, s, \tilde{s}, d$ (where $d = s \times \tilde{s}$) with twists

$$\theta_1 = \theta_d = 1 \ , \quad \theta_s = i \ , \quad \theta_{\tilde{s}} = -i \ . \tag{2.7}$$

In the first two cases, $K_C = \mathbb{Q}$, and the corresponding Galois group is trivial. Therefore Toric Code and the 3-Fermion Model are Galois invariant and are not related to each other by a Galois action. In the Double Semion case (2.7), we see that $K_C = \mathbb{Q}(i)$ and so $\text{Gal}(K_C) = \mathbb{Z}_2$ has a non-trivial element implementing complex conjugation. However, complex conjugation exchanges the twists $\theta_s \leftrightarrow \theta_{\tilde{s}}$ while leaving the rest of the data invariant. Therefore, the Double Semion theory is mapped to itself. In summary, all three of these theories share the same fusion rules, but they are unrelated by a Galois action.

More generally, we will have theories that transform non-trivially under Galois actions (and we will see many such examples in the next section). In fact, it is possible that Galois actions supplemented by some other procedure act transitively on the solutions of the Pentagon and Hexagon equations. For example, Galois conjugations along with a particular change of $F$ symbols act transitively on all MTCs with the same fusion rules as $SU(N)_k$ Chern-Simons theory [14].

Given our discussion of Galois conjugation, we would like to study how these operations interact with global properties of TQFT. In particular, our immediate goal is to understand how Galois conjugation affects the subcategory structure of $C$ (and therefore $\mathcal{T}$).

To that end, note that a proper subset of anyons in $C$ may close under fusion and therefore form a braided fusion subcategory whose $F$ and $R$ symbols are given by the restriction of the $F$ and $R$ symbols of $C$ onto that subcategory. Since Galois conjugation preserves fusion rules, it is clear that it preserves the braided fusion subcategory structure of a modular tensor category. This observation will play a crucial role in our analysis of discrete gauge theories.

The braiding in a subcategory may or may not be degenerate (i.e., the corresponding modular $\tilde{S}$ matrix may or may not be degenerate). If it is non-degenerate, then a general result of Müger [23] guarantees that the subcategory factorizes from the rest of the theory (i.e., anyons in the subcategory braid trivially with anyons outside the subcategory). In this case, our TQFT has a product structure

$$\mathcal{T} = \boxtimes_i \mathcal{T}_i \ , \tag{2.8}$$

where each $\mathcal{T}_i$ has a corresponding MTC $C_i$.[15] The decomposition in (2.8) is called a "prime decomposition" into prime factors $\mathcal{T}_i$ (each factor braiding trivially with other factors).

---

[15]The symbol "$\boxtimes$" denotes the so-called "Deligne" product and is an appropriate categorical generalization of a direct product.

A basic question is then to understand the Galois action on the prime factors:

**Theorem 2.1:** The space of prime TQFTs is closed under Galois action.

**Proof:** Consider a non-prime TQFT, $\mathcal{T}$. The associated MTC, $C$, has a modular sub-category, $K$. The set of anyons in $K$ label a modular sub-matrix, $\tilde{S}_K$, of the $\tilde{S}$ matrix of $C$. Suppose $\mathbb{Q}(\xi_N)$ is the cyclotomic field containing the elements of the $\tilde{S}$ matrix, where $\xi_N := \exp(2\pi i/N)$. The cyclotomic field $\mathbb{Q}(\xi_N)$ has a cyclotomic subfield $\mathbb{Q}(\xi_{N_K})$ which contains the elements of the matrix $\tilde{S}_K$. Any element of $\text{Gal}(\mathbb{Q}(\xi_N))$ restricts to a Galois action on $\mathbb{Q}(\xi_{N_K})$. Hence, under a Galois conjugation of the $\tilde{S}$ matrix, the modular sub-matrix $\tilde{S}_K$ gets transformed into another modular matrix. Therefore, the set of anyons in $K$ forms a modular subcategory of the Galois-conjugated theory. As a result, Galois conjugation of a non-prime TQFT results in a non-prime TQFT. From invertibility of the Galois action it is clear that the space of prime TQFTs is closed under Galois conjugation.[16] □

As a result, the Galois action on $\mathcal{T}$ in (2.8) can be obtained from the Galois action on the prime theories, $\mathcal{T}_i$. The notion of primeness is independent of whether the TQFT is unitary or not. Note that the prime factorization of TQFTs into Deligne products described above is not always unique. Even the number of prime TQFTs in a prime factorization is not always unique. For example, Toric Code $\boxtimes$ Semion = Semion $\boxtimes$ $\overline{\text{Semion}}$ $\boxtimes$ Semion.

If a TQFT, $\mathcal{T}$, transforms non-trivially under Galois action, then it is clear that at least one of its prime factors should transform non-trivially under it. However, non-trivial Galois transformation of the prime factors of a TQFT may act trivially on the full TQFT. Indeed, this is the case in the Double Semion example discussed previously since it turns out that

$$\text{Double Semion} = \text{Semion} \boxtimes \overline{\text{Semion}} \ . \tag{2.9}$$

As we saw above, both the Semion and $\overline{\text{Semion}}$ models transform non-trivially under Galois action (the twists $\theta_s = i$ from the Semion model and $\theta_{\tilde{s}} = -i$ from $\overline{\text{Semion}}$ are complex conjugated). However, since these two factors transform into each other under Galois action, the Double Semion model is invariant under all Galois conjugations. As we will see in Section 3, the Galois invariance of Double Semion model can also be explained using the Galois invariance of its gapped boundary.

---

[16]This result can also be seen from the fact that Galois conjugations preserve invertibilty of a matrix.

*2.1. Unitary Galois Orbits*

The preceding discussion was very general and applies to all types of Galois transformations, including those that transform unitary theories into non-unitary ones (and vice-versa). However, on physical grounds, it is important to understand the conditions under which Galois transformations preserve unitarity. To that end, in this section we will obtain a sufficient condition for unitarity preservation.

Let us begin by building up to the extra constraints that a unitary MTC should satisfy. We will call a fusion category unitary if there exists a gauge in which the $F$ symbols are unitary.[17][18] A braided fusion category is unitary if there exists a gauge in which both the $F$ and $R$ symbols are unitary. The condition on $R$ is not an extra constraint, since any set of consistent $R$ symbols obtained from unitary $F$ symbols is unitary [26]. Therefore, every braided fusion category defined over a unitary fusion category is unitary. Note that to define quantum dimensions, we need to add a ribbon structure to a braided fusion category. In general, there is more than one inequivalent choice for the ribbon structure. However, there is a unique choice which guarantees that all the quantum dimensions are positive [26]. A ribbon fusion category is called unitary if there exists a gauge in which the $F$ and $R$ symbols are unitary and if all quantum dimensions are positive.

Given this discussion, we see that a sufficient and necessary condition for an MTC, $C$, to be unitary is that it has a gauge in which the $F$ and $R$ symbols are unitary and the quantum dimensions satisfy $d_a > 0$, $\forall a \in C$ (here $a$ is an anyon of the TQFT or a simple object of $C$) [27]. [19][20]

From a unitary MTC, we can always construct a unitary TQFT by choosing the total quantum dimension to be positive (i.e., $\mathcal{D} > 0$). Indeed, the corresponding TQFT inner product is then positive definite [12]. On the other hand, starting from a non-unitary MTC, we cannot construct a unitary TQFT.

Since making unitarity manifest requires choosing a particular gauge, it is useful to check that this choice does not clash with the gauge choice required for the MTC data to belong to a finite field extension, $K_C$. In fact, the authors of [27] showed that there is a gauge in which both can be achieved simultaneously.

---

[17]There is a gauge-independent definition of unitarity of a fusion category. But we will use the definition in terms of the $F$ matrices since both are equivalent [24].

[18]Given a set of labels and its fusion rules, a necessary condition for a unitary fusion category with these fusion rules to exist is given in [25].

[19]In an MTC without unitarity, the $F$ and $R$ symbols are defined only up to gauge transformations. In a unitary MTC, the unitary $F$ and $R$ matrices are defined only up to unitary gauge transformations. Moreover, if the $F$ and $R$ matrices can be made unitary in two different gauges, then they are unitarily gauge equivalent [28]. Therefore, the unitary structure on an MTC is unique.

[20]Since all anyons have a dual, $d_a > 0 \implies d_a \geq 1$

Now, given a finite field extension, $K_C$, in which the $F$ symbols of $C$ are unitary, determining whether a Galois conjugation of a unitary MTC results in a unitary MTC depends on how the $F$ symbols and quantum dimensions get transformed under the Galois action. We will call a Galois action which takes a unitary MTC to a unitary MTC a unitarity-preserving Galois action. This construction also maps a unitary TQFT to a unitary TQFT since we can always supplement our MTC action with a choice of $\mathcal{D} > 0$. Before looking at the $F$ symbols, let us study the action of a unitarity-preserving Galois action on the quantum dimensions.

**Lemma 2.2:** A unitarity-preserving Galois action acts trivially on the quantum dimensions.

**Proof:** Consider a unitary TQFT, $\mathcal{T}$, with associated unitary MTC, $C$, having defining number field $K_C$. Let $q(C)$ be a unitary MTC (with corresponding unitary TQFT, $q(\mathcal{T})$), where $q(C)$ is the Galois conjugate of $C$ with respect to some $q \in \text{Gal}(K_C)$. Since $C$ is unitary, the quantum dimension, $d_a$, of an anyon $a \in C$ is equal to the corresponding Frobenius-Perron dimension and is positive. We denote $q(d_a)$ as the quantum dimension of the corresponding anyon in $q(C)$. Since $q(C)$ is unitary, $q(d_a)$ are also positive. By proposition 3.3.4 of [29],

$$|q(d_a)| \leq d_a . \qquad (2.10)$$

Suppose $q(d_a) < d_a$ for some anyon $a$. Using the inverse Galois action, we have $\bar{q}(d_a) > d_a$. This contradicts (2.10). Therefore, we must have

$$g(d_a) = d_a \ \forall a . \qquad (2.11)$$

□

As a result of the above lemma, invariance of the quantum dimesions under Galois action is necessary for preserving unitarity. However, this is not sufficient. To see this, let us study how Galois conjuation changes the $F$ symbols. In general, Galois conjugation does not preserve unitarity of a matrix. To understand this statement, suppose we have some unitary matrix, $U$, such that the elements of the matrix belong to an algebraic number field, $K$. $U$ satisfies $U^\dagger U = I$. Galois conjugating this relation which respect to some $q \in \text{Gal}(K)$ gives

$$q(U^\dagger U = I) \implies q(U^\dagger)q(U) = I . \qquad (2.12)$$

If complex conjugation commutes with $q$, then the above equation simplifies to

$$q(U)^\dagger q(U) = I \ . \tag{2.13}$$

Therefore, the Galois conjugated matrix is still unitary. However, it often happens that complex conjugation does not commute with the Galois action. In this case, $q(U)$ is non-unitary.

All MTCs conjecturally have a gauge in which the defining number field is cyclotomic [14], and in this case any Galois conjugation commutes with complex conjugation. However, a unitary TQFT in such a gauge may not have unitary $F$ symbols. The simplest example of this is Galois conjugation of the Fibonacci model to get the Yang-Lee model. In the Fibonacci model, there is a basis in which $F$ and $R$ are unitary. However, in this basis, $F$ and $R$ symbols belong to a field extension which has a non-abelian Galois group. On the other hand, if we choose a gauge in the $F$ and $R$ symbols are in a cyclotomic field, $F$ symbols become non-unitary. This example is studied in detail in [15, 27].

It is clear from this discussion that if there is a gauge in which the unitary $F$ and $R$ symbols of a unitary TQFT are real, then any Galois conjugation will result in unitary $F$ and $R$ matrices. In this case, the defining number field, $K_C$, is called "totally real." A more general statement holds if the defining number field is a CM field (note: all cyclotomic fields are CM fields, although the converse is not true). A CM field is a quadratic extension of a totally real field. In other words, a CM field, $K$, is of the form $H(\alpha)$, where $H$ is a totally real field such that $K$ is complex (i.e., it cannot be embedded as a subfield of $\mathbb{R}$). A simple example is the cyclotomic field (appearing in the Double Semion discussed above), $\mathbb{Q}(i)$, which contains numbers of the form $a + ib$ where $a, b \in \mathbb{Q}$. A CM field has the property that complex conjugation is in the center of the Galois group. In fact, any number field with complex conjugation in the center of the Galois group should either be a totally real field (in which case complex conjugation acts trivially) or a CM field [30]. This discussion leads to the following result:

**Theorem 2.3:** Let $C$ be a unitary MTC, and let $K_C$ be its defining number field. Let $K_F$ be the Galois field obtained from the normal closure of the $F$ symbols added to the rationals. If there is a gauge in which the $F$ symbols are unitary and $K_F$ is a totally real field or a CM field, then any Galois conjugation which acts trivially on the quantum dimensions results in a unitary TQFT.

**Proof:** If $K_F$ is a totally real field, then complex conjugation acts trivially on the $F$ symbols. Any Galois conjugation $q \in \text{Gal}(K_C)$ takes unitary $F$ symbols to unitary $F$ symbols. Therefore, the Galois conjugate TQFT has unitary $F$ symbols. From [26], the $R$ matrices of the Galois conjugate TQFT should be unitary. Now suppose the Galois conjugation acts trivially on the quantum dimensions, then the Galois conjugate TQFT has positive quantum dimensions. It follows that the resulting TQFT is unitary.

If $K_F$ is a CM field, then complex conjugation is in the center of the Galois group $\text{Gal}(K_F)$. Therefore, the unitarity of the $F$ symbols is preserved under Galois action with respect to any $q \in \text{Gal}(K_C)$. If the quantum dimensions are invariant under Galois action, then the Galois conjugate TQFT has positive quantum dimensions. It follows that the resulting MTC is unitary, and we can therefore also take the corresponding TQFT to be unitary (we must choose positive total quantum dimension). □

For TQFTs described by integral MTCs, the quantum dimensions are integers and hence Galois invariant. Therefore, any Galois action which preserves the unitarity of the $F$ symbols gives us a unitary Galois conjugate MTC and hence (by a choice of $\mathcal{D}$) a unitary TQFT. For example, in abelian TQFTs, there exists a gauge in which the $F$ symbols belong to $\{\pm 1\}$ [22]. Therefore, in this case $K_F = \mathbb{Q}$ is a totally real field and any such Galois action preserves unitarity.

In [13], Wang conjectures that a ribbon fusion category (and hence an MTC) with positive quantum dimensions is unitary. If this conjecture is true, then our Lemma 2.2 alone is enough to characterize unitary Galois orbits. That is, any Galois action of the type we consider which acts trivially on the quantum dimensions of a unitary TQFT results in a unitary TQFT.

In the next subsection, we will study Galois actions on abelian TQFTs. These provide the simplest example of unitary Galois orbits.

## 2.2. Abelian TQFTs and Unitary Galois Orbits

In this section we study abelian TQFTs (i.e., theories whose fusion rules are those of a finite abelian group) and the corresponding Galois orbits. As is well-known, a TQFT is abelian if and only if the quantum dimensions of all anyons are equal to 1. Since Galois conjugation preserves integers, abelian TQFTs are closed under this action. Moreover, since the $F$ and $R$ matrices of an abelian theory are phases [22], any abelian MTC is unitary and has a cyclotomic defining number field (by choosing $\mathcal{D} > 0$ as described above, we restrict our attention to unitary abelian TQFTs). Therefore, Galois conjugation of such an abelian TQFT always preserves unitarity, and so we will leave the unitary nature of these theories implicit in what follows.

Our strategy below consists of noting that general abelian TQFTs can be written as Deligne products of prime abelian TQFTs. Galois conjugation of an abelian TQFT can thus be reduced to describing the Galois conjugation of prime TQFTs. Moreover, by the discussion in footnote 14 and the surrounding text, for abelian theories the defining number field can be taken to be the cyclotomic field of the twists.

Table 1 gives the classification of prime abelian TQFTs [31] in one particular description of the underlying defining twist data. As we will see when we study Galois actions on these theories, there can be dual descriptions as well.

| Theory | Fusion rules | Twists | $K^{-1}$ |
|---|---|---|---|
| $A_{p^r}$ | $\mathbb{Z}_{p^r}$ | $\theta_a = e^{\frac{2\pi i 2 a^2}{p^r}}$ | $\begin{pmatrix} \alpha & 2^{-1} & & & & \\ 2^{-1} & 2^{-1}a_1 & 2^{-1} & & & \\ 0 & 2^{-1} & 2^{-1}a_2 & 2^{-1} & & \\ & & & \ddots & & \\ & & & & 2^{-1}a_{k-1} & 2^{-1} \\ & & & & 2^{-1} & 2^{-1}a_{k-1} \end{pmatrix}$ where $\alpha = p^{-r}$ and $a_i$ have to be determined using Wall's algorithm. |
| $B_{p^r}$ | $\mathbb{Z}_{p^r}$ | $\theta_a = e^{\frac{2\pi i a^2}{p^r}}$ | Same as above with $\alpha = 2^{-1} p^{-r}$ |
| $A_{2^r}$ | $\mathbb{Z}_{2^r}$ | $\theta_a = e^{\frac{2\pi i a^2}{2^{r+1}}}$ | $\left(2^{-(r+1)}\right)$ |
| $B_{2^r}$ | $\mathbb{Z}_{2^r}$ | $\theta_a = e^{\frac{-2\pi i a^2}{2^{r+1}}}$ | $\left(-2^{-(r+1)}\right)$ |
| $C_{2^r}$ | $\mathbb{Z}_{2^r}$ | $\theta_a = e^{\frac{2\pi i 5 a^2}{2^{r+1}}}$ | Same as $A_{p^r}$ with $\alpha = 5 \times 2^{-(r+1)}$ |
| $D_{2^r}$ | $\mathbb{Z}_{2^r}$ | $\theta_a = e^{\frac{-2\pi i 5 a^2}{2^{r+1}}}$ | Same as $A_{p^r}$ with $\alpha = -5 \times 2^{-(r+1)}$ |
| $E_{2^r}$ | $\mathbb{Z}_{2^r} \times \mathbb{Z}_{2^r}$ | $\theta_{(m,n)} = e^{\frac{2\pi i m n}{2^r}}$ | $\begin{pmatrix} 0 & 2^{-(r+1)} \\ 2^{-(r+1)} & 0 \end{pmatrix}$ |
| $F_{2^r}$ | $\mathbb{Z}_{2^r} \times \mathbb{Z}_{2^r}$ | $\theta_{(m,n)} = e^{\frac{2\pi i (m^2 + n^2 + mn)}{2^r}}$ | $\begin{pmatrix} 2^{-r} & 2^{-(r+1)} & 0 & 0 \\ 2^{-(r+1)} & 2^{-r} & 2^{-1} & 0 \\ 0 & 2^{-1} & 3^{-1}(2^r + (-1)^{r-1}) & 2^{-1} \\ 0 & 0 & 2^{-1} & (-1)^{r-1} \end{pmatrix}$ |

**Table 1:** Classification of prime abelian TQFTs and associated K matrices.

Since we know that the space of abelian TQFTs and the space of prime TQFTs is closed under Galois action (Theorem 2.1), a prime abelian TQFT should either be invariant or get transformed into another prime abelian TQFT under a Galois conjugation. Consider the $A_{p^r}$ prime abelian TQFT. Since the twists are $p^r$-roots of unity, the cyclotomic field containing the data of this theory is $\mathbb{Q}(\xi_{p^r})$ with Galois group $\mathbb{Z}_{p^r}^\times$ (the multiplicative group of integers mod $p^r$).

Under a Galois action corresponding to some $q \in \mathbb{Z}_{p^r}^\times$ there are two possibilities, $A_{p^r}$ remains invariant or $A_{p^r} \to B_{p^r}$. We can consider two cases. Suppose $q \bmod p^r = \alpha^2$ for some $\alpha$. Then,

$$\theta_a = e^{\frac{2\pi i 2 a^2}{p^r}} \to e^{\frac{2\pi i 2 \alpha^2 a^2}{p^r}} = e^{\frac{2\pi i 2 (\alpha a)^2}{p^r}} = \theta_{\alpha a \bmod p^r} \ . \tag{2.14}$$

Hence, the Galois conjugation in this case can be interpreted as a permutation of anyons in the theory given by $\alpha a \bmod p^r$. Moreover, this permutation preserves the fusion rules. Therefore, this is a dual description of the same theory. Now suppose $q$ is not a quadratic residue mod $p^r$, then it is clear that $4q$ is also not a quadratic residue mod $p^r$. As a result, we have

$$\theta_a = e^{\frac{2\pi i 2 a^2}{p^r}} \to e^{\frac{2\pi i 2 q a^2}{p^r}} \ , \tag{2.15}$$

where the resulting twists are those of the $B_{p^r}$ theory (since any integer which is not a quadratic residue mod $p^r$ defines the same theory). So we can summarize the Galois action on $A_{p^r}$ as follows

$$\alpha^2 = q \bmod p^r : A_{p^r} \to A_{p^r} \ , \quad \alpha^2 \neq q \bmod p^r : A_{p^r} \to B_{p^r} \ , \tag{2.16}$$

for some integer $\alpha$.

**Example:** For the $A_5$ theory, we have Galois conjugations corresponding to $q = 1, 2, 3, 4$ (the Galois group is $\mathbb{Z}_4$). $q = 4$ is a duality, while for $q = 2, 3$ we have $A_5 \to B_5$.

Now let us consider Galois conjugation of the $A_{2^r}$ theory. Since the roots are $2^{r+1}$-roots of unity, the cyclotomic field containing all the data of the theory is $\mathbb{Q}(\xi_{2^{r+1}})$ with Galois group $\mathbb{Z}_{2^{r+1}}^\times = \{\text{all odd integers} < 2^{r+1}\}$. Before discussing the general pattern of Galois action on $A_{2^r}$ theory, let us discuss an example.

**Example:** Consider the $A_4$ theory. We have Galois conjugations corresponding to $q = 1, 3, 5, 7$ constituting the Klein four-group. We have the following transformations forming the edges of a tetrahedron:

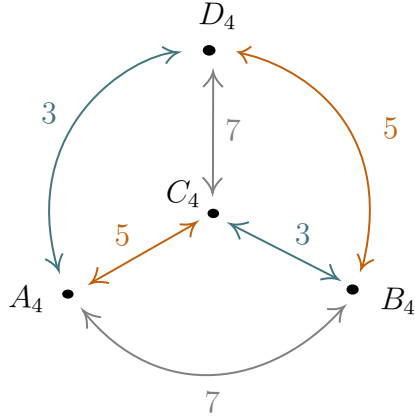

The Galois action on a general $A_{2^r}$ theory by some $q \in \mathbb{Z}^\times_{2^{r+1}}$ depends on the nature of $q$ mod $2^{r+1}$. Suppose $\alpha^2 = q$ mod $2^{r+1}$. Using Hensel's lemma, this has a solution if and only if $q = 1$ mod $8$. Since $q$ is odd, $\alpha$ has to be odd, and $\gcd(\alpha, 2^r) = 1$. Therefore, if $\alpha^2 = q$ mod $2^{r+1}$, then this Galois action acts as an automorphism of the fusion rules $\mathbb{Z}_{2^r}$ given by $a \to \alpha a$ mod $2^r$, $a \in \mathbb{Z}_{2^r}$. Similarly, if $-1\alpha^2 = q$ mod $2^{r+1}$, then the Galois action transforms the twists of $A_{2^r}$ to $B_{2^r}$ family of prime abelian theories up to an automorphism of the fusion rules given by $\alpha$.

Most generally, Galois conjugations permute the prime theories $A_{2^r}, B_{2^r}, C_{2^r}$, and $D_{2^r}$ as follows:

$$\alpha^2 = q \text{ mod } 2^{r+1}: A_{2^r} \to A_{2^r}, B_{2^r} \to B_{2^r}, C_{2^r} \to C_{2^r}, D_{2^r} \to D_{2^r} \;, \tag{2.17}$$

$$-1\alpha^2 = q \text{ mod } 2^{r+1}: A_{2^r} \to B_{2^r}, B_{2^r} \to A_{2^r}, C_{2^r} \to D_{2^r}, D_{2^r} \to C_{2^r} \;, \tag{2.18}$$

$$5\alpha^2 = q \text{ mod } 2^{r+1}: A_{2^r} \to C_{2^r}, B_{2^r} \to D_{2^r}, C_{2^r} \to A_{2^r}, D_{2^r} \to B_{2^r} \;, \tag{2.19}$$

$$-5\alpha^2 = q \text{ mod } 2^{r+1}: A_{2^r} \to D_{2^r}, B_{2^r} \to C_{2^r}, C_{2^r} \to B_{2^r}, D_{2^r} \to A_{2^r} \;. \tag{2.20}$$

Now let us consider Galois action on $E_{2^r}$ theories. From the twists, it is clear that the defining number field is $\mathbb{Q}(\xi_{2^r})$ with Galois group $\mathbb{Z}^\times_{2^r} = \{\text{all odd integers} < 2^r\}$. Under a Galois action corresponding to $q$, we have

$$\theta_{(m,n)} = e^{\frac{2\pi i m n}{2^r}} \to e^{\frac{2\pi i q m n}{2^r}} = e^{\frac{2\pi i (qm)n}{2^r}} = \theta_{(qm \text{ mod } 2^r, n)} \;. \tag{2.21}$$

So Galois conjugation with respect to any $q$ can be interpreted as a permutation of the anyons given by $(m,n) \to (qm \text{ mod } 2^r, n)$. In fact, this is an automorphism of the fusion rules, $\mathbb{Z}_{2^r} \times \mathbb{Z}_{2^r}$. Therefore, we see that the Galois conjugate of $E_{2^r}$ corresponds to a dual description of the same theory.

The Galois invariance of $E_{2^r}$ can also be deduced from the existence of a Lagrangian subcategory. To understand this statement, first note that, given an abelian group $G$, we can construct an abelian TQFT with fusion rules $G \times \hat{G}$. Here the anyons are labelled by $(g, \chi)$ where $g \in G$, and $\chi \in \hat{G}$ is a character of an irreducible representation belonging to the character group $\hat{G}$ of $G$. The twist of the anyon $(g, \chi)$ is $\theta_{(g,\chi)} = \chi(g)$. In fact, this construction gives the untwisted discrete gauge theory based on the abelian group $G$. Indeed, the $E_{2^r}$ family of prime abelian TQFTs are untwisted $\mathbb{Z}_{2^r}$ discrete gauge theories.

Now, the Lagrangian subcategory arises as follows: we have the anyons $(0, g)$ for any $g \in \mathbb{Z}_{2^r}$ which are all bosons. These form a subcategory of $E_{2^r}$ equivalent to the symmetric tensor category $\text{Rep}(\mathbb{Z}_{2^r})$ (a symmetric subcategory is characterized by completely trivial braiding). Moreover, note that $\dim(\text{Rep}(\mathbb{Z}_{2^r}))^2 = \dim(E_{2^r})$ (a subcategory of bosons satisfying this constraint is called a Lagrangian subcategory). A Galois conjugation of $E_{2^r}$ must result in a prime TQFT with a $\text{Rep}(\mathbb{Z}_{2^r})$ Lagrangian subcategory. However, the $E_{2^r}$ TQFTs are the only prime abelian TQFTs with a $\text{Rep}(\mathbb{Z}_{2^r})$ Lagrangian subcategory. Hence, $E_{2^r}$ TQFTs are mapped to themselves under Galois conjugation (i.e., they are unitary Galois fixed point TQFTs).

It is clear that $F_{2^r}$ theories are also unitary Galois fixed point TQFTs. This invariance follows from the fact that the only possibility for $F_{2^r}$ to transform to another prime theory is for it to get transformed into $E_{2^r}$ theory. However, $E_{2^r}$ and $F_{2^r}$ have different numbers of bosons. Indeed, an anyon, $(m, n)$, of $F_{2^r}$ theory is a boson if and only if it satisfies

$$\theta_{(m,n)} = 1 \implies m^2 + n^2 + mn = 0 \mod 2^r . \tag{2.22}$$

It is clear that if $m^2 = 0 \mod 2^r$, then $(0, m)$ and $(m, 0)$ are bosons. In fact, $(m, n)$ is a boson in $F_{2^r}$ if and only if $m^2 = 0 \mod 2^r$ and $n^2 = 0 \mod 2^r$. Note that for $(m, n)$ to be a boson, both $m$ and $n$ should be even. Let $m = 2m_1$ and $n = 2n_1$ for some integers $m_1, n_1$. If $(m, n)$ is boson, then

$$m^2 + n^2 + mn = 0 \mod 2^r \implies m_1^2 + n_1^2 + m_1 n_1 = 0 \mod 2^{r-2} . \tag{2.23}$$

This constraint is satisfied only if $m_1$ and $n_1$ are even. Therefore, we can choose $m_1 = 2m_2$ and $n_1 = 2n_2$ for some integers $m_2, n_2$. Iterating this process, we find that both $m^2$ and $n^2$ should be multiples of $2^r$. Given a boson $(m, n)$ of $F_{2^r}$ TQFT, note that it is also a boson of $E_{2^r}$ theory, since $mn = \mod 2^r$. However, there are clearly more bosons in $E_{2^r}$ than in $F_{2^r}$. For example, $(0, m)$ for any $m$ is a boson in $E_{2^r}$ TQFT while this is true for $F_{2^r}$ only if $m^2 = 0 \mod 2^r$. Since Galois conjugations preserve the number of bosons, $F_{2^r}$ cannot transform into an $E_{2^r}$ theory. Note that a boson $(m, n)$ in $F_{2^r}$ theory has order (under fusion) strictly less than $2^r$. Hence, $F_{2^r}$ does not have a $\text{Rep}(\mathbb{Z}_{2^r})$ Lagrangian subcategory.

Therefore, we find that $E_{2^r}$ and $F_{2^r}$ are the only prime abelian TQFTs which are invariant under Galois conjugation. We found that the Galois invariance of these theories can be explained using their bosonic substructure. Discrete gauge theories, generalizing the abelian $E_{2^r}$ cases, are another class of TQFTs largely determined by their bosonic substructure. In the next section, we will explore Galois actions on these theories. We will find that, similarly to the $E_{2^r}$ and $F_{2^r}$ prime abelian TQFTs, the transformation of discrete gauge theories under Galois conjugation is heavily constrained by the presence of certain bosons.

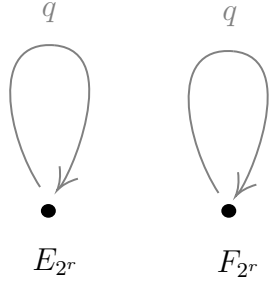

**Fig. 4:** The $E_{2^r}$ and $F_{2^r}$ families of prime abelian TQFTs are invariant under Galois conjugation.

*2.3. Discrete Gauge Theories*

Since Galois conjugation fixes rational numbers, it is clear that the space of integral MTCs (i.e., theories whose anyons all have integer quantum dimensions) is closed under it. An important class of integral MTCs are (twisted) discrete gauge theories (see [32] for a recent discussion of these theories, their subcategory structure, and their fusion rules). Since there are integral MTCs that are not (twisted) discrete gauge theories [33], we might naively imagine that these theories mix with discrete gauge theories under Galois conjugation. We will argue below that this is not the case and that the space of discrete gauge theories is therefore closed under Galois conjugation.

Before discussing discrete gauge theories, let us recall some notions which will be useful for our discussion. Two anyons $a$ and $b$ are said to centralize each other if $S_{ab} = \frac{1}{\mathcal{D}} d_a d_b$. This is the statement that the braiding between $a$ and $b$ is trivial (the Hopf link can be replaced by two disjoint circles labelled by $a$ and $b$). This notion can be used to define the centralizer for a fusion subcategory $D$ as the fusion subcategory, $D'$, where any $b \in D'$ centralizes any $a \in D$. It is clear that the fusion subcategory $D$ is symmetric (i.e., has completely trivial braiding) if and only if $D \subseteq D'$. A fusion subcategory is called isotropic if all its anyons are bosons. An isotropic subcategory $D \subset C$ is called Lagrangian if $D' = D$, or equivalently $\dim(D)^2 = \dim(C)$.

A twisted discrete gauge theory, $\mathcal{Z}(\text{Vec}_G^\omega)$, with Dijkgraaf-Witten twist, $\omega \in H^3(G, U(1))$, has $\text{Rep}(G)$ as a fusion subcategory. $\text{Rep}(G)$ has the irreducible representations of $G$ as its simple objects, the representation semi-ring of $G$ as the fusion rules, and $F$ symbols given by the $6j$ symbols. $\text{Rep}(G)$ is a symmetric fusion subcategory where we have

$$R_{ab}^c R_{ba}^c = 1; \quad S_{ab} = \frac{1}{\mathcal{D}} d_a d_b; \quad \theta_a = 1 \quad \forall \ a, b, c \in \text{Rep}(G) \ . \tag{2.24}$$

In fact, if a fusion subcategory only has bosons in it, it is a symmetric fusion category and is gauge equivalent to $\text{Rep}(H)$ for some group $H$ [34]. An important property of $\text{Rep}(G)$ that will be crucial for our discussion is that it is Lagrangian. We have $\dim(\text{Rep}(G))^2 = |G|^2 = \dim(\mathcal{Z}(\text{Vec}_G^\omega))$. It is clear that under Galois conjugation of $\mathcal{Z}(\text{Vec}_G^\omega)$, $\text{Rep}(G)$ will transform into a symmetric fusion category. Given anyons $a$ and $b$ in a modular tensor category $C$ that centralize each other, the corresponding anyons in the Galois conjugate theory also centralize each other. Therefore, Galois conjugation of a discrete gauge theory results in a modular tensor category that has a Lagrangian subcategory. Hence, we have the following result:

**Lemma 2.4:** The space of twisted discrete gauge theories is closed under Galois conjugation.

This result holds because an MTC corresponds to a discrete gauge theory if and only if it has a Lagrangian subcategory [29]. The invertibility of Galois conjugation then implies that the Galois conjugation of a non-discrete gauge theory (for example, those originating from quantum groups) should result in a non-discrete gauge theory.[21] To determine how Galois conjugation affects the gauge group and twist of a discrete gauge theory, we have to study the behavior of $\text{Rep}(G)$ under Galois action.

*2.3.1. Galois Conjugation of Rep(G)*

The discussion in the previous subsection shows that the Galois action on $\text{Rep}(G)$ results in a symmetric tensor category, $\text{Rep}(H)$, for some finite group $H$. We will find that $G \cong H$ follows from the algebraic nature of the Tannaka-Krein reconstruction theorem.

In Tannaka-Krein reconstruction, the group is reconstructed from a subgroup of the group of endomorphisms of the vector spaces on which the representations act. More

---

[21] There are some discrete gauge theories that have a dual description in terms of a Chern-Simons theory with a continuous Lie gauge group. For example, the Toric code, which is a $\mathbb{Z}_2$ discrete gauge theory, can also be described as $\text{Spin}(16)_1$ Chern-Simons theory. By a non-discrete gauge theory, we mean a TQFT which is not equivalent to a (twisted) discrete gauge theory.

precisely, consider a fiber functor (i.e., a monoidal functor to Vec)

$$F : \text{Rep}(G) \to \text{Vec} \ , \tag{2.25}$$

$$\pi \to V_{\pi'} \ , \tag{2.26}$$

where $\pi$ is a representation of $G$ (in general, reducible). This map can be thought of as forgetting all information about the category $\text{Rep}(G)$ except for the vector spaces on which the irreducible representations act. $F$ is a monoidal functor. That is, there exists a natural transformation

$$\mu_{\pi,\pi'} : F(\pi) \otimes F(\pi') \to F(\pi \otimes \pi') \ \forall \ \pi, \pi' \in \text{Rep}(G) \ , \tag{2.27}$$

which is consistent with associativity of $\text{Rep}(G)$. $\mu_{\pi,\pi'}$ are simply the basis transformation matrices between the isomorphic vector spaces $V_\pi \otimes V_{\pi'}$ and $V_{\pi \otimes \pi'}$. The former has a natural tensor product basis, and the latter has a basis given by the decomposition of $\pi \otimes \pi'$ into irreducible representations. In other words, $\mu_{\pi,\pi'}$ are determined by the 3j symbols.

Recall that two functors can be related by a natural transformation. A natural automorphism is a natural isomorphism between the same functor; it can be seen as a symmetry of the functor. Automorphisms of the functor, $F$, defined above are given by a collection of invertible matrices, $\{U_\pi\}$, that act on the vector spaces, $V_\pi$. These actions should commute with any intertwiners between $V_\pi$ and $V_{\pi'}$. This requirement implies that $U_\pi$ is completely specified by its action on the vector spaces of the irreps of $G$. Therefore, the symmetry group of the monoidal functor $F$ is

$$\text{Aut}(F) = \prod_i GL(V_{\pi_i}) \ , \tag{2.28}$$

where $V_{\pi_i}$ are the vector spaces corresponding to the $\pi_i$ irreps of $G$.

The finite group $G$ can be reconstructed from $\text{Rep}(G)$ by picking a particular subgroup of $\text{Aut}(F)$. This subgroup is specified by the following extra condition on the $\{U_\pi\}$

$$\begin{array}{ccc} F(\pi \otimes \pi') & \xrightarrow{U_{\pi \otimes \pi'}} & F(\pi \otimes \pi') \\ \downarrow{\mu_{\pi,\pi'}} & & \downarrow{\mu_{\pi,\pi'}} \\ F(\pi) \otimes F(\pi') & \xrightarrow{U_\pi \otimes U_{\pi'}} & F(\pi) \otimes F(\pi') \end{array} \ ,$$

which is same as the constraint

$$\begin{array}{ccc} \bigoplus_i F(\pi_i) & \xrightarrow{\bigoplus_i U_{\pi_i}} & \bigoplus_i F(\pi_i) \\ \downarrow{\mu_{\pi,\pi'}} & & \downarrow{\mu_{\pi,\pi'}} \\ F(\pi) \otimes F(\pi') & \xrightarrow{U_\pi \otimes U_{\pi'}} & F(\pi) \otimes F(\pi') \end{array} \ , \tag{2.29}$$

where $\pi_i$ are the irreducible representations into which the representation $\pi \otimes \pi'$ decomposes. The matrices $\{U_\pi\}$ which satisfy this constraint are called, "tensor-preserving automorphisms." We can also define a conjugation operation on $U_\pi$ given by

$$\overline{U}_\pi(x) := \overline{U_{\overline{\pi}}(\overline{x})} \;, \tag{2.30}$$

where $\overline{\pi}$ is the conjugate representation of $\pi$, $x \in V_\pi$ and $\bar{x} \in V_{\overline{\pi}}$. Let $\mathrm{Aut}^\otimes(F) \subset \mathrm{Aut}(F)$ be the set of self-conjugate ($\overline{U}_\pi = U_\pi$) tensor-preserving automorphisms. It is clear that, given some element $g \in G$, there is a canonical map

$$L: \quad G \to \mathrm{Aut}^\otimes(F) \;, \tag{2.31}$$
$$g \to U^{(g)} \;, \tag{2.32}$$

where $U^{(g)}$ acts on the vector space $V_\pi$ through $\pi(g)$. The non-trivial result of Tannaka-Krein is that the canonical map $L$ defined above is in fact an isomorphism [35]. Therefore, the automorphisms of the fiber functor $F$, along with the tensor-preserving and self-conjugation constraints, give us all the representations of the group. We can then reconstruct the group.

Consider $\mathrm{Rep}(G)$ defined over some finite field extension $K_{\mathrm{Rep}(G)}$.[22] In other words, the $3j, 6j$ symbols, and the R-matrices belong to $K_{\mathrm{Rep}(G)}$. In particular, the matrices $\mu_{\pi,\pi'}$ belong to $K_{\mathrm{Rep}(G)}$. Therefore, (2.29) gives a set of polynomial constraints on the elements of the $U_\pi$ matrices. The coefficients of the polynomial belong to the field $K_{\mathrm{Rep}(G)}$. Therefore, these polynomials are defined over $\bar{\mathbb{Q}}$, and hence there exists a solution belonging to a number field (up to gauge choices). Therefore, every Galois action with respect to some element of $\mathrm{Gal}(K_{\mathrm{Rep}(G)})$ induces a Galois action on $U_\pi$. If

$$U_\pi^{(g)} U_\pi^{(h)} = U_\pi^{(k)} \;, \tag{2.33}$$

for some $g, h, k$, then this relation does not change under a Galois action on $U_\pi^{(g)}$. As a result, the group $\mathrm{Aut}^\otimes(G)$ is invariant under Galois action. Hence, the representation category of a group is invariant under Galois conjugation.

In a discrete gauge theory $\mathcal{Z}(\mathrm{Vec}_G^\omega)$, $\mathrm{Rep}(G)$ is a Lagrangian subcategory. Moreover, since there is a gauge in which the data of $\mathcal{Z}(\mathrm{Vec}_G^\omega)$ is in a finite field extension, the data of the subcategory $\mathrm{Rep}(G)$ is in the same finite field extension. Therefore, our discussion above applies, and we find that the gauge group of a discrete gauge theory is invariant under Galois conjugation.

---

[22]In fact, from Brauer's Theorem [36], we can choose $K_{\mathrm{Rep}(G)}$ to be cyclotomic, but the exact nature of the field won't be important for our argument.

All that is left to study is how the Galois group acts on the Dijkgraaf-Witten twist. The cyclotomic field containing the elements of the $S$ and $T$ matrices of the discrete gauge theory, $\mathcal{Z}(\text{Vec}_G^\omega)$, is $\mathbb{Q}(\xi_{ne(G)})$, where $n$ is the order of the 3-cocycle $\omega \in H^3(G, U(1))$, $e(G)$ is the exponent of the group $G$ [37], and $\xi_{ne(G)}$ is a primitive $ne(G)^{\text{th}}$ root of unity. In particular, the 3-cocycle $\omega$ is contained in this cyclotomic field. Suppose $K_C$ is the defining number field containing the full data of $\mathcal{Z}(\text{Vec}_G^\omega)$ in some gauge. $\mathbb{Q}(\xi_n)$ is a cyclotomic subfield of $K_C$. If $q \in \text{Gal}(K_C)$ acts on $\mathcal{Z}(\text{Vec}_G^\omega)$, then it acts on the 3-cocycle $\omega$ as $q|_{\mathbb{Q}(\xi_n)}(\omega)$. Moreover, since $K_C$ and $\mathbb{Q}(\xi_n)$ are Galois extensions, for every Galois action $q' \in \text{Gal}(\mathbb{Q}(\xi_n))$ there exists a $q \in \text{Gal}(K_C)$ such that $q|_{\mathbb{Q}(\xi_n)} = q'$. Therefore, any Galois conjugation of the MTC $\mathcal{Z}(\text{Vec}_G^\omega)$ acts as a Galois conjugation on the 3-cocycle $\omega$. We get the following results:

**Theorem 2.5:** Let $K_C$ be the number field containing the MTC data of $\mathcal{Z}(\text{Vec}_G^\omega)$. Galois conjugation with respect to $q \in \text{Gal}(K_C)$ results in the discrete gauge theory $\mathcal{Z}(\text{Vec}_G^{q|_{\mathbb{Q}(\xi_n)}(\omega)})$.

**Corollary 2.6:** The untwisted discrete gauge theory $\mathcal{Z}(\text{Vec}_G)$ is invariant under Galois conjugation.

**Corollary 2.7:** Every Galois conjugation of $\mathcal{Z}(\text{Vec}_G^\omega)$ acts as a Galois conjugation on the gapped boundary described by $\text{Vec}_G^\omega$.

Suppose we have the fusion category $\text{Vec}_G^\omega$. The cyclotomic field containing the MTC data of this category is $\mathbb{Q}(\xi_n)$, where $n$ is the order of $\omega \in H^3(G, U(1))$. Therefore, after a Galois conjugation, we get $\text{Vec}_G^{\omega^q}$ for some $q$ co-prime to $n$. Taking the Drinfeld center before and after the Galois conjugation gives us the discrete gauge theories, $\mathcal{Z}(\text{Vec}_G^\omega)$ and $\mathcal{Z}(\text{Vec}_G^{\omega^q})$, respectively. Since these discrete gauge theories are related by a Galois conjugation, we see that Galois conjugation commutes with taking the Drinfeld center of $\text{Vec}_G^\omega$. In Section 3, we will generalize this result to Drinfeld centers of general spherical fusion categories.

Note that a TQFT can have multiple Lagrangian subcategories. In particular, if a TQFT has Lagrangian subcategories $\text{Rep}(G)$ and $\text{Rep}(H)$, where $G$ is not isomorphic to $H$, then it can be seen as a discrete gauge theory based on the gauge group $G$ or the gauge group $H$. That is, the gauge group is not duality invariant.[23]

---

[23]This is different from $3+1$D discrete gauge theories whose gauge group is invariant under such dualities.

Therefore, Galois invariance of the gauge group of the discrete gauge theory is more precisely stated as follows: Given a discrete gauge theory $\mathcal{Z}(\text{Vec}_G^\omega)$, all of its Galois conjugates are $G$ gauge theories up to dualities. In fact, the dualities of a discrete gauge theory are determined by the Lagrangian subcategories in the theory and they have been classified in [38]. Since the number of Lagrangian subcategories does not change under Galois conjugation, the duality structure of Galois conjugate discrete gauge theories is the same.

While Galois conjugation of a discrete gauge theory $\mathcal{Z}(\text{Vec}_G^\omega)$ may act non-trivially on the 3-cocycle $\omega$, the resulting discrete gauge theory is not guaranteed to be distinct. This is because for a given group $G$, distinct 3-cocycles in $H^3(G, U(1))$ can give the same TQFT. In the next subsection, we will explore how this happens for discrete gauge theories with abelian gauge groups.

### 2.3.2. Discrete gauge theories with abelian gauge group

In this subsection we will study (twisted) discrete gauge theories with abelian gauge groups more carefully (we already encountered many of these theories when we discussed abelian TQFTs previously). This discussion will help us to understand Galois action on the 3-cocycle $\omega$ implied by Theorem 2.5 more explicitly.

Note that discrete gauge theories based on abelian gauge groups need not be abelian. Indeed, the quantum dimension of an anyon $([g], \pi_g^\omega)$ in a general discrete gauge theory with gauge group $G$ and 3-cocycle $\omega$ is

$$d_{([g],\pi_g^\omega)} = |[g]|\dim(\pi_g^\omega) \;, \tag{2.34}$$

where $[g]$ is a conjugacy class in $G$, and $\pi_g^\omega$ is a projective representation of the centralizer of $g$, say $N_g$, determined by the 2-cocycle

$$\gamma_g(h, k) = \frac{\omega(g, h, k)\omega(h, k, g)}{\omega(h, g, k)} \;. \tag{2.35}$$

In an abelian discrete gauge theory, all anyons have quantum dimension 1. Therefore, all conjugacy classes should have only a single element. Hence, the gauge group $G$ should be abelian. Therefore, the centralizer of each element is $G$ itself. Moreover, we also require the representation $\pi_g^\omega$ to be 1-dimensional. Since projective representations are necessarily

---

This is because in $3+1$D *all* line operators braid trivially with each other, and they are described by $\text{Rep}(G)$, where $G$ is the gauge group of the $3+1$D discrete gauge theory. If there were a dual description based on a gauge group $H$, then the line operators would be described by $\text{Rep}(H)$. However, $\text{Rep}(G) \cong \text{Rep}(H)$ if and only if $G \cong H$ from Tannaka-Krein reconstruction.

higher dimensional, for an abelian discrete gauge theory, the 3-cocycle $\omega$ should be such that $\gamma_g(h,k) \in H^2(G, U(1))$ is trivial for all $g \in G$. Therefore, a discrete gauge theory is abelian if and only if the gauge group is abelian with CT (cohomologically trivial) twisting [37]. This is a stronger constraint than having abelian gauge groups.

For an abelian group $G$, a general 3-cocycle $\omega \in H^3(G, U(1))$ is generated by the following 3-cocycles [39]

$$\omega^{(i)}(\vec{g}, \vec{h}, \vec{k}) = e^{\frac{2\pi i p^{(i)}}{n_i^2}(g_i(h_i + k_i - h_i + k_i \bmod n_i))}, 1 \leq i \leq n, \tag{2.36}$$

$$\omega^{(i,j)}(\vec{g}, \vec{h}, \vec{k}) = e^{\frac{2\pi i p^{(i,j)}}{n_i n_j}(g_i(h_j + k_j - h_j + k_j \bmod n_j))}, 1 \leq i < j \leq n, \tag{2.37}$$

$$\omega^{(i,j,l)}(\vec{g}, \vec{h}, \vec{k}) = e^{\frac{2\pi i p^{(i,j,l)}}{\gcd(n_i, n_j, n_l)}(g_i h_j k_l)}, 1 \leq i < j < l \leq n, \tag{2.38}$$

where $G \cong \mathbb{Z}_{n_1} \otimes \cdots \otimes \mathbb{Z}_{n_N}$, $\vec{g}, \vec{h}, \vec{k} \in G$. Here $p^{(i)}$ is an integer defined modulo $n_i$, $p^{(i,j)}$ is an integer defined modulo $\gcd(n_i, n_j)$, and $p^{(i,j,l)}$ is an integer defined modulo $\gcd(n_i, n_j, n_l)$. We will refer to these as Type I,II, and III generators respectively.

Consider the action of $\alpha \in \text{Aut}(G)$ on the group $G$. This induces an action on $\omega(\vec{g}, \vec{h}, \vec{k})$ as $\omega(\vec{g}, \vec{h}, \vec{k}) \to \omega(\alpha(\vec{g}), \alpha(\vec{h}), \alpha(\vec{k}))$. Suppose $\alpha$ acts on the group elements through the $N \times N$ matrix $M$. We have $g_i = \sum_a M_{ia} g_a$. Using the explicit expressions for the 3-cocycle generators above, we get

$$\omega^{(i)}(\alpha(\vec{g}), \alpha(\vec{h}), \alpha(\vec{k})) = e^{\frac{2\pi i p^{(i)}}{n_i^2}(\sum_{a,b} M_{ia} M_{ib} g_a (h_b + k_b - h_b + k_b \bmod n_i))} = \prod_{a,b} (\omega^{(a,b)}(\vec{g}, \vec{h}, \vec{k}))^{M_{ia} M_{ib}}. \tag{2.39}$$

Similarly, we get

$$\omega^{(i,j)}(\alpha(\vec{g}), \alpha(\vec{h}), \alpha(\vec{k})) = \prod_{a,b} (\omega^{(a,b)}(\vec{g}, \vec{h}, \vec{k}))^{M_{ia} M_{jb}}, \tag{2.40}$$

$$\omega^{(i,j,l)}(\alpha(\vec{g}), \alpha(\vec{h}), \alpha(\vec{k})) = \prod_{a,b,c} (\omega^{(a,b,c)}(\vec{g}, \vec{h}, \vec{k}))^{M_{ia} M_{jb} M_{lc}}. \tag{2.41}$$

Note that the discrete gauge theory $\mathcal{Z}(\text{Vec}_G^\omega)$ is uniquely specifed by the fusion category $\text{Vec}_G^\omega$ [40]. Since $\text{Vec}_G^\omega$ and $\text{Vec}_G^{\alpha(\omega)}$ are equivalent as fusion categories when $\alpha \in \text{Aut}(G)$, the discrete gauge theories $\mathcal{Z}(\text{Vec}_G^\omega)$ and $\mathcal{Z}(\text{Vec}_G^{\alpha(\omega)})$ are also equivalent.

The modular data of a discrete gauge theory lies in the cyclotomic field $\mathbb{Q}(\xi_{ne})$ where $n$ is the order of $\omega$ and $e$ is the exponent of $G$. Consider a Galois action corresponding to some $q \in \mathbb{Z}_{ne}^\times$. Then the 3-cocycle generators transform as

$$\begin{aligned}
\omega^{(i)}(\vec{g}, \vec{h}, \vec{k}) &\to (\omega^{(i)}(\vec{g}, \vec{h}, \vec{k}))^q \ , \\
\omega^{(i,j)}(\vec{g}, \vec{h}, \vec{k}) &\to (\omega^{(i,j)}(\vec{g}, \vec{h}, \vec{k}))^q \ , \\
\omega^{(i,j,l)}(\vec{g}, \vec{h}, \vec{k}) &\to (\omega^{(i,j,l)}(\vec{g}, \vec{h}, \vec{k}))^q \ .
\end{aligned} \quad (2.42)$$

Consider a general 3-cocycle $\omega$

$$\omega = \prod_{a=1}^{N_I} \omega^{(i_a)} \prod_{b=1}^{N_{II}} \omega^{(j_b, l_b)} \prod_{c=1}^{N_{III}} \omega^{(m_c, r_c, o_c)} \ , \quad (2.43)$$

with $N_I$ type I generators, $N_{II}$ type II generators, and $N_{III}$ type III generators. Here $i_a, j_b, l_b, m_c, r_c, o_c$ are all integers valued in $\{1, ..., N\}$. Without loss of generality we can assume that $i_a$ is distinct for each $a$ in the product (and similarly for $(j_b, l_b)$ and $(m_c, r_c, o_c)$).

This Galois action coincides with the transformation of the 3-cocycle under an automorphism of $G$ if the following conditions are satisfied

$$M_{i_a x} = 0 \text{ for } i_a \neq x \text{ and } M_{i_a i_a}^2 = q \bmod n_{i_a} \ \forall a \ , \quad (2.44)$$

$$M_{j_b x} M_{l_b y} = 0 \text{ for } j_b \neq x \text{ or } l_b \neq y \text{ and } M_{j_b j_b} M_{l_b l_b} = q \bmod n_{j_b} n_{l_b} \ \forall b \ , \quad (2.45)$$

$$M_{m_c x} M_{r_c y} M_{o_c z} = 0 \text{ for } m_c \neq x \text{ or } r_c \neq y \text{ or } o_c \neq z \text{ and} \quad (2.46)$$

$$M_{m_c m_c} M_{r_c r_c} M_{o_c o_c} = q \bmod \gcd(n_{m_c}, n_{r_c}, n_{o_c}) \ . \quad (2.47)$$

If these conditions are satisfied, the Galois action is an automorphism of the gauge group $G$, and the discrete gauge theory is Galois invariant. However, all Galois actions (2.42) need not correspond to automorphisms of the gauge group.

**Example:** Consider the $\mathbb{Z}_N$ discrete gauge theory with some twist $\omega \in H^3(\mathbb{Z}_N, U(1))$. In this case, $\omega$ has the explicit expression

$$\omega(g, h, k) = e^{\frac{2\pi i p}{N^2}(g(h+k-h+k \bmod N))} \ . \quad (2.48)$$

Since $H^2(\mathbb{Z}_N, U(1))$ is trivial, $\mathbb{Z}_N$ discrete gauge theory for any twist $\omega$ is abelian. Consider the action of $\alpha \in \text{Aut}(\mathbb{Z}_N) \cong \mathbb{Z}_N^\times$ given by $g \to \alpha g \bmod N$, $g \in \mathbb{Z}_N$. Then $\omega$ transforms as

$$\omega_p(g, h, k) \to \omega_p(\alpha g, \alpha h, \alpha k) = \omega_p(g, h, k)^{\alpha^2} = \omega_{\alpha^2 p}(g, h, k) \ . \quad (2.49)$$

A Galois conjugation with respect to some $q$ coprime to $N$ transforms the 3-cocycle as

$$\omega_p(g, h, k) \to \omega_p(g, h, k)^q = \omega_{qp}(g, h, k) \ . \quad (2.50)$$

Therefore, a Galois conjugation w.r.t. $q$ is an automorphism of the gauge group $G$ only if $\alpha^2 = q \mod N$.

As a particularly concrete example, consider $N = 5$ and the 3-cocycle with $p = 1$. Then $2 \mod 5 \neq \alpha^2$ for any $\alpha \in \text{Aut}(\mathbb{Z}_5) \cong \mathbb{Z}_5^\times = \{1, 2, 3, 4\}$. Therefore, Galois conjugation w.r.t. to 2 takes us from the discrete gauge theory $\mathcal{Z}(\text{Vec}_{\mathbb{Z}_5}^{\omega_1})$ to $\mathcal{Z}(\text{Vec}_{\mathbb{Z}_5}^{\omega_2})$. In fact, $\mathcal{Z}(\text{Vec}_{\mathbb{Z}_5}^{\omega_1})$ is the prime abelian theory $B_{25}$ and $\mathcal{Z}(\text{Vec}_{\mathbb{Z}_5}^{\omega_2})$ is the prime abelian theory $A_{25}$. From our discussion in section 2.2, we know that there are non-trivial Galois conjugations which take us between these theories, and we know that our discussion in this section is consistent.

Note that while automorphisms of the group naturally lead to equivalence of discrete gauge theories based on different twists, this is not the only way in which equivalences arise. Even after taking the automorphisms of the gauge group $G$ and its action on the 3-cocycle into account, labelling discrete gauge theories by the gauge group and the orbits of the automorphism group action on $H^3(G, U(1))$ is not faithful. For example, consider the group $\mathbb{Z}_2 \times D_8$. There exists two 3-cocycles for this group, not related by group automorphisms, which give the same discrete gauge theory [41].[24]

### 2.4. Weakly Integral Modular Categories

Until now, we studied theories that only have integer quantum dimensions. We saw that in these theories any Galois action on a unitary TQFT results in a unitary TQFT. Now we look at TQFTs described by weakly integral MTCs. These theories have quantum dimensions of the form $d_a = \sqrt{n_a}$, for some integer $n_a$. As a result such MTCs have Galois conjugations that take a unitary TQFT to a non-unitary one. Before looking at the general case, let us consider the specific case of the Ising model and its Galois conjugates.

#### 2.4.1. The Ising$^{(\nu)}$ Model

The Ising$^{(\nu)}$ family of theories is specified by the following data. There are three anyons $\{I, \sigma, \psi\}$ satisfying the fusion rules

$$\psi \otimes \psi = I, \ \sigma \otimes \sigma = I + \psi , \tag{2.51}$$

where $I$ is an boson, $\psi$ is a fermion, and $\sigma$ is an anyon with twist $e^{\frac{2\pi i \nu}{16}}$. They have quantum dimensions $d_I = 1, d_\psi = 1, d_\sigma = \sqrt{2}$. Here $\nu$ is an odd integer modulo 16. The Ising model corresponds to $\nu = 1$. Note that the $\nu$ parameter here only classifies the unitary MTCs with the same fusion rules as the Ising model.

---

[24]In [42], the authors conjecture that equivalence classes of $3+1$D discrete gauge theories based on gauge group $G$ are classified by $H^4(G, U(1))$ up to group automorphisms. We note that the $2+1$D version of this conjecture is not true because of this counter-example.

The full MTC data belongs to the cyclotomic field $\mathbb{Q}(\xi_{16})$. Therefore, we have the Galois group $\mathbb{Z}_{16}^\times = \{1, 3, 5, 7, 9, 11, 13, 15\} \cong \mathbb{Z}_4 \times \mathbb{Z}_2$. If we start with any of the above family of Ising$^{(\nu)}$ models, a unitarity preserving Galois action should not change the quantum dimension of $\sigma$ to $-\sqrt{2}$. Therefore, the unitarity preserving Galois actions correspond to $q = 1, 7, 9, 15$. These form the Klein four-group. Under these Galois actions, the Ising$^{(\nu)}$ family of models transform as

$$\text{Ising}^{(\nu)} \to \text{Ising}^{(q\nu \bmod 16)} . \tag{2.52}$$

### 2.4.2. Metaplectic Modular Categories

Let us now discuss unitary Galois orbits in the more general family of metaplectic modular categories (of which Ising$^{(\nu)}$ are examples). These are categories for which the fusion rules are the same as those of the Spin$(N)_2$ theories. In general, metaplectic categories have strictly weakly integral anyons. However, certain metaplectic categories are integral (e.g., Spin$(8)_2$). As shown in [43], integral metaplectic categories are group theoretical; hence, they belong to the class of theories discussed in the previous section.

Therefore, we can focus on strictly weakly integral metaplectic categories. Even though it is an extremely hard problem to solve the Pentagon and Hexagon equations for large rank theories, amazingly, for metaplectic categories, the MTC data can be found. Moreover, they play an important role (along with discrete gauge theories) in the classification of weakly integral categories.

By examining the explicit expressions for the $F$ and $R$ matrices for Spin$(N)_2$ metaplectic modular categories for odd $N$ in [44], we find the following cyclotomic field extensions

$$K_{\tilde{S},T} = \mathbb{Q}(\xi_{\text{lcm}(2N,8)}) , \tag{2.53}$$

$$K_R = \mathbb{Q}(\xi_{\text{lcm}(2N,16)}) , \tag{2.54}$$

$$K_{F,R} = \mathbb{Q}(\xi_{\text{lcm}(2N,16)}) . \tag{2.55}$$

Since the metaplectic modular categories are multiplicity free, the $R$ matrices are all phases. Some are $2N^{\text{th}}$ roots of unity while others are $16^{\text{th}}$ roots of unity. The F-matrices consists of $2N^{\text{th}}$ roots of unity, $\sqrt{2}$, and $\sqrt{N}$. Note that $\sqrt{N}$ belongs to the cyclotomic field $\mathbb{Q}(\xi_{\text{lcm}(N,4)}) \subset \mathbb{Q}(\xi_{\text{lcm}(2N,16)})$. Therefore, the $F$ and $R$ symbols belong to the cyclotomic field $\mathbb{Q}(\xi_{\text{lcm}(2N,16)})$.

The $F$ matrices given in [44] are real and unitary. As a result, Galois conjugation is guaranteed to result in unitary $F$ matrices. Also, since the $R$ matrices are phases, they remain unitary under Galois conjugation. However, the quantum dimensions need not remain positive. Therefore, the resulting theory need not be unitary. This is a generalization of what happens in Ising$^{(\nu)}$ models that we discussed above. However, we know from [26] that braided fusion categories with unitary $F$ and $R$ symbols have a unique spherical structure which makes it a unitary MTC. Therefore, even though Galois conjugation of a metaplectic theory need not land us on a unitary TQFT, we can always choose a spherical structure to make the theory unitary (we must also choose $\mathcal{D} > 0$). This statement is in fact true for any weakly group theoretical modular tensor category from the following result

**Theorem 2.8 [45]:** Every weakly group theoretical fusion category is unitary.

Therefore, any Galois conjugate of a given unitary weakly group-theoretical modular tensor category can be made unitary by the choice of a unique spherical structure. All known weakly integral categories are weakly group theoretical. If all weakly integral categories can be shown to be weakly group theoretical, then any Galois conjugate of a unitary weakly integral modular tensor category can be made unitary by the choice of a unique spherical structure.

Let us discuss another example of a metaplectic modular category that we will come back to in our further discussoins. The Spin(5)$_2$ Chern-Simons theory has 6 anyons labelled by $\{1, \epsilon, \phi_1, \phi_2, \psi_+, \psi_-\}$ with quantum dimensions $\{1, 1, 2, 2, \sqrt{5}, \sqrt{5}\}$, respectively. The fusion rules are given by

$$\epsilon \otimes \epsilon = 1 \ , \quad \epsilon \otimes \phi_i = \phi_i \ , \quad \epsilon \otimes \psi_\pm = \psi_\mp \ , \quad \phi_i \otimes \phi_i = 1 \oplus \epsilon \oplus \phi_{\min(2i, 5-2i)} \ , \quad \phi_1 \otimes \phi_2 = \phi_1 \oplus \phi_2 \ ,$$

$$\phi_i \otimes \psi_\pm = \psi_\pm \oplus \psi_\mp \ , \quad \psi_\pm \otimes \psi_\pm = 1 \oplus \phi_1 \oplus \phi_2 \ , \quad \psi_\pm \otimes \psi_\mp = \epsilon + \phi_1 \oplus \phi_2 \ ,$$

where $i = 1, 2$. The twists of the anyons are

$$\theta_\epsilon = 1, \ \theta_{\phi_1} = e^{\frac{4\pi i}{5}}, \ \theta_{\phi_2} = e^{-\frac{4\pi i}{5}}, \ \theta_{\psi_\pm} = \pm i \ . \tag{2.56}$$

Therefore, the twists belong to the cyclotomic field $\mathbb{Q}(\xi_{20})$. All MTCs with the same fusion rules as Spin(5)$_2$ Chern-Simons theory can be distinguished using the $T$ matrix alone [44]. Therefore, we only need to consider the Galois action on the twists to study the Galois action on the whole theory. The Galois group acting on the twists is $\mathbb{Z}_{20}^\times = \{1, 3, 7, 9, 11, 13, 17, 19\}$. For unitary Galois orbits, we should consider Galois actions which leave $d_{\psi_\pm} = \sqrt{5}$ invariant. These are $\{1, 9, 11, 19\}$. Under the action of 9 we get the twists

$$\theta_\epsilon = 1 \ , \ \theta_{\phi_1} = e^{-\frac{4\pi i}{5}} \ , \ \theta_{\phi_2} = e^{\frac{4\pi i}{5}} \ , \ \theta_{\psi_\pm} = \pm i \ . \tag{2.57}$$

This theory is the same as Spin(5)$_2$ under the permutation of the anyons $\phi_1 \leftrightarrow \phi_2$. Under the action of 19 we get the twists

$$\theta_\epsilon = 1 \ , \ \theta_{\phi_1} = e^{-\frac{4\pi i}{5}} \ , \ \theta_{\phi_2} = e^{\frac{4\pi i}{5}} \ , \ \theta_{\psi_\pm} = \mp i \ . \tag{2.58}$$

Therefore, acting with 19 complex conjugates the theory. This Galois action can be inverted using the permutation of the anyons $\phi_1 \leftrightarrow \phi_2$ and $\psi_+ \leftrightarrow \psi_-$. This is because Spin(5)$_2$ Chern-Simons theory is time-reversal invariant. Under the action of 11 we get the twists

$$\theta_\epsilon = 1 \ , \ \theta_{\phi_1} = e^{\frac{4\pi i}{5}} \ , \ \theta_{\phi_2} = e^{-\frac{4\pi i}{5}} \ , \ \theta_{\psi_\pm} = \mp i \ . \tag{2.59}$$

It is clear that this theory is same as Spin(5)$_2$ because of the time-reversal symmetry and the symmetry of the fusion rules under $\phi_1 \leftrightarrow \phi_2$.

Therefore, we find that the Spin(5)$_2$ Chern-Simons theory is invariant under all unitarity preserving Galois actions (recall we fix $\mathcal{D} > 0$).

## 3. Gapped Boundaries and Galois Conjugation

In section 2.3, we found that Galois conjugation of the gapped boundary of a discrete gauge theory induces a Galois action on the bulk TQFT and vice-versa. In this section, we will explore this connection further. First, we will revisit discrete gauge theories using the classification of bosonic gapped boundaries. Then we will look at bosonic gapped boundaries of more general TQFTs by studying the properties of their Lagrangian algebras. Finally, we will discuss how Galois conjugation and taking the Drinfeld center of a spherical fusion category interact with each other. This will give us a general result relating the Galois action of the bosonic gapped boundary and the bulk TQFT.

### 3.1. Gapped boundaries of discrete gauge theories

An abelian TQFT is described by a so-called "pointed" MTC. As a fusion category, a pointed MTC is equivalent to Vec$_G^\omega$ for some abelian group $G$ and some $\omega \in H^3(G, U(1))$. The bosonic gapped boundaries of this theory correspond to Lagrangian subgroups of $G$ [46, 47]. A Lagrangian subgroup of $L \subset G$ is a subgroup such that the fusion subcategory Vec$_L^{\omega|_L}$ is Lagrangian. This discussion immediately implies that in order for an abelian theory to have bosonic gapped boundaries, it should necessarily originate from a discrete gauge theory. Since we argued that the number of Lagrangian subcategories is invariant under Galois conjugation, it is clear that Galois conjugate abelian theories have the same number of bosonic gapped boundaries.[25]

Now let us study a non-abelian discrete gauge theory, $\mathcal{Z}(\text{Vec}_G^\omega)$. The gapped boundaries of this theory are classified by the pair $(L, \eta)$, where $L$ is a subgroup up to conjugation of $G$ such that $\omega|_L$ is trivial in cohomology, and $\eta \in H^2(L, U(1))$ [49]. From our previous discussion, we know that the Galois conjugate of $\mathcal{Z}(\text{Vec}_G^\omega)$ is $\mathcal{Z}(\text{Vec}_G^{q|_{\mathbb{Q}(\xi_n)}(\omega)})$ for some $q \in \text{Gal}(K_C)$. Moreover, if $\omega|_L$ is cohomologically trivial, so is $(q|_{\mathbb{Q}(\xi_n)}(\omega))|_L$. Therefore, the number of gapped boundaries of Galois conjugate twisted discrete gauge theories is the same.

*3.2. Gapped boundaries of general TQFTs*

In a general TQFT described by a modular tensor category $C$, a gapped boundary corresponds to the condensation of a subset of anyons in $C$ which admits the structure of a Lagrangian algebra [47]. Therefore, in order to study the nature of bosonic gapped boundaries of Galois conjugate TQFTs, we have to study the behavior of a Lagrangian algebra under Galois conjugation. To that end, consider the following two theorems that characterize a Lagrangian algebra.

**Theorem 3.1 [47]:** $\mathcal{A}$ is a commutative algebra in a modular tensor category $C$ if and only if the object $\mathcal{A}$ decomposes as $\mathcal{A} = \oplus n_i a_i$ into simple objects $a_i \in C$ and $\theta_{a_i} = 1$ for all $i$ such that $n_i \neq 0$.

**Theorem 3.2 [47]:** A commutative connected algebra $\mathcal{A} = \oplus n_i a_i$ in a unitary modular tensor category $C$ with $\dim(\mathcal{A})^2 = \dim(C)$ is Lagrangian if and only if

$$n_i n_j \leq N_{ik}^k n_k \;, \tag{3.1}$$

for all $i, j, k$.

It is clear that if a set of anyons satisfies the constraints in theorem 3.1 and 3.2 in an MTC, then the same holds after a unitarity-preserving Galois conjugation. Hence, if a set of anyons form a Lagrangian algebra, then the same set of anyons form a Lagrangian algebra in the Galois conjugate unitary theory. Therefore, a set of condensable anyons remains condensable under unitarity preserving Galois conjugation.

---

[25]Bosonic TQFTs can have gapped boundaries sensitive to the spin structure; such boundaries are obtained from fermion condensation [48]. We only discuss gapped boundaries obtained from condensation of bosons.

Suppose we have a bulk excitation $a$. Under condensation, this anyon "splits" into several anyons to give excitations on the boundary.

$$a = \sum_x W_{ax} x \, , \qquad (3.2)$$

where $W_{ax}$ is an integer matrix. Note that even though this is called "splitting" in the literature (for example see [50]), it may be that the anyon $a$ gets identified with other anyons to produce some boundary excitation. The matrix $W$ determines the relationship between bulk and boundary excitations. The $W$ matrix plays a crucial rule in determining the fusion rules of boundary excitations. From [50], we have

$$n_{xy}^z = \sum_w \frac{V_{xw} V_{yw} V_{wz}^{-1}}{S_{0w}} \, , \qquad (3.3)$$

where $V_{xw} := \sum_a \overline{S_{ax}} W_{aw}$ and $S_{ax}$ is the bulk S-matrix (note that in this formula, the normalization of $S$ does not matter). If we substitute for the $V_{xw}$ matrix in (3.3), it becomes an equation in $n_{xy}^z$, $W$, and the bulk S-matrix $S$. If the $W$ matrix is invariant under Galois conjugation, then it is clear that the integer $n_{xy}^z$ being a combination of $S$ and $W$ is also invariant under Galois conjugation. Note that even though the $S$ matrix can change non-trivially under Galois conjugation, $n_{xy}^z$ is an integer given by a combination of $S$ matrix elements, and hence it is preserved under Galois conjugation.[26]

For the above picture to hold, we have to show that the integer matrix $W$ is invariant under Galois conjugation. Given a Lagrangian algebra $\mathcal{A}$ in an MTC $C$, the relationship between bulk and boundary excitations is found by constructing the pre-quotient category $\tilde{Q} = C/\mathcal{A}$. The simple objects of the canonical idempotent completion, $Q$, of $C/\mathcal{A}$ are the boundary excitations. The details of the construction of these categories are not relevant to our discussion. The crucial point is that the construction of the simple elements of $Q$ and their relationship to bulk anyons depend only on the fusions rules of the bulk theory and the choice of the anyons forming the Lagrangian algebra $\mathcal{A}$ [47, 51]. Hence, it follows that the $W$ matrix is invariant under Galois conjugation. As a result, we have the following theorem:

**Theorem 3.3:** The fusion rules of the boundary excitations are invariant under a unitarity-preserving Galois conjugation of the bulk TQFT.

---

[26]Moreover, since the S-matrix belongs to a cyclotomic field, any Galois conjugation acting on the S-matrix commutes with complex conjugation.

The full data of the gapped boundary is encoded in a spherical fusion category. The above theorem guarantees that the fusion rules of this spherical fusion category are invariant under Galois conjugation. However, the $F$ matrices of the boundary theory can change.

Suppose we have a discrete gauge theory $\mathcal{Z}(\text{Vec}_G^\omega)$. This theory always allows for a gapped boundary described by $\text{Vec}_G^\omega$ whose fusion rules are simply the group multiplication in $G$. The invariance of the fusion rules of the boundary excitations under Galois conjugation implies that, under a Galois conjugation, the gapped boundary described by $\text{Vec}_G^\omega$ changes at most by a difference in the twist $\omega$. That is, Galois conjugation of the bulk theory, $\mathcal{Z}(\text{Vec}_G^\omega)$, results in a new theory with gapped boundary described by $\text{Vec}_G^{\omega'}$ for some $\omega'$ which may not be equal to $\omega$. After Galois conjugation, the bulk theory is given by $\mathcal{Z}(\text{Vec}_G^{\omega'})$. This statement agrees with our discussion of Galois conjugation of discrete gauge theories.

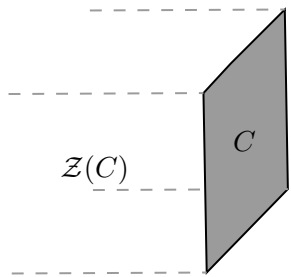

**Fig. 5:** The bulk TQFT is the Drinfeld center of the spherical fusion category describing the boundary excitations.

*3.3. Galois Conjugation and the Drinfeld Center*

In this section we will explore how the Galois action on a spherical fusion category affects its Drinfeld center. To that end, suppose we have a spherical fusion category $C$. Using the $F$ symbols of $C$, we can construct an algebraic field extension, $\mathbb{Q}(F)$, by adjoining the elements of the $F$ symbols to the rationals. Let $K_C$ be the Galois closure of $\mathbb{Q}(F)$. This is the defining number field of $C$ that we will work with. The Galois group, $\text{Gal}(K_C)$, acting on $C$ gives us other spherical fusion categories.

Now consider the Drinfeld center, $\mathcal{Z}(C)$, of $C$, which, on general grounds, is an MTC [52]. Let $K_{\mathcal{Z}(C)}$ be the Galois closure of the number field obtained by adjoinig the $F$ and $R$ symbols of $\mathcal{Z}(C)$ to the rationals. We can then act on $\mathcal{Z}(C)$ with the elements of $\text{Gal}(K_{\mathcal{Z}(C)})$ to get other MTCs.

If $x$ is an object in $C$, the objects of $\mathcal{Z}(C)$ are of the form $(x, e_x)$ where $e_x(y) \in$

Hom$(xy, yx)$ is a half-braiding which satisfies the constraint [53]

$$\alpha^{-1}(y, z, x) \circ (1 \otimes e_x(z)) \circ \alpha_{y,x,z} \circ (e_x(y) \otimes 1) \circ \alpha_{x,y,z}^{-1} = e_x(yz) , \tag{3.4}$$

where $e_x(1)$ is normalized to be the identity map, and $\alpha_{x,y,z}$ is the associativity map of the spherical fusion category. The Hom spaces, tensor product of objects, and braiding of the resulting modular tensor category are given by [52]

$$\text{Hom}((x, e_x), (y, e_y)) = \{f \in \text{Hom}(x, y) | 1 \otimes f \circ e_x(z) = e_y(z) \circ f \otimes 1 \ \forall \ z \in C\} , \tag{3.5}$$

$$(x, e_x) \otimes (y, e_y) = (x \otimes y, e_{xy}), \text{ where } e_{xy} = (e_x \otimes id_y) \circ (id_x \otimes e_y) , \tag{3.6}$$

$$c((x, e_x), (y, e_y)) = e_x(y) . \tag{3.7}$$

Therefore, we see that the braidings in the bulk are determined by the half-braidings. Note that given a simple object, $(x, e_x) \in \mathcal{Z}(C)$, $x \in C$ need not be simple. Indeed, we have to use (3.5) to identify the simple objects in the bulk using the fact that

$$\text{Hom}((x, e_x), (x, e_x)) \simeq \mathbb{C} , \tag{3.8}$$

if and only if $(x, e_x)$ is simple.

Note that the MTC data of $\mathcal{Z}(C)$ is determined by the data of $C$ along with the half-braidings. We can choose a basis for the fusion spaces and solve for the half-braidings by solving some multi-variable polynomials with coefficients in the field $\mathbb{Q}(F)$ obtained by adding the $F$ symbols of $C$ to the rationals (the constraints are given explicitly in equation (48) of [54]). Also, determining the full data of $C$ describing the boundary of the bulk TQFT corresponding to $\mathcal{Z}(C)$ involves a series of steps. First we have to determine the multiplication of the Lagrangian algebra in $\mathcal{Z}(C)$ corresponding to the gapped boundary. Representations of this algebra form the fusion category $C$. Therefore, to determine the boundary $F$ symbols, we have to find the $6j$ symbols for these representations [40, 48]. Though tedious, the constraints to be solved are algebraic in the data defining the bulk and boundary theory.

Given a Galois action on $C$ by some element of $q \in \text{Gal}(K_C)$, we have some corresponding Galois action $q' \in \text{Gal}(K_{\mathcal{Z}(C)})$ obtained as follows.[27] Let $g_1, \cdots, g_n$ be a basis of $K_C$ as a vector space, where $n$ is the finite degree of the field extension. The Galois action by some element $q \in \text{Gal}(K_C)$ on $K_C$ is completely specified by its action on the $g_i$. Similarly,

---

[27]Since the $F$ symbols of $C$ belong to a number field $K_C$, the equations which define the data of $\mathcal{Z}(C)$ are polynomials over $K_C$. Therefore, we can always find a solution to these polynomials which belongs to a number field (up to gauge choices).

let $h_1, \cdots, h_{n'}$ be a basis of $K_{\mathcal{Z}(C)}$ as a vector space, where $n'$ is the finite degree of the field extension $K_{\mathcal{Z}(C)}$. Then we can choose some $q' \in \text{Gal}(K_{\mathcal{Z}(C)})$ such that the action of $q$ and $q'$ on $\{g_1, \cdots, g_n\} \cap \{h_1, \cdots, h_{n'}\}$ agree.[28] If $K_C$ and $K_{\mathcal{Z}(C)}$ are distinct, this choice is not unique. Galois action by $q'$ on the $F$ and $R$ symbols of $\mathcal{Z}(C)$ results in the MTC which is the Drinfeld center of the spherical fusion category obtained by Galois conjugating $C$ with respect to $q$. This leads to the following result:

**Theorem 3.4:** Corresponding to every Galois action, $q(C)$, on a spherical fusion category, $C$, where $q \in K_C$, there exists a Galois action $q' \in K_{\mathcal{Z}(C)}$ such that

$$\mathcal{Z}(q(C)) = q'(\mathcal{Z}(C)) \, , \tag{3.9}$$

and vice-versa.

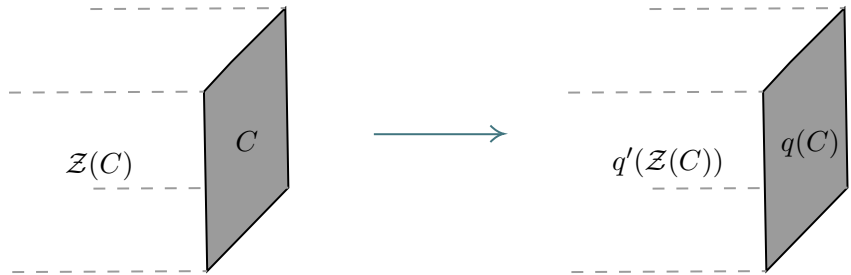

**Fig. 6:** Galois conjugation on the bulk induces a Galois action on the boundary and vice-versa.

Note that it is possible for $K_C$ to be a non-abelian field extension and $K_{\mathcal{Z}(C)}$ to be abelian. For example, the data of the fusion category, $H$, obtained from the principal even part of the Haagerup subfactor, cannot be contained in a cyclotomic field [18]. Therefore, by the Kronecker-Weber theorem, $K_H$ for this category is necessarily a non-abelian extension. It is also known that the MTC data of the Drinfeld center $\mathcal{Z}(H)$ belongs to a cyclotomic number field. Therefore, we can choose $K_{\mathcal{Z}(H)}$ to be an abelian extension.

We immediately get an application of (3.9) as follows. Recall that the Drinfeld center of a spherical fusion category is unique. Moreover, Morita equivalent spherical fusion categories have the same Drinfeld center. Therefore, (3.9) implies:

**Corollary 3.5:** The number of distinct Galois conjugates of $\mathcal{Z}(C)$ is a lower bound on the number of non-Morita equivalent Galois conjugates of $C$.

---

[28]These can be thought of as Galois actions on the composite extension obtained from $K_C$ and $K_{\mathcal{Z}(C)}$. See appendix A.

**Corollary 3.6:** The number of distinct Galois conjugates of $C$ is an upper bound on the number of distinct Galois conjugates of $\mathcal{Z}(C)$.

As a result, if $C$ is Galois invariant, so is $\mathcal{Z}(C)$. That is, the Galois invariance of the $1+1$D boundary implies that the bulk TQFT is Galois invariant. Similarly, if $\mathcal{Z}(C)$ is Galois invariant, all Galois conjugates of $C$ should be Morita equivalent to $C$.

It follows that given a bulk TQFT $\mathcal{Z}(C)$ with boundary described by the spherical fusion category $C$, the Galois conjugate $q'(\mathcal{Z}(C))$ admits the boundary condition $q(C)$. This agrees with our result above that Galois action on a Lagrangian algebra results in a Lagrangian algebra. Using Galois actions arising from the cyclotomic field containing the modular data, this was argued recently in [10].

## 4. Symmetries, Gauging and Galois Fixed Point TQFTs

Symmetries are, of course, a duality-invariant feature of quantum field theories. However, there is a priori, no guarantee that they are also Galois invariant. We therefore wish to study the question of how symmetries transform under Galois actions.

To that end, recall that the main observables in $2+1$D TQFTs are line operators. These naturally lead to 1-form symmetries. One can also define surface operators which act on these line operators and permute them. These are 0-form symmetries. Sometimes the 0-form symmetry and 1-form symmetry can form a 2-group. Therefore, $2+1$D TQFTs have a rich symmetry structure.[29] In this section, we will study the relationship between the symmetries of Galois conjugate TQFTs. The case of abelian TQFTs is the simplest to analyse. After that we will study symmetries of non-abelian Galois conjugate TQFTs. Following this, we will look at gauging the 0-form symmetry and how Galois conjugation of the TQFT affects the gauging procedure.

### 4.1. Symmetries of a TQFT

Given the set of anyons, $\{a, b, \cdots\}$, of a TQFT, the subset of abelian anyons corresponds to some abelian group. This is the 1-form symmetry group, $\mathcal{A}$, of a TQFT. Moreover, we can define an automorphism group of the set of anyons, $G$, which preserves the MTC data (up to conjugation for anti-unitary symmetries). This is the 0-form symmetry group of the TQFT. These symmetries can lead to a natural 2-Group structure. For a given MTC, there are certain permutations of the anyons that leave all the gauge-invariant data unchanged. These form the intrinsic symmetry of the TQFT. The gauge-invariant data is left unchanged up to a conjugation for anti-unitary symmetries.

---

[29]If we allow for topological point operators, then we can also have 2-form symmetries.

Given an MTC, the 2-group structure is define by the quadruple $(G, \mathcal{A}, \rho, [\beta])$. Here, $\rho$ is the action of the 0-form symmetry group on the 1-form charges, $\rho : G \to \text{Aut}(\mathcal{A})$, and $[\beta] \in H^3_\rho(G, \mathcal{A})$. To understand how this 2-group structure arises, let us define how the 0-form symmetry acts on the MTC. Let $g \in G$. As alluded to before, $G$ acts on the anyons through a permutation. Hence, $g(a) = a'$. For it to be symmetry, the gauge-invariant quantities should be invariant under it. For example:

$$g(N^c_{ab}) = N^{g(c)}_{g(a)g(b)} = N^c_{ab} , \tag{4.1}$$

$$g(\theta_a) = K^g \theta_a K^g , \tag{4.2}$$

$$g(S_{ab}) = K^g S_{g(a)g(b)} K^g , \tag{4.3}$$

where $K^g$ is an operator which complex conjugates the quantity in between if $g$ is an anti-unitary symmetry. The gauge-dependent quantities should change only up to a gauge transformation. Since $G$ acts on all anyons, its restriction to the abelian anyons, $\mathcal{A}$, specifies the map, $\rho : G \to \text{Aut}(\mathcal{A})$.

The action of $g$ on the fusion space is

$$g(|a, b, c; \mu\rangle) = |a', b', c'; \mu\rangle . \tag{4.4}$$

For our convenience, we would like to define a map which leaves even the gauge-dependent quantities invariant. For this, we will redefine the action of the above map on the fusion space as

$$g(|a, b, c; \mu\rangle) = \sum_{\mu'} U_g(a', b', c')_{\mu,\mu'} K^g |a', b', c'; \mu'\rangle , \tag{4.5}$$

where $U_g(a', b'; c')_{\mu,\mu'}$ is a unitary matrix, and $K^g$ is an operator introduced above so that the quantities sandwitched between two $K^g$'s are complex conjugated if $g$ is an anti-unitary symmetry. This changes the F and R-matrices as follows

$$U_g(g(b), g(a), g(c)) R^{g(c)}_{g(a)g(b)} U_g(g(a), g(b), g(c))^{-1} \tag{4.6}$$

$$U_g(g(a), g(b), g(e)) U_g(g(e), g(c), g(d)) (F^{g(d)}_{g(a)g(b)g(c)})^{g(f)}_{g(e)}$$
$$\times U_g(g(b), g(c), g(f))^{-1} U_g(g(a), g(f), g(d))^{-1} , \tag{4.7}$$

where $a \otimes b = e$, $b \otimes c = f$, and we have supressed the indices labelling the basis vectors of the fusion spaces. For $g \in G$ to be a symmetry, we require

$$g(R_{ab}^c) = U_g(g(b), g(a), g(c))R_{g(a)g(b)}^{g(c)}U_g(g(a), g(b), g(c))^{-1} = K^g R_{ab}^c K^g , \tag{4.8}$$

$$g((F_{abc}^d)_e^f) = U_g(g(a), g(b), g(e))U_g(g(e), g(c), g(d))(F_{g(a)g(b)g(c)}^{g(d)})_{g(e)}^{g(f)}$$
$$\times U_g(g(b), g(c), g(f))^{-1}U_g(g(a), g(f), g(d))^{-1} = K^g(F_{abc}^d)_e^f K^g , \tag{4.9}$$

where $a \otimes b = e$, and $b \otimes c = f$. This definition of $g$ ensures the invariance of even gauge dependent quantities under its action. Hence, the action of $g$ on a category can be seen as a permutation of the anyons along with a gauge transformation. Among such maps, there are those that act on the labels and fusion spaces as follows

$$\Upsilon(a) = a; \quad \Upsilon(|a, b, c; \mu\rangle) = \frac{\gamma_a \gamma_b}{\gamma_c} |a, b, c; \mu\rangle , \tag{4.10}$$

for some phases, $\gamma_a$. By definition, such maps don't permute the anyons, and they leave all the data invariant. The $\Upsilon$ are called natural isomorphisms. Note that these are gauge transformations, where the unitary gauge transformation matrix acting on the fusion space is $\frac{\gamma_a \gamma_b}{\gamma_c}\delta_{\mu\mu'}$. The 0-form symmetry group of the theory, $G$, is the set of maps, $g$, modulo natural isomorphisms. Hence, the group elements are equivalence classes, $[g]$. For $[g], [h], [k] \in G$ the group multiplication is given by

$$[g] \cdot [h] = [k] \iff \Upsilon_1 \cdot g \cdot \Upsilon_2 \cdot h = \Upsilon_3 \cdot k \implies k = \kappa_{g,h} \cdot g \cdot h , \tag{4.11}$$

where $\kappa_{g,h} = \Upsilon_3^{-1} \cdot \Upsilon_1 \cdot g \cdot \Upsilon_2 \cdot g^{-1}$. Here $\kappa_{g,h}$ is a natural isomorphism which can be written in terms of phases as

$$\kappa_{g,h}(a, b, c)_{\mu\nu} = \frac{\gamma_a(g, h)\gamma_b(g, h)}{\gamma_c(g, h)}\delta_{\mu,\nu} . \tag{4.12}$$

The phases in $\gamma_a(g, h)$ look arbitrary, but they obey some consistency conditions. In fact, they can be extracted from the TQFT data. In the language of symmetry defects, $U_g(a, b, c)$ represents the action of a symmetry defect on a fusion vertex, and the $\gamma_a(g, h)$ phases represent the difference in the action of $g$ and then $h$ on an anyon compared to the action of $g \cdot h$ (see Fig. 7 and Fig. 8).

To respect the freedom to add or remove identity lines, we should impose

$$\gamma_1(g, h) = \gamma_a(e, h) = \gamma_a(g, e) = 1 , \tag{4.13}$$

$$U_e(a, b, c) = U_g(1, b, c) = U_g(a, 1, c) = 1 , \tag{4.14}$$

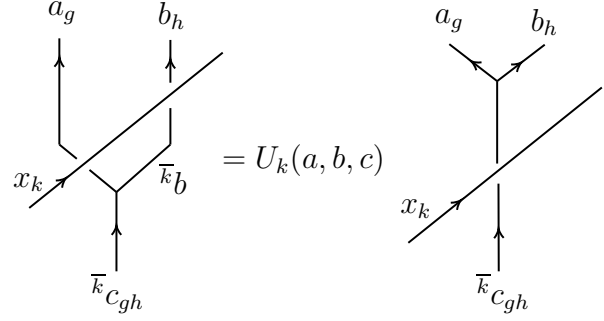

**Fig. 7:** Diagrammatic definition of $U_k(a,b,c)$

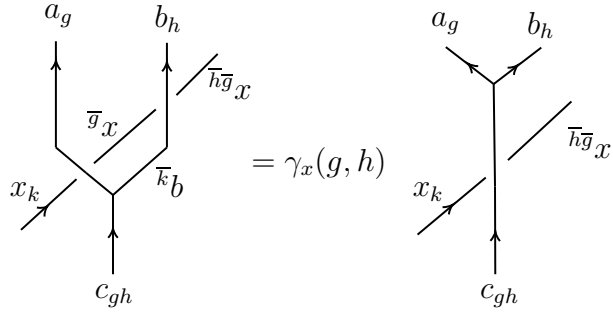

**Fig. 8:** Diagrammatic definition of $\gamma_k(g,h)$

where 1 respresents the vacuum, and $e$ is the identity in $G$. Using $\kappa_{g,h} g \cdot h = k$ and the action of the symmetries in the fusion spaces, we get the following expression for $\kappa_{g,h}$.

$$\kappa_{g,h} = \frac{\gamma_a(g,h)\gamma_b(g,h)}{\gamma_c(g,h)} = U_g(a,b,c)^{-1}(K^g U_h^{-1}(g^{-1}(a), g^{-1}(b), g^{-1}(c))K^g) U_{gh}(a,b,c) \ . \quad (4.15)$$

Let us look at the 0-form symmetry group element $g \cdot h \cdot k$

$$g \cdot h \cdot k = \kappa_{g,hk} \cdot g \cdot (h \cdot k) \tag{4.16}$$

$$= \kappa_{g,hk} \cdot g \cdot \kappa_{h,k} \cdot h \cdot k \tag{4.17}$$

$$= \kappa_{g,hk} \cdot g \cdot \kappa_{h,k} \cdot g^{-1} \cdot g \cdot h \cdot k \ . \tag{4.18}$$

We also have

$$g \cdot h \cdot k = \kappa_{gh,k} \cdot (g \cdot h) \cdot k \tag{4.19}$$

$$= \kappa_{gh,k} \cdot \kappa_{g,h} \cdot g \cdot h \cdot k \ . \tag{4.20}$$

Hence, we find the following consistency condition

$$\kappa_{g,hk} \cdot g \cdot \kappa_{h,k} \cdot g^{-1} = \kappa_{gh,k} \cdot \kappa_{g,h} \ . \tag{4.21}$$

Action of $\kappa_{g,hk}$ on the fusion spaces gives

$$K^g \frac{\gamma_{g^{-1}(a)}(h,k)\gamma_{g^{-1}(b)}(h,k)}{\gamma_{g^{-1}(c)}(h,k)} K^g \frac{\gamma_a(g,hk)\gamma_b(g,hk)}{\gamma_c(g,hk)} = \frac{\gamma_a(gh,k)\gamma_b(gh,k)}{\gamma_c(gh,k)} \frac{\gamma_a(g,h)\gamma_b(g,h)}{\gamma_c(g,h)} . \quad (4.22)$$

Sometimes the 0-form and 1-form symmetries form a non-trivial 2-group. This is determined by a 3-cocycle, $[\beta]$, sometimes called the "Postnikov class," and it belongs to the cohomology group $H^3_{[\rho]}(G, \mathcal{A})$, where $\rho : G \to \mathrm{Aut}(\mathcal{A})$ specifies the action of the 0-form symmetry group $G$ on the 1-form symmetry group $\mathcal{A}$. To determine this class, let us define the phase

$$\Omega_a(g,h,k) := \frac{K^g \gamma_{g^{-1}(a)}(h,k) K^g \gamma_a(g,hk)}{\gamma_a(gh,k)\gamma_a(g,h)} . \quad (4.23)$$

From this definition, it follows that

$$\frac{K^g \Omega_{g^{-1}(a)}(h,k,l) K^g \Omega_a(g,hk,l) \Omega_a(g,h,k)}{\Omega_a(gh,k,l)\Omega_a(g,h,kl)} = 1 . \quad (4.24)$$

This result can be shown by brute-force substitution and simplification. Using (4.22), we can show that

$$\Omega_a(g,h,k)\Omega_b(g,h,k) = \Omega_c(g,h,k) , \quad (4.25)$$

whenever $N^c_{ab} \neq 0$. Then,

$$d_a \Omega_a(g,h,k) d_b \Omega_b(g,h,k) = \sum_c N^c_{ab} d_c \Omega_c(g,h,k) . \quad (4.26)$$

Hence, $d_a \Omega_a(g,h,k)$ forms a 1-dimensional representation of the fusion rules and should be equal to $\frac{S_{ae}}{S_{1e}}$ for some charge $e$. As a result, we have

$$\Omega_a(g,h,k) = \frac{S_{ae}S_{11}}{S_{1e}S_{1a}} = M^*_{ae} . \quad (4.27)$$

Since, for a given $e$, $\Omega_a(g,h,k)$ is a phase for all $a$, the label $e$ is abelian in the sense that its quantum dimension satisfies $d_e = 1$. This fact can be shown using the following argument.

$$d_e^2 = \sum_b \left| \frac{d_e d_b}{D} \right|^2 = \sum_b \left| \frac{d_e d_b}{D} M_{be} \right|^2 = \sum_b \left| \frac{d_e d_b}{D} \frac{S_{be}S_{00}}{S_{0e}S_{0b}} \right|^2 = \sum_b |S_{be}|^2 = 1 . \quad (4.28)$$

Hence, $e = \beta(g,h,k)$ is a map $\beta(g,h,k) : G \times G \times G \to \mathcal{A}$. It is a 3-cochain $\beta(g,h,k) \in$

$C^3(G, \mathcal{A})$. Let us use (4.27) to simplify (4.24).

$$
\begin{aligned}
1 &= \frac{\Omega_{g^{-1}(a)}(h,k,l)\Omega_a(g,hk,l)\Omega_a(g,h,k)}{\Omega_a(gh,k,l)\Omega_a(g,h,kl)} \\
1 &= M^*_{g^{-1}(a)\beta(h,k,l)} M^*_{a\beta(g,hk,l)} M^*_{a\beta(g,h,k)} M_{a\beta(gh,k,l)} M_{a\beta(g,h,kl)} \\
&= M^*_{ag(\beta(h,k,l))} M^*_{a\beta(g,hk,l)} M^*_{a\beta(g,h,k)} M^*_{a\overline{\beta(gh,k,l)}} M^*_{a\overline{\beta(g,h,kl)}} \\
&= M^*_{ag(\beta(h,k,l))\cdot\beta(g,hk,l)\cdot\beta(g,h,k)\cdot\overline{\beta(gh,k,\varphi_4)}\cdot\overline{\beta(g,h,kl)}} \ .
\end{aligned} \tag{4.29}
$$

Since this logic holds for all $a$, we have

$$
g(\beta(h,k,l)) \cdot \beta(g,hk,l) \cdot \beta(g,h,k) \cdot \overline{\beta(gh,k,\varphi_4)} \cdot \overline{\beta(g,h,kl)} = 0 \ . \tag{4.30}
$$

This shows that $\beta$ is a 3-cocycle. In particular, $\beta \in Z^3_{[\rho]}(G, \mathcal{A})$, where the subscript, $\rho$, indicates a twisted cohomology group due to the non-trivial action of G on $\mathcal{A}$.

In fact, we can say more. Indeed, there is some freedom in decomposing natural isomorphisms in terms of phases (in (4.12)). More specifically, we have the freedom to choose

$$
\gamma_a(g,h) \quad \text{or} \quad v_a(g,h)\gamma_a(g,h) \ , \tag{4.31}
$$

where $v_a$ are phases that satisfy $v_a v_b = v_c$ whenever $N^c_{ab} \neq 0$. It is easy to see that either choice leads to the same $\kappa_{g,h}$ in (4.12). However, the latter will change $\beta \in Z^3_{[\rho]}(G, \mathcal{A})$ by to an exact cocycle. Hence, what defines a 2-group are actually equivalence classes $[\beta] \in H^3_{[\rho]}(G, \mathcal{A})$.

**Example:** Recall the Spin$(5)_2$ Chern-Simons theory that we discussed previously. This theory has a time reversal symmetry given by the permutation $\phi_1 \leftrightarrow \phi_2$ and $\psi_+ \leftrightarrow \psi_-$. Hence, it has a $\mathbb{Z}_2 = \{e, z\}$ 0-form symmetry. The modular data can be used to fix the possible values for the Postnikov class. For a non-unitary symmetry, we have

$$
\Omega_a(g,h,k) = \frac{K^g \gamma_{g^{-1}(a)}(h,k) K^g \gamma_a(g,hk)}{\gamma_a(gh,k)\gamma_a(g,h)} \ , \tag{4.32}
$$

where $K^g$ is an operator which complex conjugates the element in between if $g$ is a non-unitary symmetry. The only non-trivial $\Omega_a(g,h,k)$ in our case is

$$
\Omega_a(z,z,z) = \frac{\gamma^*_{z(a)}(z,z)}{\gamma_a(z,z)} \ . \tag{4.33}
$$

From, (4.27) we know that the only non-trivial $\beta(\cdots)$ is given by $\beta(z,z,z)$. Since the relevant cohomology group is $H^3(\mathbb{Z}_2, \mathbb{Z}_2) = \mathbb{Z}_2$, $\beta(z,z,z)$ should be an order 2 abelian anyon. The only options are $\beta(z,z,z) = 1, \epsilon$. The relation (4.27) is trivially satisfied for $\beta(z,z,z) = 1$. For, $\beta(z,z,z) = \epsilon$ we have

$$\frac{\gamma^*_{z(a)}(z,z)}{\gamma_a(z,z)} = \frac{S_{a\epsilon}}{S_{a1}} \ . \tag{4.34}$$

Using this equation, we can derive some relations among the $\gamma_a(z,z)$ phases. In particular

$$\gamma_e(z,z) = \gamma^*_e(z,z) \ , \quad \gamma_{\phi_1}(z,z) = \gamma^*_{\phi_2}(z,z) \ , \quad \gamma_{\psi_+}(z,z) = -\gamma^*_{\psi_-}(z,z) \ . \tag{4.35}$$

If these relations are satisfied, then $\beta(z,z,z) = \epsilon$ is a valid choice. Note that since $\epsilon$ is not a quadratic residue, this choice corresponds to a non-trivial Postnikov class. On the other hand, for $\beta(z,z,z) = 1$, the quantities in (4.35) satisfy

$$\gamma_e(z,z) = \gamma^*_e(z,z) \ , \quad \gamma_{\phi_1}(z,z) = \gamma^*_{\phi_2}(z,z) \ , \quad \gamma_{\psi_+}(z,z) = \gamma^*_{\psi_-}(z,z) \ . \tag{4.36}$$

Now, the values for the $F$ and $R$ matrices for $\text{Spin}(5)_2$ can be used to constrain $U_z(a,b,c)$, which, in turn, will put several constraints on $\gamma_a(z,z)$. Using (4.15) we have the equation

$$\frac{\gamma_a(z,z)\gamma_{a^*}(z,z)}{\gamma_1(z,z)} = U_z(a,a^*,1)^{-1} U_z(z(a), z(a^*), 1) \ . \tag{4.37}$$

It follows that

$$\gamma_{a^*}(z,z) = \gamma^*_a(z,z) \ , \tag{4.38}$$

for anyon, $a$, satisfying $z(a) = a$. Also, since all anyons in this theory are self conjugate, (4.38) implies that $\gamma_a(z,z)$ with $z(a) = a$ are real. This discussion restricts the quantities in (4.38) to be $\pm 1$.

Now let us make the choice $a = \psi_+, b = \psi_-, c = \epsilon$ in (4.15)

$$\frac{\gamma_{\psi_+}(z,z)\gamma_{\psi_-}(z,z)}{\gamma_\epsilon(z,z)} = U_z(\psi_+, \psi_-, \epsilon)^{-1} U_z(\psi_-, \psi_+, \epsilon) \ . \tag{4.39}$$

We would like to substitute for $\gamma_\epsilon(z,z)$ to write $\gamma_{\psi_+}(z,z)\gamma_{\psi_-}(z,z)$ purely in terms of $U_z(a,b,c)$ phases. Let us choose $a = \epsilon, b = \phi_1, c = \phi_1$ in (4.15)

$$\frac{\gamma_\epsilon(z,z)\gamma_{\phi_1}(z,z)}{\gamma_{\phi_1}(z,z)} = U_z(\epsilon, \phi_1, \phi_1)^{-1} U_z(\epsilon, \phi_2, \phi_2) \tag{4.40}$$

$$\implies \gamma_\epsilon(z,z) = U_z(\epsilon, \phi_1, \phi_1)^{-1} U_z(\epsilon, \phi_2, \phi_2) \ . \tag{4.41}$$

Then we get,

$$\gamma_{\psi_+}(z,z)\gamma_{\psi_-}(z,z) = U_z(\psi_+, \psi_-, \epsilon)^{-1} U_z(\psi_-, \psi_+, \epsilon) U_z(\epsilon, \phi_1, \phi_1)^{-1} U_z(\epsilon, \phi_2, \phi_2) \ . \tag{4.42}$$

The R-matrix, $R^\epsilon_{\psi_+\psi_-}$, transforms under the symmetry in the following way

$$z(R^\epsilon_{\psi_+\psi_-}) = U_z(\psi_-, \psi_+, \epsilon) R^\epsilon_{\psi_-\psi_+} U_z(\psi_+, \psi_-, \epsilon)^{-1} = (R^\epsilon_{\psi_+\psi_-})^* \ . \tag{4.43}$$

From the MTC data of Spin(5)$_2$ (see [44] for the full MTC data), we have $R^\epsilon_{\psi_+\psi_-} = R^\epsilon_{\psi_-\psi_+} = 1$. It follows that

$$U_z(\psi_-, \psi_+, \epsilon) U_z(\psi_+, \psi_-, \epsilon)^{-1} = 1 \ . \tag{4.44}$$

Also, the F-matrix $(F^{\phi_1}_{\epsilon\phi_2\phi_1})^{\phi_1}_{\phi_2}$ transforms under the symmetry action as

$$U_z(\epsilon, \phi_1, \phi_1) U_z(\phi_1, \phi_2, \phi_2)(F^{\phi_2}_{\epsilon\phi_1\phi_2})^{\phi_2}_{\phi_1} U_z(\phi_1, \phi_2, \phi_2)^{-1} U_z(\epsilon, \phi_2, \phi_2)^{-1} = ((F^{\phi_1}_{\epsilon\phi_2\phi_1})^{\phi_1}_{\phi_2})^* \ . \tag{4.45}$$

Using, $(F^{\phi_1}_{\epsilon\phi_2\phi_1})^{\phi_1}_{\phi_2} = -1$ and $(F^{\phi_2}_{\epsilon\phi_1\phi_2})^{\phi_2}_{\phi_1} = 1$, we have

$$U_z(\epsilon, \phi_1, \phi_1) U_z(\epsilon, \phi_2, \phi_2)^{-1} = -1 \ . \tag{4.46}$$

From these relations, we have

$$\gamma_{\psi_+}(z, z) \gamma_{\psi_-}(z, z) = -1 \ . \tag{4.47}$$

This agrees with the constraints on $\gamma_a(z, z)$ set by $\beta(z, z, z) = \epsilon$. Hence, Spin(5)$_2$ MTC has a non-trivial Postnikov class.[30]

## 4.2. Symmetries of Abelian TQFTs and Galois Conjugation

In abelian TQFTs, all anyons are abelian and hence the 1-form symmetry group coincides with the fusion rules. We know that Galois conjugation relates different solutions of the Pentagon and Hexagon equations. Hence, it preserves the fusion rules. Thus, the 1-form symmetry group is invariant under Galois conjugation. We would like to find the relationship between 0-form symmetries of Galois conjugate abelian theories. To that end, let $\mathcal{A}$ be the set of anyons of the theory. As alluded to previously, an abelian TQFT is determined completely by this set and the topological spin function

$$\theta : \mathcal{A} \to U(1) \ . \tag{4.48}$$

The automorphism group of $\mathcal{A}$, denoted Aut($\mathcal{A}$), is a subset of the permutation group acting on $\mathcal{A}$. For it to be a symmetry, $G$, of the TQFT, it has to preserve the topological spins. That is, if $g \in G$ we require

$$\theta_{g(a)} = \theta_a \text{ (up to conjugation for anti-unitary symmetries) } . \tag{4.49}$$

---

[30]This theory and its non-trivial Postnikov class are discussed in [55] with the equivalent name USp(4)$_2$ Chern-Simons theory.

The symmetry group, G, is a subgroup of Aut($\mathcal{A}$). The topological spins are of the form $\theta_a = e^{2\pi i h_a}$, where $h_a$ is a rational number. Hence, the topological spins are roots of unity, and we can write $h_a = \frac{f(a)}{N}$ for some integer N. The condition for $g$ to be a symmetry can be written as

$$h_{g(a)} = \pm \; h_a \quad \text{mod 1 (minus sign for anti-unitary symmetries)} \tag{4.50}$$

$$\implies f(g(a)) = \pm f(a) \text{ mod } N . \tag{4.51}$$

Under Galois conjugation by some $q$ coprime to N,

$$h_a \to q h_a . \tag{4.52}$$

We can see that the condition (4.50) becomes

$$q f(g(a)) = \pm q f(a) \text{ mod } N \implies f(g(a)) = \pm f(a) \text{ mod } N . \tag{4.53}$$

Since the set of labels $\mathcal{A}$ do not change under Galois conjugation, neither does the automorphism group. We have seen above that the condition (4.50) which restricts the symmetry group G to a subgroup of Aut($\mathcal{A}$) also doesn't change under Galois conjugation. Hence, the 0-form symmetry group is Galois invariant.

In summary, we have found that Galois conjugate symmetries have the same 0-form and 1-form symmetries. What about the 2-group structure? It is widely believed that all abelian TQFTs have trivial 2-group. This is in agreement with all known cases. A proof for unitary symmetries was given in [56]. Assuming this result extends to all 0-form symmetries, it is trivially true that the 2-group symmetry is invariant under Galois conjugation. However, let us give an alternate proof which does not rely on this conjecture.

For an abelian TQFT, the $F$ and $R$ symbols are phases. Moreover, the $U_g(a,b)$ can be taken to belong to a cyclotomic field. This statement follows from the fact that for abelian TQFTs, we can choose a gauge in which all the $F$ symbols are valued in $\pm 1$ [22] . For abelian TQFTs, the symmetry transformation of the $F$ symbols (4.9) can be written as

$$U_g(g(a),g(b))U_g(g(a)+g(b),g(c))U_g(g(b),g(c))^{-1}U_g(g(a),g(b)+g(c))^{-1}F(g(a),g(b),g(c))$$
$$= K^g F(a,b,c) K^g . \tag{4.54}$$

If we are in a gauge in which the $F$ symbols are valued in $\pm 1$, we get

$$U_g(g(a),g(b))^2 U_g(g(a)+g(b),g(c))^2 U_g(g(b),g(c))^{-2} U_g(g(a),g(b)+g(c))^{-2} = 1 . \tag{4.55}$$

This result shows that the phases $U_g(g(a),g(b))^2$ should form a 2-cocycle. Moreover, since $U_g(g(a),g(b))^2$ is defined only up to symmetry gauge transformations, $U_g(g(a),g(b))^2$ should be an element of $H^2(G, U(1))$. These quantities can always be chosen to be $|G|^{\text{th}}$ roots of unity. Therefore, the phases $U_g(g(a), g(b))$ are at most $2|G|^{\text{th}}$ roots of unity.

Therefore, a Galois conjugation of $F$ and $R$ induces a Galois conjugation on $U_g(a, b)$ which acts on these phases as

$$F(a, b, c) \to F(a, b, c)^q \ , \quad R(a, b) \to R(a, b)^q \ , \quad U(a, b) \to U(a, b)^q \ , \tag{4.56}$$

for some integer $q$ coprime to the order of the cyclotomic field. As a result, in the Galois conjugate theory, the symmetry acts on the fusion spaces as $U_g(a, b)^q$. The phases satisfying (4.15) are $\gamma_a(g, h)^q$. Therefore, in the Galois conjugate theory, using (4.23) and (4.27), we get

$$\omega_a(g, h, k)^q = \frac{(S_{a\beta(g,h,k)})^q}{(S_{a1})^q} \ , \tag{4.57}$$

where $(\tilde{S}_{ab})^q$ is an element of the (un-normalized) S-matrix of the Galois conjugate theory. Hence, the Galois conjugate theory has the same Postnikov class.

### 4.3. Symmetries of non-abelian TQFTs and Galois conjugation

As a more gentle starting point, we first consider the case of multiplicity free non-abelian TQFTs (i.e., theories with non-invertible anyons where each fusion product appears at most once in a given fusion) with a cyclotomic defining number field. We then proceed to the general non-abelian case.

#### 4.3.1. $N_{ab}^c = 0, 1$ and cyclotomic defining number field

Let us consider a multiplicity-free MTC, $C$, with MTC data denoted by $R_{ab}^c, F_{abc}^d$. Let $K_C$ be the defining number field of $C$. Before considering the possibility of a more general defining number field, it is useful to consider the case when $K_C$ is a cyclotomic field.

In this case, the Galois action on the MTC data, as well as its effect on the $U_g(a, b, c)$ and $\gamma_a(g, h)$ phases, can be described explicitly. Therefore, let $K_C = \mathbb{Q}(\xi_N)$, where $N$ is some integer. Let us consider the MTC, $q(C)$, which is obtained by Galois conjugating this data with respect to some $q \in \text{Gal}(\mathbb{Q}(\xi_N))$. All quantities in $q(C)$ will have a hat on top, and so the MTC data of the Galois conjugated theory is $\widehat{R}_{ab}^c, \widehat{F}_{abc}^d$. $C$ has a 1-form symmetry group $\mathcal{A}$ and 0-form symmetry group $G$. $G$ acts on the anyons in the theory as $g(a)$ and permutes them.

The gauge-invariant quantities of the theory should be invariant under the symmetry action. For example, we have $S_{ab} = S_{g(a)g(b)}$. This relation hold holds even after Galois conjugation. Therefore $\widehat{S}_{ab} = \widehat{S}_{g(a)g(b)}$. As a result, the zero-form symmetry group, $G$, of the initial TQFT, $\mathcal{T}_1$ is isomorphic to the symmetry group of the gauge-invariant data of the Galois-conjugated TQFT, $\mathcal{T}_2$.

The symmetry acts on the fusion spaces of $C$ through the unitary matrix, $U_g$. Because we have a multiplicity free theory, the $U_g$'s are just phases. By definition, we have the following equalities

$$U_g(g(b), g(a), g(c)) R^{g(c)}_{g(a)g(b)} U_g(g(a), g(b), g(c))^{-1} = K^g R^c_{ab} K^g , \tag{4.58}$$

$$U_g(g(a), g(b), g(e)) U_g(g(e), g(c), g(d)) (F^{g(d)}_{g(a)g(b)g(c)})^{g(f)}_{g(e)}$$
$$\times U_g(g(b), g(c), g(f))^{-1} U_g(g(a), g(f), g(d))^{-1} = K^g (F^d_{abc})^f_e K^g . \tag{4.59}$$

From these equations we have

$$U_g(g(b), g(a), g(c)) U_g(g(a), g(b), g(c))^{-1} = K^g R^c_{ab} K^g (R^{g(c)}_{g(a)g(b)})^{-1} , \tag{4.60}$$

$$U_g(g(a), g(b), g(e)) U_g(g(e), g(c), g(d)) U_g(g(b), g(c), g(f))^{-1} U_g(g(a), g(f), g(d))^{-1}$$
$$= K^g (F^d_{abc})^f_e K^g ((F^{g(d)}_{g(a)g(b)g(c)})^{g(f)}_{g(e)})^{-1} . \tag{4.61}$$

Since $R^{g(c)}_{g(a)g(b)}$ and $(F^{g(d)}_{g(a)g(b)g(c)})^{g(f)}_{g(e)}$ belong to $\mathbb{Q}(\xi_N)$, $U_g(g(b), g(a), g(c)) U_g(g(a), g(b), g(c))^{-1}$ and $U_g(g(a), g(b), g(e)) U_g(g(e), g(c), g(d)) U_g(g(b), g(c), g(f))^{-1} U_g(g(a), g(f), g(d))^{-1}$ are both phases in $\mathbb{Q}(\xi_N)$. Note that even though the above combinations of the $U_g$ phases are guaranteed to be in the cyclotomic field of the MTC data, we do not assume that the individual phases themselves belong to a cyclotomic field. Galois conjugating both sides of the above equations by $q \in \mathrm{Gal}(\mathbb{Q}(\xi_N))$, we get

$$q(U_g(g(b), g(a), g(c)) U_g(g(a), g(b), g(c))^{-1}) = q(K^g R^c_{ab} K^g (R^{g(c)}_{g(a)g(b)})^{-1}) \tag{4.62}$$
$$= K^g (R^c_{ab})^q K^g (R^{g(c)}_{g(a)g(b)})^{-q}) \tag{4.63}$$
$$= K^g \widehat{R}^c_{ab} K^g (\widehat{R}^{g(c)}_{g(a)g(b)})^{-1}) . \tag{4.64}$$

In writing down the equations above, we used the fact that the $R^c_{ab}$ are phases for a multiplicity-free theory and that $\widehat{R}^c_{ab} = (R^c_{ab})^q$. Also, since $U_g(g(b), g(a), g(c)) U_g(g(a), g(b), g(c))^{-1}$ is a phase in $\mathbb{Q}(\xi_N)$, Galois conjugating it by $q$ amounts to taking its $q^{\mathrm{th}}$ power. Note that we can commute the complex conjugation and Galois conjugation operation on the RHS of the above equation since we have chosen a gauge in which the MTC data is in a cyclotomic field (the Galois group in this case is abelian). If we had chosen another basis in which the MTC data belongs to a field extension with non-abelian Galois group, complex conjugation might not commute with a general Galois conjugation. We have

$$(U_g(g(b), g(a), g(c)) U_g(g(a), g(b), g(c))^{-1})^q \widehat{R}^{g(c)}_{g(a)g(b)} = K^g \widehat{R}^c_{ab} K^g . \tag{4.65}$$

Following the same arguments, from the action of the symmetry on the $F^d_{abc}$, we obtain

$$(U_g(g(a), g(b), g(e))U_g(g(e), g(c), g(d))U_g(g(b), g(c), g(f))^{-1}U_g(g(a), g(f), g(d))^{-1})^q$$
$$\times (\widehat{F}^{g(d)}_{g(a)g(b)g(c)})^{g(f)}_{g(e)} = K^g (\widehat{F}^d_{abc})^f_e K^g \ . \tag{4.66}$$

Note that since the $F$ matrix elements need not be phases, their Galois conjugation does not usually correspond to taking a $q^{\text{th}}$ power. However, we have only used $\widehat{F}^d_{abc} = q(F^d_{abc})$ in writing down the above equations.

Let us define phases $\widehat{U}_g(g(a), g(b), g(c))$ as follows

$$\widehat{U}_g(g(a), g(b), g(c)) := U_g(g(a), g(b), g(c))^q \ . \tag{4.67}$$

Then, we have

$$\widehat{U}_g(g(b), g(a), g(c))\widehat{U}_g(g(a), g(b), g(c))^{-1}\widehat{R}^{g(c)}_{g(a)g(b)} = K^g \widehat{R}^c_{ab} K^g \ , \tag{4.68}$$

and

$$\widehat{U}_g(g(a), g(b), g(e))\widehat{U}_g(g(e), g(c), g(d))\widehat{U}_g(g(b), g(c), g(f))^{-1}\widehat{U}_g(g(a), g(f), g(d))^{-1}$$
$$\times (\widehat{F}^{g(d)}_{g(a)g(b)g(c)})^{g(f)}_{g(e)} = K^g (\widehat{F}^d_{abc})^f_e K^g \ . \tag{4.69}$$

This argument shows that $q(C)$, with MTC data $\widehat{R}^c_{ab}, \widehat{F}^d_{abc}$, has an isomorphic symmetry group, $G$, which acts on its anyons as $g(a)$, but now with an action on the fusion spaces given by $\widehat{U}_g(a, b, c)$. This discussion implies that Galois conjugation preserves the 0-form symmetry of the theory.[31]

To understand what happens to the Postnikov class, let us also define the phases $\widehat{\gamma}_a(g, h)$

$$\widehat{\gamma}_a(g, h) := (\gamma_a(g, h))^q \ , \tag{4.70}$$

where $\gamma_a(g, h)$ are phases satisfying (4.15). It is clear that we have,

$$\frac{\widehat{\gamma}_a(g, h)\widehat{\gamma}_b(g, h)}{\widehat{\gamma}_c(g, h)} = \widehat{U}_g(a, b, c)^{-1}(K^g \widehat{U}_h^{-1}(g^{-1}(a), g^{-1}(b), g^{-1}(c))K^g)\widehat{U}_{gh}(a, b, c) \ , \tag{4.71}$$

If $\beta(g, h, k)$ is the Postnikov class of $C$, it satisfies (from (4.27))

$$\Omega_a(g, h, k) = \frac{S_{a\beta(g,h,k)}}{S_{a1}} \ . \tag{4.72}$$

---

[31]More precisely, what we have shown is that the 0-form symmetry of $C$ maps to a subgroup of that of $q(C)$. But, using the invertibility of the Galois action, we can run the above argument starting from $q(C)$ proving that their 0-form symmetry groups are indeed isomorphic.

Here $\frac{S_{a\beta(g,h,k)}}{S_{a1}}$ is a phase for an abelian anyon, $\beta(g,h,k)$. Hence, Galois conjugation by $q$ corresponds to taking its $q^{th}$ power. So we have,

$$\frac{\widehat{S}_{a\beta(g,h,k)}}{\widehat{S}_{a1}} = q\left(\frac{S_{a\beta(g,h,k)}}{S_{a1}}\right) = \left(\frac{S_{a\beta(g,h,k)}}{S_{a1}}\right)^q . \quad (4.73)$$

Also, from the relation between $\widehat{\gamma}_a(g,h)$ and $\gamma_a(g,h)$, we have

$$\widehat{\Omega}_a(g,h,k) = (\Omega_a(g,h,k))^q , \quad (4.74)$$

where $\widehat{\Omega}_a(g,h,k)$ is defined similarly to (4.23), but now with $\widehat{\gamma}_a(g,h)$.

Using (4.72) we have,

$$\widehat{\Omega}_a(g,h,k) = \frac{\widehat{S}_{a\beta(g,h,k)}}{\widehat{S}_{a1}} . \quad (4.75)$$

Hence, the Postnikov class, $\widehat{\beta}(g,h,k)$, of $q(C)$ is the same as that of $C$. This discussion shows that Galois conjugation preserves the complete 2-group symmetry of a multiplicity-free TQFT.

In the next subsection, we will extend the argument in this section to TQFTs with multiplicity in its fusion rules.

*4.3.2. General TQFTs*

Let us consider a general MTC, $C$, with defining number field, $K_C$ (i.e., we do not impose a restriction on multiplicity or take $K_C$ to necessarily be cyclotomic). In this case, the transformation laws for the $F$ and $R$ matrices under the symmetry action are more complicated.

$$\sum_{\mu'\nu'}[U_g(g(b),g(a),g(c))]_{\mu\mu'}(R^{g(c)}_{g(a)g(b)})_{\mu'\nu'}[U_g(g(a),g(b),g(c))^{-1}]_{\nu'\nu} = K^g(R^c_{ab})_{\mu\nu}K^g , \quad (4.76)$$

$$\sum_{\alpha'\beta',\mu',nu'} [U_g(g(a),g(b),g(e))]_{\alpha\alpha'}[U_g(g(e),g(c),g(d))]_{\beta\beta'}(F^{g(d)}_{g(a)g(b)g(c)})^{(g(f),\mu',\nu')}_{(g(e),\alpha',\beta')}$$
$$\times [U_g(g(b),g(c),g(f))^{-1}]_{\mu'\mu}[U_g(g(a),g(f),g(d))^{-1}]_{\nu'\nu} = K^g(F^d_{abc})^{(f,\mu,\nu)}_{(e,\alpha,\beta)}K^g . \quad (4.77)$$

Note that the above equations form a set of polynomial equations for $U_g(a,b,c)$ with coefficients belonging to $K_C$. Hence, if the $U_g(a,b,c)$'s belong to a finite field extension, then it has to be an extension over $K_C$. The following Lemma shows that the $U_g(a,b,c)$'s belong to a finite field extension:

**Lemma 4.1:** [17] Algebraic points of a complex affine algebraic variety defined over $\overline{\mathbb{Q}}$ are dense in the Zariski topology.

We know that there is a gauge in which $F$ and $R$ matrices are given in an algebraic number field. Any algebraic number field is a subfield of $\overline{\mathbb{Q}}$. Hence, $U_g(a, b, c)$ are solutions of polynomials with coefficients in $\overline{\mathbb{Q}}$. Using the Lemma above, it is clear that there is a gauge in which $U_g(a, b, c)$ belongs to an algebraic field, say $K'_U$. Let $K_U$ be the normal closure of $K'_U$. This procedure defines a Galois field, and $K_U$ is, in general, a field extension of $K_C$.

We expect the equations (4.9) and (4.8) to give a unique solution up to symmetry gauge transformations.[32] Hence, any element $p \in \text{Gal}(K_U/K_C)$ acts on $U_g(a, b, c)$ to relate it to another set of solutions which is gauge equivalent to the one we started with.

The existence of the Galois field $K_U$ shows that we have an action of $\text{Gal}(K_U)$ on $F$, $R$, and $U_g$. Therefore, we have a map from MTC data with symmetry $g$ and symmetry action $U_g$ on the fusion spaces to another such system. Consider the Galois action on the $F$ and $R$ matrices corresponding to some $q \in \text{Gal}(K_C)$. We know that there exists some $\sigma \in \text{Gal}(K_U)$ such that the restriction of the action of $\sigma$ to $K_C$ is equal to $q$. Hence, $\sigma(U_g(a, b, c))$ is a solution for the equations (4.77) and (4.76) where the $F$ and $R$ matrices are replaced by $\sigma(F) = q(F)$ and $\sigma(R) = q(R)$.

Note that the equations (4.77) and (4.76) are not algebraic. For anti-unitary symmetries, we have a complex conjugation action on the $F$ and $R$ symbols which may not commute with the Galois action. If $F$ and $R$ belongs to a CM field, then we know that any Galois conjugation commutes with complex conjugation. Therefore, we get the following result:

**Theorem 4.2:** A TQFT and its Galois conjugates have isomorphic unitary and anti-unitary 0-form symmetries provided there is a gauge in which the $F$ and $R$ symbols of the TQFT belong to a CM field.

For unitary symmetries, the equations (4.77) and (4.76) are algebraic. Therefore, we get the corollary

**Corollary 4.3:** A TQFT and its Galois conjugates have isomorphic 0-form unitary symmetries.

---

[32]This statement has been proven in the case with no multiplicity [56], but it is an open problem in the general case.

In order to check whether the whole 2-group is invariant under Galois conjugation, we have to show that the Postnikov class remains invariant under it. In order to find the Postnikov class, we have to solve the constraint

$$\frac{\gamma_a(g,h)\gamma_b(g,h)}{\gamma_c(g,h)}\delta_{\mu\nu} = \sum_{\alpha,\beta}[U_g(a,b;c)^{-1}]_{\mu\alpha}K^{q(g)}[U_h[\overline{g}(a),\overline{g}(b),\overline{g}(c)]_{\alpha\beta}K^{q(g)}[U_{gh}(a,b,c)]_{\beta\nu} \ . \tag{4.78}$$

Using the same arguments as we used in analyzing the Galois action on the $U_g(a,b,c)$, we can define a Galois field, $K_\gamma$, containing $\gamma_a(g,h)$, that is, in general, a field extension of $K_U$. Corresponding to every element $q \in \text{Gal}(K_C)$, where $K_C$ is the Galois field containing the $F$ and $R$ symbols, we have some $\sigma \in \text{Gal}(K_\gamma)$ such that $\sigma|_{K_C} = q$. The phases $\sigma(\gamma_a(g,h))$ satisfy the constraint (4.78) with $U_g(a,b,c)$ replaced by $\sigma(U_g(a,b,c))$ if Galois action on the $U_g$ matrices commutes with complex conjugation.

Therefore, we find that if $g$ is a unitary symmetry of an MTC, $C$, with symmetry action phases $U_g(a,b,c)$ and $\gamma_a(g,h)$ satisfying (4.78), then the Galois conjugate theory $q(C)$ for some $q \in \text{Gal}(K_C)$ has symmetry $g$ with symmetry action phases $\sigma(U_g(a,b,c))$ and $\sigma(\gamma_a(g,h))$ where $\sigma \in \text{Gal}(K_\gamma)$ and $\sigma|_{K_C} = q$. If $q$ is anti-unitary, then the same is true if $K_C$ and $K_U$ are CM fields.

If $K_U$ is a cyclotomic field extension, we can show that the $\gamma_a(g,h)$ also belong to a cyclotomic field. Indeed, suppose we have $K_U = \mathbb{Q}(\xi_M)$ for some integer $M$ to which $U_g$ belongs to. Since the RHS of (4.78) is a phase, it should have an order which divides $M$. Hence, we have

$$\left(\frac{\gamma_a(g,h)\gamma_b(g,h)}{\gamma_c(g,h)}\right)^M = \frac{\gamma_a(g,h)^M\gamma_b(g,h)^M}{\gamma_c(g,h)^M} = 1 \ , \tag{4.79}$$

whenever $N_{ab}^c \neq 0$. Therefore, we can perform the $\nu$-gauge transformation

$$\gamma_a(g,h)^M \to \gamma_a(g,h)^M \nu_a(g,h) \ , \tag{4.80}$$

where $\nu_a(g,h) = \gamma_a(g,h)^{-M}$ to set $\gamma_a(g,h)^M = 1$ for all anyons $a$ and $g,h \in G$. This shows that there exists a $\nu$-gauge in which the phases $\gamma_a(g,h)$ all belong to $\mathbb{Q}(\xi_M)$. Hence, given $U_g$ matrices, the solutions to (4.78) belong to $\mathbb{Q}(\xi_M)$.

To complete our discussion, note that we have the relation

$$\Omega_a(g,h,k) = \frac{S_{a\beta(g,h,k)}}{S_{a1}} \ , \tag{4.81}$$

where $\Omega_a(g,h,k)$ is defined in (4.23). Under the action of any $\sigma \in \text{Gal}(K_\gamma)$, we have

$$\sigma(\Omega_a(g,h,k)) = \sigma\left(\frac{S_{a\beta(g,h,k)}}{S_{a1}}\right) \ . \tag{4.82}$$

Since $\widehat{\Omega}_a(g,h,k) = \sigma(\Omega_a(g,h,k))$ and $\widehat{S}_{ab} = \sigma(S_{ab})$ are the respective quantities in the Galois conjugate theory, we have

$$\widehat{\Omega}_a(g,h,k) = \frac{\widehat{S}_{a\beta(g,h,k)}}{\widehat{S}_{a1}} \ . \tag{4.83}$$

The actions of such $\sigma$'s exhaust all possible Galois conjugations of $F$ and $R$.

In summary, we have the following result:

**Theorem 4.4:** A TQFT and its Galois conjugates have isomorphic 2-group symmetry provided that there is a gauge in which the $F$ and $R$ symbols as well as the $U_g(a,b,c)$ belong to a CM field.

For unitary symmetries, all the constraints involved are algebraic. Therefore, we get the corollary

**Corollary 4.5:** A TQFT and its Galois conjugates have the same unitary 2-group symmetry.

By a unitary 2-group symmetry, we mean a 2-group symmetry in which the 0-form symmetry is a group of unitary symmetries. Note that the set of TQFTs with the same fusion rules shares the same 1-form symmetry group. However, they may not share the same 0-form, and consequently the same 2-group symmetry. For example, the Toric code has $\mathbb{Z}_2$ 0-form symmetry while the 3-fermion model has an $S_3$ 0-form symmetry group. However, our results above show that Galois orbits should contain TQFTs with the same 0-form and 2-group symmetries (up to a mild assumption for anti-unitary symmetries).

*4.4. Gauging and Galois Conjugation*

In previous sections, we studied how Galois conjugation acts on the space of TQFTs and how it acts on specific families of TQFTs within it. Gauging is another way to move through the space of TQFTs.

If a TQFT $\mathcal{T}$ has a 0-form symmetry given by some finite group $G$, it can be gauged to obtain a new TQFT, $\mathcal{T}/G$. We will somewhat unconventionally refer to $\mathcal{T}$ and $\mathcal{T}/G$ as the *magnetic and electric theories*, respectively.[33] The $\mathcal{T}/G$ TQFT does not have the 0-form symmetry $G$. Instead, it has a $\text{Rep}(G)$ fusion subcategory. We can go from $\mathcal{T}/G$ to $\mathcal{T}$ by condensing Rep($G$) [57]. When $G$ is abelian, this condensation is the same as gauging the 1-form symmetry, Rep($G$) [58]. Therefore, Rep($G$) condensation is the inverse of $G$ gauging.

---
[33]Let $C^G$ be the MTC corresponding to the TQFT $\mathcal{T}/G$, where $C$ is the MTC corresponding to $\mathcal{T}$. We will use the notation $C^G$ and $\mathcal{T}/G$ more or less interchangeably to denote the electric theory.

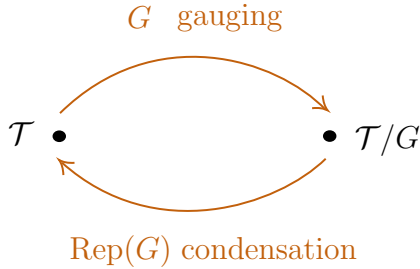

**Fig. 9:** Gauging and condensation are inverses.

We would like to understand how gauging a 0-form symmetry interacts with Galois conjugation. Our discussion of the Galois action on Rep($G$) reveals the following result:

**Theorem 4.6:** If a TQFT, $\mathcal{T}$, is obtained from gauging a symmetry $G$ of another TQFT, so are all of its Galois conjugates.

**Proof:** Since $\mathcal{T}$ is obtained from another TQFT by gauging $G$, it contains a fusion subcategory, Rep($G$). Under a Galois conjugation of $\mathcal{T}$, the resulting theory, $\mathcal{T}'$, also has a fusion subcategory Rep($G$). This statement holds because Rep($G$) is invariant under Galois conjugation. Now, we can condense Rep($G$) $\subset \mathcal{T}'$ to obtain another TQFT with 0-form symmetry $G$. Therefore, $\mathcal{T}'$ is also obtained from gauging a symmetry $G$ of some TQFT. $\square$

In fact, Galois conjugation of a TQFT, $C^G$, with a Rep($G$) subcategory can be related to Galois conjugation of the TQFT, $C$, obtained by condensing the Rep($G$) subcategory. To obtain $C$ from $C^G$ through condensation of Rep($G$), we don't have to keep track of all the anyons in $C^G$. In fact, the anyons in $C$ correspond to the anyons in the subcategory, $L \subset C^G$, which braid trivially with all condensing anyons. That is, $L$ is the centralizer of Rep($G$) in $C^G$

$$L = \{c \in C^G | S_{ca} = \frac{1}{\mathcal{D}} d_c d_a \; \forall a \in \text{Rep}(G)\} \; . \tag{4.84}$$

The twists of the anyons in $C$ are completely determined by the twists of the anyons in $L$. Moreover, the quantum dimensions of the anyons in $C$ are the same as those in $L$, up to some integer factors. Therefore, the cyclotomic field containing the ($\tilde{S}, T$) modular data of $L \subset C^G$, $\mathbb{Q}(\xi_M)$, is the same as the cyclotomic field containing the corresponding modular data of the anyons in $C$, $K_M$.[34] Also, from Theorem 3.4, we know that the fusion rules of the boundary excitations are invariant under Galois action of the bulk TQFT. Condensation of anyons is a more general procedure, where we have a domain wall between two phases instead of a gapped boundary. In fact, the TQFT obtained after condensation is described by a modular subcategory of the category of representations of the connected commutative separable algebra describing the boundary excitations [40]. Therefore, the fusion rules of $C$ obtained after condensation are invariant under Galois conjugation of $C^G$. Since the (un-normalized) $S$ matrix of $C$ is determined by the twists and quantum dimensions of $L$ along with the fusion rules of $C^G$, we have the following result:

**Theorem 4.7:** Galois conjugation of $C^G$ with respect to $q \in \text{Gal}(K_{C^G})$, where $K_{C^G}$ is the defining Galois field of $C^G$, induces a Galois action on the modular data of $C$ by $q|_{K_M}$, where $K_M$ is the subfield containing the $(\tilde{S}, T)$ modular data of $C$.

**Proof:** Let $K_{C^G}$ be the algebraic field extension containing the data of the MTC $C^G$. Consider the Galois conjugation of $C^G$ by some $q \in \text{Gal}(K_{C^G})$. The cyclotomic field $K_M$ is a subfield of $K_{C^G}$, which is a normal extension of $\mathbb{Q}$. Therefore, the restriction $q|_{K_M}$, where $q \in \text{Gal}(K_{C^G})$ acts on $K_M$ as Galois action on the field and this restriction is surjective. Since the modular data of $C$ is determined by twists and quantum dimensions of $L$, as well as the fusion rules of $C^G$, $q \in \text{Gal}(K_{C^G})$ action on $C^G$ induces a $q|_{K_M}$ action on the modular data of $C$. $\square$

In fact, the results above can be generalized due to the algebraic nature of 0-form symmetry gauging. To understand this statement, consider a fusion category, $C$, with a $G$-action. Gauging $G$ amounts to constructing the category of $G$-equivariant objects. A $G$-equivariant object is a pair, $(x, u_g)$, for all $g \in G$ and $x$ an object in $C$. Here, $u_g$ are isomorphisms, $u_g : g(x) \to x$, such that the following constraint is satisfied for all $g, h \in G$

$$u_{gh} \circ \gamma_a(g, h) = u_g \circ g(u_h) \ , \tag{4.86}$$

where $\gamma_a(g, h)$ is the isomorphism $g(h(a)) \to gh(a)$. This discussion is analogous to how we go from a global symmetry acting on a Hilbert space, which acts non-trivially on the states, to a gauged theory where the physical states are invariant under the gauge group. In the $G$-equivariant object $(x, u_g)$, $u_g$ is the isomorphism which tells us that $g(x)$ is the same as

$$\tilde{S}_{ab} = \sum_c N_{ab}^c \frac{\theta_c}{\theta_a \theta_b} d_c \ . \tag{4.85}$$

---

[34]This statement follows from re-writing the un-normalized $S$ matrix as [59]

$x$. Since $C$ is a tensor category, we also have isomorphisms, $\psi_g(a,b): g(a) \otimes g(b) \to g(a \otimes b)$. The morphisms between $G$-equivariant objects are

$$\text{Hom}((x, u_g), (y, v_g)) = \{f \in \text{Hom}(x,y) | v_g \circ g(f) = f \circ u_g \; \forall g \in G\} \;. \tag{4.87}$$

The tensor product of objects is

$$(x, u_g) \otimes (y, v_g) = (x \otimes y, w_g) \;, \tag{4.88}$$

where $w_g = u_g v_g \circ \psi_g^{-1}(a,b)$. The $G$-equivariant objects form a fusion category $C^G$ (for a detailed discussion of this construction see [60, 61]).

Given a TQFT with a 0-form symmetry $G$, we have a corresponding MTC, $C$, with a $G$ action. Provided that certain obstructions vanish, we can construct a G-crossed braided category, $C_G$, from $C$ with a $G$-crossed braiding [62]

$$c_{x,y}: x \otimes y \to g(y) \otimes x \text{ where } x \in C_g, g \in G, y \in C \;. \tag{4.89}$$

This amounts to adding the data of the symmetry defects. In a somewhat more field theoretical language, this step can be thought of as coupling the theory to background gauge fields prior to gauging [63]. Gauging the symmetry, $G$, then amounts to constructing the category of $G$-equivariant objects of $C_G$. The G-crossed braiding in $C_G$ can be used to endow $C^G$ with a braiding as follows [60]

$$b_{(x,u_g),(y,v_g)} = (v_g \otimes id_{x_g}) \circ c_{x_g,y} \;, \tag{4.90}$$

where $x = \oplus_g \; x_g$. Note that the braided fusion category, $C^G$, is modular if and only if $C$ is modular, and the grading in $C_G$ is faithful (recall that since $C$ is modular, it has a spherical structure, so $C^G$ is also spherical) [64].

Since the data of $C^G$ and $C_G$ are related algebraically, every Galois action on $C_G$ leads to a Galois conjugated $C^G$ and vice-versa. We can also use our discussion on Galois action and Drinfeld center to obtain this result. Indeed, suppose we have some MTC, $C$, with 0-form symmetry, $G$. Let us also suppose that the obstructions to gauging vanishes and we have a G-crossed braided category, $C_G$. Let $C^G$ be the TQFT obtained after gauging the symmetry $G$. These theories are related in the following way [65]

$$C \boxtimes \bar{C}^G = \mathcal{Z}(C_G) \;. \tag{4.91}$$

Here, $\bar{C}^G$ is a modular tensor category with braiding given by $\bar{c}_{x,y} = c_{y,x}^{-1}$, where $c_{x,y}$ is the braiding of $C^G$.

To understand this relation, it is useful to examine two special cases. When $C$ is the trivial TQFT, then $C_G$ is equivalent to $\text{Vec}_G^\omega$, and the above relation becomes $\bar{C}^G = \mathcal{Z}(\text{Vec}_G^\omega)$, which is the familiar result that taking the Drinfeld center of $\text{Vec}_G^\omega$ is the same as gauging a natural isomorphism of the trivial TQFT (up to inverted braiding). When the group $G$ is trivial, this relation becomes $C \boxtimes \bar{C} = Z(C)$, which shows that the Drinfeld center of an MTC is a Deligne product of that MTC with itself up to inverted braiding. Equation (4.91) implies that the MTC data of the various TQFTs appearing in (4.91) are related via

$$F_C \otimes F_{\bar{C}^G} = F_{\mathcal{Z}(C_G)}, \ R_C \otimes R_{\bar{C}^G} = R_{\mathcal{Z}(C_G)} . \tag{4.92}$$

Therefore, the MTC data of $C^G$ can be determined in terms of the data of $C$ and $\mathcal{Z}(C_G)$.

Suppose we Galois conjugate $C_G$ w.r.t. some $q \in \text{Gal}(K_{C_G})$. $C$ is a modular subcategory of $C_G$. Therefore, $q$ acts on $C$. From (3.9) we have some $q' \in \text{Gal}(K_{\mathcal{Z}(C_G)})$ such that

$$\mathcal{Z}(q(C_G)) = q'(\mathcal{Z}(C_G)) . \tag{4.93}$$

Hence, we get

$$\mathcal{Z}(q(C_G)) = q'(\mathcal{Z}(C_G)) = q'(C \boxtimes \bar{C}^G) = q(C) \boxtimes q''(C^G) , \tag{4.94}$$

where $q'' \in \text{Gal}(K_{C^G})$, and in the last equality above we have used the fact that any Galois action on a Deligne product can be written as a Galois action on the individual TQFTs. We have also used the fact that $q$ acts on $C$ when $q$ acts on $C_G$. Therefore, we find that Galois action on the G-crossed braided theory induces a Galois action on the gauged theory.

As a consequence, similarly to Theorem 3.4, we obtain the following:

**Theorem 4.8:** Corresponding to every $q \in \text{Gal}(K_{C_G})$, there exists a $q' \in \text{Gal}(K_{C^G})$ such that

$$(q(C_G))^G = q'(C^G) , \tag{4.95}$$

where $(q(C_G))^G$ denotes gauging the $G$ symmetry after Galois acion on $C_G$.[35]

Given an MTC, $C$, with symmetry, $G$, there is a cohomological classification of unitary G-crossed braided categories that can be constructed from $C$ [57, 62]. We can use this classification to describe the Galois action on $C_G$ more explicitly. To gauge a symmetry $G$ of $C$, we should have trivial Postnikov class. This is because a non-trivial Postnikov class leads to a coupling between gauge transformations of the 1-form and 0-form symmetry background gauge fields [63]. Therefore, the 0-form symmetry alone cannot be gauged, though if the 0-form and 1-form 't Hooft anomaly vanishes, the full 2-group can be gauged. If the Postnikov class vanishes, then the classification follows from a choice of the fractionalization class, $\eta$, which forms a torsor over $H^2_{[\rho]}(G, A)$, where $[\rho]$ indicates that we have a twisted cohomology group due to the $G$ group action on the abelian anyons, $A$, in $C$. Given a fractionalization class, it determines element of the group $H^4(G, U(1))$ which is the 't Hooft anomaly of the symmetry $G$ (which is also sometimes called the defectification obstruction). If the 't Hooft anomaly vanishes, we can gauge the symmetry $G$.

---

[35]See [66] for a similar result in the context of gauging symmetries of certain VOAs.

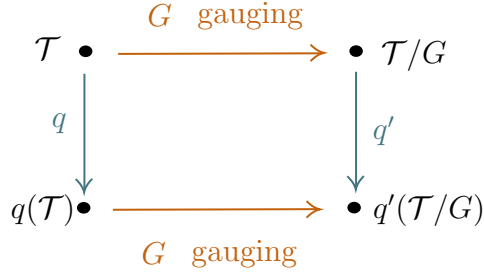

**Fig. 10:** Galois conjugation of $\mathcal{T}$ induces a Galois action on $\mathcal{T}/G$ and vice-versa.

However, before gauging the symmetry, we have the freedom to stack an SPT. That is, given the G-crossed braided theory $C_G$, we can form the Deligne product

$$C_G \boxtimes_G \mathrm{Vec}_G^\omega \ , \tag{4.96}$$

where $\omega \in H^3(G, U(1))$. The subscript $G$ on the Deligne product indicates that we should take a product of the $C_g$ sector of $C_G$ with the $g$ anyon in $\mathrm{Vec}_G^\omega$. We will denote a $G$-crossed braided theory obtained from these choices as $C_G(\eta, \alpha)$. The phase $\eta_a(g, h)$, which is the fractionalization class when $a$ is a genuine anyon, enters into the Heptagon equations; these equations need to be solved in order to construct $C_G$. Therefore, a Galois action on $C_G$ by some $q \in \mathrm{Gal}(K_{C_G})$ should act on $\eta$ as[36]

$$\eta_a(g, h) \to q(\eta_a(g, h)) \ . \tag{4.97}$$

Similarly, since $\omega \in H^3(G, U(1))$ enters into the gauging procedure through stacking by an SPT, under Galois conjugation we get[37]

$$\omega(g, h, k) \to q(\omega(g, h, k)) \ . \tag{4.98}$$

---

[36]If $\eta_a(g, h)$ is a root of unity, then the Galois action will act on it by raising it to a power co-prime to the order of $\eta_a(g, h)$.

[37]Since $\omega(g, h, k)$ can always be chosen to be a root of unity, the Galois action on it can also be written

Therefore, Galois conjugations which take a unitary G-crossed braided MTC to a unitary G-crossed braided MTC are completely specified by their action on $C$, $\eta_a(g,h)$, and $\omega(g,h,k)$. In particular, if $C$, $\eta_a(g,h)$, and $\omega(g,h,k)$ are invariant under Galois action, then there are no unitarity preserving non-trivial Galois actions on $C_G$. Therefore, the corresponding gauged theory, $C^G$, is not related through Galois conjugations to other unitary theories.

### 4.4.1. Example: Spin$(k)_2$ Chern-Simons Theory

We have already seen that the Spin$(5)_2$ theory has a non-trivial Postnikov class. This theory can be obtained from the $A_5$ abelian TQFT by gauging the charge-conjugation symmetry. The $A_5$ theory also has a time-reversal symmetry given by

$$T : j \to 2j \ , \tag{4.99}$$

where $T^2 = C$, and $C$ is the charge-conjugation symmetry. We can generalize this procedure to generate an infinite family of theories with non-trivial Postnikov class and then explicitly analyze the Galois action.

Let us consider a general abelian TQFT with fusion rules forming the group $\mathbb{Z}_k$. For $\mathbb{Z}_k$ fusion rules, there are several gauge-inequivalent solutions to the Pentagon and Hexagon equations labelled by $p = 0, \cdots, k-1$. The twists of the anyons in the $\mathbb{Z}_k$ MTC, corresponding to a choice of $p$, are

$$\theta_a = e^{\frac{2\pi i p a^2}{k}} \ . \tag{4.100}$$

We have the set of anyons $0, 1, \cdots, k-1$. Irrespective of $k$, we always have the charge conjugation symmetry

$$C : j \to -j \bmod k \ . \tag{4.101}$$

However, the TQFT has a time-reversal symmetry if and only if $k$ satisfies $1+l^2 = 0 \bmod k$ for some integer $l$ [67]. We will assume that $k$ is odd. The time-reversal symmetry is given by

$$T : j \to lj \bmod k \ . \tag{4.102}$$

It is clear that $T^2 = C$. Hence, we have a $\mathbb{Z}_4 = \{e, z, c, cz\}$ time-reversal symmetry and a $\mathbb{Z}_2 = \{e, c\}$ charge conjugation symmetry. The idea is to gauge this charge conjugation symmetry. To that end, we have to first construct the $\mathbb{Z}_2$-crossed braided category $(\mathbb{Z}_k)_{\mathbb{Z}_2}$. We have

$$(\mathbb{Z}_k)_{\mathbb{Z}_2} = C_e \oplus C_c \ , \tag{4.103}$$

---

as $\omega(g,h,k) \to \omega(g,h,k)^p$ where $p$ is an integer co-prime to the order of $\omega(g,h,k)$ specified by the restriction of $q \in \text{Gal}(K_C)$ to the cyclotomic field containing $\omega$.

where $C_e$ contains the anyons $0, ..., k - 1$. For odd $k$, vaccum is the only element invariant under charge conjugation. Hence, $C_c$ contains only a single defect $\psi$. Along with the fusion rules of the anyons in $\mathbb{Z}_k$, the $\mathbb{Z}_2$-crossed braided theory has the fusion rules

$$\psi \otimes j = \psi , \quad \psi \otimes \psi = 0 \oplus ... \oplus k - 1 , \tag{4.104}$$

which implies that $d_\psi = \sqrt{k}$.

It is easy to verify that the $H^2_{[\rho]}(\mathbb{Z}_2, \mathbb{Z}_2)$ group is trivial. Therefore, there is a unique fractionalization class. Moreover, $H^4(\mathbb{Z}_2, U(1)) \cong \mathbb{Z}_1$, and the $\mathbb{Z}_2$ charge conjugation symmetry does not have a 't Hooft anomaly. As a result, this symmetry can be gauged. To obtain the anyons in the gauged theory, we need the $\mathbb{Z}_2$ orbits and their stabilizers. We have the following orbits: $[0]$, $[1]$, $\cdots$, $[\frac{k-1}{2}]$, $[\psi]$. The $[1]$, $\cdots$, $[\frac{k-1}{2}]$ orbits have trivial stabilizers, while $[0]$ and $[\psi]$ have a $\mathbb{Z}_2$ stabilizer group. The representations of $\mathbb{Z}_2$ can be labelled by $[+], [-]$, where $[+]$ is the trivial representation. We have the following anyons in the gauged theory

$$([0], [+]) , \ ([0], [-]) , \ ([1], \mathbb{1}) , \cdots , \ \left(\left[\frac{k-1}{2}\right], \mathbb{1}\right) , \ ([\psi], [+]) , \ ([\psi], [-]) . \tag{4.105}$$

We will denote the first two anyons as $1, \epsilon$, the last two as $\psi_+, \psi_-$, and the rest by $\phi_j$. For different $p$, the fusion rules of the gauged theory remain the same, however the MTC data of the gauged theory changes. For $p = \frac{k-1}{2}$, it was shown in [57] that the resulting gauged theory has the fusion rules and MTC data of $\text{Spin}(k)_2$ Chern-Simons theory. For other values of $p$, we get theories with the same fusion rules, but different MTC data. In the discussion below, we will choose the value of $p$ to be $\frac{k-1}{2}$.

The topological twists of the anyons are[38]

$$\theta_1 = \theta_\epsilon = 1 , \quad \theta_{\phi_j} = e^{\frac{2\pi i(k-1)j^2}{2k}} , \quad \theta_{\psi_\pm} = \pm\theta_\psi = \pm e^{\frac{2\pi i(k-1)}{16}} . \tag{4.106}$$

Note that the topological twist of the symmetry defect is not invariant under gauge-transformations of the symmetry action. However, the twist of $\psi_\pm$ is given by

$$\theta_{\psi_\pm} = \theta_\psi \chi(\pi_\pm) , \tag{4.107}$$

where $\chi(\pi_a)$ is the projective character of $\pi_a$. In the gauge $\eta_a(g, h) = 1 \ \forall g, h$ we have $\chi(\pi_\pm) = \pm 1$. A symmetry action gauge transformation changes $\theta_\psi$ and $\chi(\pi_a)$ by opposite phases, resulting in gauge-invariant twists $\theta_{\psi_\pm}$.

---

[38]We can stack a non-trivial $\mathbb{Z}_2$ SPT before gauging. $\theta_{\psi_\pm}$ of the resulting gauged theory is same as the twists obtained without SPT stacking up to a factor of $-1$.

If $k$ is such that $\theta_{\psi_{\pm}}$ are complex conjugates of each other (so $k = 5$ mod 8), then we can define a time-reversal symmetry for this theory which acts on the anyons as follows

$$T : \phi_j \to \phi_{qj} \ , \quad T : \psi_+ \to \psi_- \ . \tag{4.108}$$

Since $\mathrm{Spin}(k)_2$ MTC is self-dual, it is clear that this time-reversal symmetry is a $\mathbb{Z}_2$ symmetry. Similar to our analysis of the $\mathrm{Spin}(5)_2$ theory, we can use the explicit MTC data of $\mathrm{Spin}(k)_2$ in [44] to show that this time-reversal symmetry, along with the $\mathbb{Z}_2$ 1-form symmetry generated by the anyon $\epsilon$ forms a non-trivial 2-group.

The authors of [63] describe a much simpler way to show there is a non-trivial 2-group using the sufficient conditions in [55]. Following this procedure, let us assume that the theory has a trivial Postnikov class and show that this leads to a contradiction. If the theory has a trivial Postnikov class, it is realizable at the boundary of a 4D SPT phase. The $\mathbb{RP}^4$ partition function of this 4D SPT phase is given by [55]

$$Z(\mathbb{RP}^4) = \sum_{a, a=T(a)} S_{1a} \theta_a \eta_a \ , \tag{4.109}$$

where $\eta_a$ is the fractionalization class corresponding to the time-reversal symmetry of the $\mathrm{Spin}(k)_2$ theory and $T(a)$ denotes the time-reversal symmetry action on the anyon $a$. For $\mathrm{Spin}(k)_2$, where $k$ satisfies $1 + l^2 = 0$ mod $k$ for some integer $l$ and $k = 5$ mod 8, we can calculate this as

$$Z(\mathbb{RP}^4) = S_{11} \theta_1 \eta_1 + S_{1\epsilon} \theta_\epsilon \eta_\epsilon = \mathcal{D}(1 + \eta_\epsilon) \neq \pm 1 \ . \tag{4.110}$$

However, it is known that the partition function of a time-reversal invariant 4D SPT on $\mathbb{RP}^4$ is valued in $\pm 1$. This shows that the $\mathrm{Spin}(k)_2$ theory ($k = 5$ mod 8) cannot be realized at the surface of a 4D SPT. Hence, the Postnikov class of the theory is non-trivial. Note that if $k$ is such that $\theta_{\psi_{\pm}}$ is real, then $\psi_{\pm}$ are also invariant under the symmetry. Hence, they will contribute to the above partition function. In fact, for these theories the partition function is valued in $\pm 1$. Indeed, in this case the Postnikov class is trivial.[39]

Now that we have explored the 2-group structure of $\mathrm{Spin}(k)_2$ Chern-Simons theory, let us show that the Postnikov class is invariant under Galois actions on this theory. Recall that the $\mathrm{Spin}(5)_2$ TQFT was invariant under all unitarity-preserving Galois actions. We showed this explicitly by studying the Galois action on the $T$ matrix. Alternatively, this can also be seen from the fact that $\mathrm{Spin}(5)_2$ TQFT is obtained from gauging a $\mathbb{Z}_2$ symmetry of the $A_5$ abelian TQFT. Indeed, we know that the fractionalization class is trivial, and the SPT stacking is determined by $\omega \in H^3(\mathbb{Z}_2, U(1))$ (which is valued in $\pm 1$). Therefore, unitarity-preserving Galois actions on $(A_5)_{\mathbb{Z}_2}$ cannot change the G-crossed braided structure. We also know that the $A_5$ TQFT has four Galois actions corresponding to $\mathbb{Z}_5^\times = \{1, 2, 3, 4\}$. The only non-trivial Galois action which preserves the unitarity of $(A_5)_{\mathbb{Z}_2}$ (i.e, which doesn't flip the sign of $d_\psi = \sqrt{5}$) is 4. We also know that $A_5$ is invariant under Galois action by 4. Therefore, we find that $(A_5)_{\mathbb{Z}_2}$ is invariant under all unitarity-preserving Galois actions. Therefore, using Theorem 4.8, we find that the electric theory $\mathrm{Spin}(5)_2$ is invariant under all unitarity-preserving Galois actions.

---

[39]Note that $Z(\mathbb{RP}^4)$ being valued in $\pm 1$ does not guarantee that the Postnikov class is trivial. It is only a necessary condition. But it can be checked that whenever $\psi_{\pm}$ is fixed under the symmetry action, then (4.27) forces the Postnikov class to be trivial.

More generally, the unitarity-preserving Galois actions on $(\mathbb{Z}_k)_{\mathbb{Z}_2}$ are those Galois actions of $\mathbb{Z}_k$ which preserve the quantum dimensions of all anyons and defects in $(\mathbb{Z}_k)_{\mathbb{Z}_2}$. Using Theorem 4.8, these Galois actions correspond to unitarity-preserving Galois acion on $\mathrm{Spin}(k)_2$ TQFT. Indeed, a unitarity-preserving Galois action on $(\mathbb{Z}_k)_{\mathbb{Z}_2}$ with respect to some $q$ co-prime to $k$ can be see as changing our choice of $p = \frac{k-1}{2}$ to $p = \frac{q(k-1)}{2}$. The twists of the resulting gauged theory then becomes

$$\theta_1 = \theta_\epsilon = 1 \ , \quad \theta_{\phi_j} = e^{\frac{2\pi i q(k-1)j^2}{2k}} \ , \quad \theta_{\psi_\pm} = \pm\theta_\psi = \pm e^{\frac{2\pi i q(k-1)}{16}} \ . \tag{4.111}$$

If $\theta_{\psi_+}$ and $\theta_{\psi_-}$ are complex conjugates before Galois action, the same is true after Galois action. Therefore, at the level of the $T$ matrix, $\mathrm{Spin}(5)_2$ and its Galois conjugates have the same time-reversal symmetry structure. This is in agreement with our Theorem 4.2. Morevoer, if the $Z(\mathbb{RP}^4)$ is not valued in $\pm 1$ before Galois action, the same is true after Galois action. Therefore, the Postnikov class is non-trivial before and after Galois action. Similarly, if the symmetry acts trivially on $\psi_\pm$ before Galois action, then we know that the Postnikov class is trivial. This result is also true after Galois action. These observations agree with our Theorem 4.4.

## 4.5. Galois Invariance and Gauging

Suppose $C$ is a Galois-invariant theory with symmetry $G$. It is then natural to ask if this invariance is preserved under gauging 0-form symmetries, 1-form symmetries, and more general anyon condensation. We expect any lack of invariance in the gauged / condensed theory to be due to a kind of generalized mixed 't Hooft anomaly between the Galois action and the symmetry / condensation in question. On the other hand, there may be subtler effects due to such an anomaly that we do not study here, and so the preservation of Galois invariance alone may not be sufficient to conclude that there is no generalized 't

Hooft anomaly.[40] Therefore, all we can say is that that there is a non-trivial Galois action-0-form mixed 't Hooft anomaly if gauging the symmetry $G$ results in a Galois non-invariant theory. Similarly, suppose we have an MTC, $C$, with a 1-form symmetry, $A$. We can say that there is a non-trivial Galois action-1-form mixed anomaly if gauging the symmetry $A$ results in a Galois non-invariant theory. More generally, we can say there is a Galois action-anyon condensation anomaly by replacing $A$ with a general connected commutative separable algebra and finding a non-invariant condensed theory.

Let us study the behavior of Galois invariance under gauging more carefully. To that end, suppose $C$ is Galois invariant. Using Theorem 4.8, we find that $C^G$ is Galois invariant if and only if $C_G$ is Galois invariant. This follows from the one-to-one relation between G-crossed braided categories and categories with a Rep($G$) subcategory. We therefore get the following result:

**Lemma 4.9:** Starting from a Galois invariant MTC, $C$, with 0-form symmetry, $G$, we obtain a Galois invariant theory, $C^G$, after gauging if and only if $C_G$ is Galois invariant.

In more field theoretical language, the above lemma amounts to the statement that the Galois invariance of the gauged TQFT can be determined by turning on background fields for $G$ and studying the Galois invariance of the TQFT prior to gauging. In the examples section, we will study particular TQFTs where 0-form gauging preserves the Galois invariance as well as cases where 0-form gauging violates the Galois invariance.

Next let us discuss how Galois invariance interacts with anyon condensation. To that end, suppose $C^G$ is Galois invariant, then it follows from (3.9) that $C$ is Galois invariant if and only if $C_G$ is Galois invariant. Suppose $C_G$ is not Galois invariant. Then there exists some $q \in \text{Gal}(K_{C_G})$ such that $q(C_G)$ is inequivalent to $C_G$. We have some $q' \in \text{Gal}(K_{C^G})$ such that $(q(C_G))^G = q'(C^G)$. Since $q(C_G)$ is inequivalent to $C_G$, $q'(C^G)$ has to be different from $C^G$. This contradicts the assumption that $C^G$ is Galois invariant. Therefore, $C_G$ should be Galois invariant. We get the result:

**Lemma 4.10:** If we start from a Galois-invariant theory, then the theory after anyon

---

[40]Indeed, in the more standard case of mixed 't Hooft anomalies between 0-form symmetries, gauging part of the 0-form symmetry group can sometimes lead to non-trivial 2-groups and other phenomena [68]. As a result, one may wonder if there is a generalization of this story involving Galois actions as well. As another possibility, recall that a mixed 0-form / 1-form 't Hooft anomaly can result in a non-trivial group extension for the 0-form symmetry after 1-form symmetry gauging [68, 69]. It would be interesting to study whether there is a generalization of this story to Galois group extensions under 1-form symmetry gauging / anyon condensation.

condensation is also Galois invariant.

More generally, Lemma 4.10 implies that, for every element of the Galois group that leaves the electric theory invariant, there is a (not necessarily unique) Galois action on the magnetic theory that leaves it invariant.[41] For example, consider a TQFT invariant under complex conjugation. If the TQFT has real MTC data this is of course trivially true. But if the MTC data is complex, then there exists a combination of gauge transformations and a map between the anyons of the TQFT and its complex conjugate preserving the fusion rules. Sometimes, such a map along with a gauge transformation arises from the time-reversal symmetry of the TQFT. However, there may not be a unqiue way to lift the complex conjugation Galois action to a time-reversal symmetry.

For example, consider the $A_5$ TQFT. The complex conjugation Galois action can be reversed using a permutation of the anyons $T(a) = 2a$ mod 5, which is a time-reversal symmetry. However, $T^3$ is also a time-reversal symmetry. Therefore, the complex conjugation Galois action can be reversed using $T$ or $T^3$. Note that complex conjugation Galois action is always order two, while time-reversal symmetry may not be order two (it is order four in the $A_5$ example). This discrepancy is due to the fact that Galois conjugation acts directly on the MTC data, and an order two permutation of the anyons reversing this Galois action may not preserve the fusion rules of the MTC.

We know that the $\text{Spin}(5)_2$ Chern-Simons theory can be obtained by gauging the $\mathbb{Z}_2$ charge-conjugation symmetry of $A_5$ TQFT. In this case both the electric and magnetic theories are invariant under the complex conjugation Galois action. In the electric theory, the Galois invariance can be lifted to an order-two time-reversal symmetry, while on the magnetic side, it can be lifted only to an order-four time-reversal symmetry. The origin of this order-four time-reversal symmmetry is due to the non-trivial mixed 't Hooft anomaly between the order-two time-reversal symmetry and the $\mathbb{Z}_2$ 1-form symmetry in the electric theory. The magnetic theory then has a $\mathbb{Z}_4$ time-reversal symmetry which arises from a group extension of the $\mathbb{Z}_2$ time-reversal symmetry of the electric theory by the $\mathbb{Z}_2$ charge conjugation symmetry of the magnetic theory [68].

*4.6. Galois Fixed Point TQFTs*

In this section, our goal is to better elucidate generalizations of the basic unitary Galois fixed point TQFTs we encountered earlier (i.e., the 3-Fermion Model, Toric Code, Double

---

[41]In particular, this statement is true even if there are other elements of the Galois group that do not leave the electric theory invariant.

Semion, and various other more complicated (twisted) discrete gauge theories). Of course, most TQFTs transform non-trivially under Galois conjugation. For example, consider a theory which is not integral. Such a TQFT should have at least one anyon, say $a$, with a real irrational quantum dimension, $d_a \notin \mathbb{Q}$. Then there exists a Galois conjugation which acts non-trivially on $d_a$ and results in a different TQFT. More generally, we have the following theorem:

**Lemma 4.11:** All unitary Galois-invariant TQFTs have only integer quantum dimensions.

**Proof:** Consider a unitary MTC, $C$. Recall from Lemma 2.2 that a unitarity-preserving Galois conjugation by an element, $g$, must satisfy

$$g(d_a) = d_a ,\qquad(4.112)$$

for all $d_a$. As a result, $d_a \in \mathbb{Q} \ \forall a \in C$. Since quantum dimensions are algebraic integers, the rational root theorem guarantees that all $d_a \in \mathbb{Z}$. $\square$

Note that this result does not hold for non-unitary Galois fixed point theories. Indeed, consider the following TQFT

$$\mathcal{T} = \boxtimes_{q \in \mathrm{Gal}(K_{C_0})} q(\mathcal{T}_0) ,\qquad(4.113)$$

where $\mathcal{T}_0$ is a TQFT with at least one irrational quantum dimension, and $C_0$ is the associated MTC. In (4.113), we take a product over the full Galois orbit of $\mathcal{T}_0$ (thereby rendering $\mathcal{T}$ Galois-invariant). Since there is an irrational quantum dimension, the product TQFT, $\mathcal{T}$, will contain at least one non-unitary factor and hence will be non-unitary. As an example, we can take $\mathcal{T}_0$ to be the Fibonacci theory (then there will be anyons with quantum dimension $(1 \pm \sqrt{5})/2$ in $\mathcal{T}$). Finally, note that not all integral theories are Galois invariant. For example, consider the Semion TQFT. Therefore, unitary Galois invariant TQFTs should lie in the subspace of integral TQFTs.[42]

Interestingly, all known examples of integral TQFTs are also weakly group theoretical (the converse does not hold). These latter TQFTs are under good control since they all have a Tannakian subcategory that comes from gauging a symmetry of a weakly anisotropic abelian TQFT [70]. Weakly anisotropic pointed categories are classified in [64]. The upshot is that any weakly anisotropic abelian TQFT is of the form $D \boxtimes A$, where $D$ is the discrete gauge theory, $\mathcal{Z}(\text{Vec}_G)$, where $G$ is an abelian group consisting of a direct sum of cyclic groups of prime orders, and $A$ is an anisotropic abelian TQFT.[43] These latter theories are:

1. $A_p$ TQFT

2. $B_p$ TQFT

3. $A_p \boxtimes A_p = B_p \boxtimes B_p$ TQFT

4. Semion and $\overline{\text{Semion}}$.

5. Semion $\boxtimes$ Semion and $\overline{\text{Semion}} \boxtimes \overline{\text{Semion}}$

6. 3-Fermion Model

7. $\mathbb{Z}_4$ TQFT and Galois conjugates.

8. $\mathbb{Z}_4 \boxtimes$ Semion TQFT, $\mathbb{Z}_4 \boxtimes \overline{\text{Semion}}$ TQFT and Galois conjugates.

---

[42]This discussion shows that classifying the set of Galois-invariant unitary TQFTs should be substantially easier than classifying the set of Galois-invariant non-unitary TQFTs. Indeed, classifying this latter class is *naively* as hard as classifying the full set of non-unitary TQFTs and finding their Galois orbits! On the other hand, for unitary Galois-invariant theories, integrality is already an enormous simplification. We will soon see that there are various potential additional constraints on the unitary Galois fixed point TQFTs.

Therefore, all weakly group theoretical integral MTCs should come from gauging a symmetry of $D \boxtimes A$ where $A$ is one among the TQFTs listed above. Note that the discrete gauge theory, $D$, is invariant under Galois conjugation. $A$ is invariant under Galois conjugation only if $A$ is the 3-Fermion Model or $A_p \boxtimes A_p$. This discussion leads to the following theorem:

**Theorem 4.12:** Let $C^G$ be a Galois-invariant weakly group theoretical TQFT, then $C^G$ can be obtained from gauging a symmetry of $D$, $D \boxtimes$ 3-fermion model, or $D \boxtimes A_p \boxtimes A_p$.
**Proof:** Let $C^G$ be a Galois invariant weakly group theoretical TQFT. Then it has to be integral. From Lemma 4.10, we know that if $C^G$ is Galois invariant, then the $G$-crossed braided theory $C_G$ should be Galois invariant. In particular, the MTC, $C$ (the $C_e$ component of $C_G$), should be Galois invariant. Weakly group theoretical integral TQFT $C^G$ comes from gauging a symmetry of $D \boxtimes A$ where $A$ is an anisotropic TQFT. Therefore, $C = D \boxtimes A$. $D$ is an unwisted discrete gauge theory which is invariant under Galois action. Hence, Galois invariance of $C$ implies that $A$ can be either the trivial MTC, the 3-fermion model, or $A_p \boxtimes A_p$. $\square$

---

[43]Anisotropic abelian TQFTs are abelian TQFTs without any subcategories containing only bosons.

As a simple check of this discussion, note that gauging a $\mathbb{Z}_2 \times \mathbb{Z}_2$ natural isomorphism of the 3-Fermion Model gives the $F_8$ prime abelian theory (see Section 5.2.2). Both are Galois invariant.

Lemma 4.9 shows that the Galois invariance of $C$ does not guarantee the Galois invariance of $C^G$. For example, gauging an intrinsic $\mathbb{Z}_3$ symmetry of the 3-fermion models and stacking a particular non-trivial SPT gives the $SU(3)_3$ Chern-Simons theory which has a non-trivial Galois conjugate (see Section 5.3.3). However, gauging the non-trivial $\mathbb{Z}_3$ symmetry of $SU(3)_3$ with trivial SPT stacking gives a Galois-invariant theory. This example can be generalized to the following theorem:

**Theorem 4.13:** Let the magnetic theory, $C$, be Galois invariant with an integer total quantum dimension (i.e., $\mathcal{D} := \sqrt{\sum_a d_a^2} \in \mathbb{Z}$). Suppose the symmetry $G$ acts non-trivially on all non-trivial anyons and satisfies $H^2_{[\rho]}(G, A) \cong \mathbb{Z}_1$, where $A$ is the group of abelian anyons in $C$. Assuming that the obstructions to gauging vanish, and choosing the trivial SPT stacking, the electric theory obtained from gauging is Galois invariant.

**Proof:** Since the symmetry, $G$, acts non-trivially on all non-trivial anyons, each defect sector, $C_g$, in the $G$-crossed braided extension $C_G$ has a single defect field (i.e., a single non-genuine line operator bounding the corresponding $g$ surface operator). The total quantum dimension of $C_g$ is same as that of $C$ for all $g$. Therefore, it is clear the quantum dimensions of all the defects are the same as the total quantum dimension of $C$. The quantum dimensions of the defect are integers, and using Theorem 2.7, we see that $C_G$ is a unitary spherical fusion category.

Therefore, the possible $G$-crossed braided extensions, $C_G(\eta, \alpha)$ are classified by the fractionalization class $\eta$ and possible SPT stackings determined by the 3-cocycle $\alpha$. Since $H^2_{[\rho]}(G, A)$ is trivial, there is a unique fractionalization class. Let us gauge the symmetry $G$ of $C_G(\eta, [1])$, where [1] denotes the trivial SPT. Since there is a unique fractionalization class, and since the trival SPT is Galois invariant, $C_G(\eta, [1])$ is Galois invariant. Therefore, the theory obtained from gauging $G$ symmetry of $C_G(\eta, [1])$ is also Galois invariant. □

For example, consider the charge conjugation symmetry acting on $A_p \boxtimes A_p$. All non-trivial anyons transform non-trivially under this symmetry. By explicitly computing the twisted cohomology groups, $H^3_{[\rho]}(\mathbb{Z}_2, \mathbb{Z}_p \otimes \mathbb{Z}_p)$ and $H^2_{[\rho]}(\mathbb{Z}_2, \mathbb{Z}_p \otimes \mathbb{Z}_p)$, we can check that they are trivial. Therefore, the Postnikov class vanishes and the fractionalization class is unique. The defectification obstruction vanishes because $H^4(\mathbb{Z}_2, U(1)) \cong \mathbb{Z}_1$. Therefore, gauging the charge conjugation symmetry of the $A_p \boxtimes A_p$ TQFT produces Galois invariant TQFTs (irrespective of the SPT stacking).

In Theorem 4.13, we considered a symmetry which acts non-trivially on all non-trivial anyons. This is to ensure that the defects have integer quantum dimensions. However, this is not a necessary constraint to get a Galois invariant TQFT by gauging non-trivial symmetries. For example, consider a $\mathbb{Z}_2$ permutation symmetry which exchanges the anyons in the two prime factors of the $C \boxtimes C$ TQFT. This symmetry is known to have trivial Postnikov class [57], and the defectification obstruction / 't Hooft anomaly vanishes since $H^4(\mathbb{Z}_2, U(1)) \cong \mathbb{Z}_1$. Also, it is known that there is a unique fractionalization class.[44] There are $|C|$ number of defects in each defect sector since all the anyons of the form $(a, a)$ are invariant under the permutation action. The quantum dimensions of the $x_a$ defects are given by [57]

$$d_{x_a} = |C| d_a .  \qquad (4.114)$$

If we assume that $C \boxtimes C$ is Galois invariant, then it is integral. Therefore, all the defects in the $\mathbb{Z}_2$ crossed braided theory have integer quantum dimensions. Gauging the permutation symmetry results in a Galois-invariant TQFT (irrespective of the SPT being stacked before gauging).

If every fusion category with integer Frobenius-Perron dimension is weakly-group theoretical, then any Galois invariant unitary TQFT can be obtained from gauging a symmetry of $D$, $D \boxtimes$ 3-Fermion Model, or $D \boxtimes A_p \boxtimes A_p$.[45] As shown in [72], any fusion category with Frobenius-Perron dimension, a natural number less than 1800 or an odd natural number less than 33075 is weakly-group theoretical. Moreover, if the Frobenius-Perron dimensions of all anyons in a TQFT are prime powers, then it is weakly-group theoretical [73].

## 5. Examples

Let us consider several examples to explicitly see how the Galois action interacts with taking the Drinfeld center and gauging. We will use the G-crossed braided MTC data computed in [57].

---

[44] The vanishing of the Postnikov class and defectification obstruction is true even for $S_n$ action on $C^{\boxtimes n}$. However, for $n > 2$ the fractionalization class is not unique [71].

[45] Some evidence in favor of this possibility follows from the fact that for integral theories, $c \in \mathbb{Z}$ (i.e., the topological central charge is an integer) [21]. Since topological central charge is preserved under gauging, it is easy to check that gauging the full list of weakly anisotropic abelian TQFTs above gives all possible integral central charges modulo eight.

## 5.1. Trivial magnetic theory

Let us first consider the simplest case of a trivial magnetic theory, Vec. In this case, gauging a natural isomorphism symmetry of $G$ is same as taking the Drinfeld center of $\text{Vec}_G^\omega$. The G-crossed braided theory is in fact $\text{Vec}_G^\omega$ itself. We have the Galois field $\mathbb{Q}(\omega)$ associated with this category. A Galois action by $q \in \text{Gal}(\mathbb{Q}(\omega))$ changes the theory as

$$\text{Vec}_G^\omega \to \text{Vec}_G^{\omega^q} \ . \tag{5.1}$$

Therefore, under the action of the Galois group, the Drinfeld center (which in this case is a discrete gauge theory) changes only by $\omega \to \omega^q$.

### 5.1.1. $G = \mathbb{Z}_2$

In this case, we have two possible choices for $\omega$. Since $\omega$ is valued in $\pm 1$, Galois action on $\text{Vec}_{\mathbb{Z}_2}^\omega$ does not change the $\mathbb{Z}_2$-crossed braided theory. Therefore, the Drinfeld center also shouldn't change under Galois action. This is indeed the case. For trivial twist, the Drinfeld center is the Toric Code which is invariant under the Galois action. For non-trivial twist, the Drinfeld center is the Double Semion model which has a complex conjugation Galois action. However, this action can be compensated by a time-reversal symmetry, and hence Double Semion is invariant under Galois action.

The example with non-trivial $\omega$ illustrates that the defining number field of the electric theory can be bigger than the $G$-crossed braided magnetic theory. Indeed, the $G$-crossed braided theory in this case is $\text{Vec}_{\mathbb{Z}_2}^\omega$, whose $F$ symbols are given by $\omega$, and the $R$ symbols can all be set to 1. Even though the defects in $\text{Vec}_{\mathbb{Z}_2}^\omega$ have trivial twist, the twists of the electric theory have $4^{\text{th}}$ roots of unity in them since the gauging procedure involves representations of $\mathbb{Z}_2$ with characters valued in the $4^{\text{th}}$ roots of unity.

### 5.1.2. $G = \mathbb{Z}_N$

In this case, the $\mathbb{Z}_N$-crossed braided theory is $\text{Vec}_{\mathbb{Z}_N}^\omega$ where

$$\omega(g,h,k) = e^{\frac{2\pi i p g}{N^2}(h+k-(h+k \bmod N))} \ , \tag{5.2}$$

and $p \in \mathbb{Z}_N$ parametrizes the different twists. Since $H^2(\mathbb{Z}_2, U(1))$ is trivial, for all values of $\omega$ the Drinfeld center is an abelian theory. The anyons of the Drinfeld center are labelled by $(a, m)$, where $a, m \in \{0, ..N-1\}$. The fusion rules and twists of the Drinfeld center are

$$(a,m) \otimes (b,n) = \left(a+b \bmod N, [m+n-\frac{2p}{N}(a+b-(a+b \bmod N))] \bmod N\right) , \tag{5.3}$$

and
$$\theta_{(a,m)} = e^{\frac{2\pi i}{N}am} e^{-\frac{2\pi i}{N^2}pa^2} . \tag{5.4}$$

The fusion rules form the group $\mathbb{Z}_{\gcd(2p,N)} \times \mathbb{Z}_{\frac{N^2}{\gcd(2p,N)}}$.

The Galois field of $\text{Vec}_{\mathbb{Z}_N}^{\omega}$ is the cyclotomic field $\mathbb{Q}(\xi_N)$ of $N^{\text{th}}$ roots of unity. A Galois action on the $\mathbb{Z}_N$-crossed braided theory corresponds to changing the parameter $p$ as follows

$$p \to qp , \tag{5.5}$$

where $\gcd(q, N) = 1$. After this Galois action, the Drinfeld center has fusion rules and twists

$$(a, m) \otimes (b, n) = \left( a + b \mod N, [m + n - \frac{2pq}{N}(a + b - (a + b \mod N))] \mod N \right) , \tag{5.6}$$

and

$$\theta_{(a,m)} = e^{\frac{2\pi i}{N}am} e^{-\frac{2\pi i}{N^2}pqa^2} . \tag{5.7}$$

It is clear that we have the same fusion rules since $\gcd(2p, N)=\gcd(2qp, N)$ when $\gcd(q, N) = 1$. It will be evident that the fusion rules are the same if we change the variable $m$ to $qm \mod N$. Then we get

$$(a, qm \mod N) \otimes (b, qn \mod N) = \left( a+b \mod N, [q(m+n-\frac{2p}{N}(a+b-(a+b \mod N)))] \mod N \right) . \tag{5.8}$$

The twists become

$$\theta_{(a,qm \mod N)} = e^{\frac{2\pi i}{N}qam} e^{-\frac{2\pi i}{N^2}pqa^2} . \tag{5.9}$$

Therefore, the twist of the anyon $(a, qm \mod N)$ in $\mathcal{Z}(\text{Vec}_{\mathbb{Z}_N}^{\omega^q})$ is the Galois conjugate of the twist of the anyon $(a, m)$ in $\mathcal{Z}(\text{Vec}_{\mathbb{Z}_N}^{\omega})$.

## 5.2. Non-trivial magnetic theory with trivial symmetry

Let us consider some cases of a non-trivial magnetic theory with natural isomorphism symmetry.

### 5.2.1. Ising MTC with $\mathbb{Z}_2$ symmetry

From section 2.4.1, we know that the unitarity preserving Galois actions on the Ising$^{(\nu)}$ family correspond to $q = 1, 7, 9, 15$. Under these Galois actions, the Ising$^{(\nu)}$ family of models transform as

$$\text{Ising}^{(\nu)} \to \text{Ising}^{(q\nu \mod 16)} . \tag{5.10}$$

The Ising model ($\nu = 1$) does not have any non-trivial intrinsic symmetries. Therefore, the $\mathbb{Z}_2$ group has to act as a natural isomorphism. We know that the Postnikov class vanishes. The fractionalization class is specified by an element in $\eta \in H^2(\mathbb{Z}_2, \mathbb{Z}_2) \cong \mathbb{Z}_2$. Since $H^4(\mathbb{Z}_2, U(1))$ is trivial, defectification obstruction vanishes. The choice of stacking a $\mathbb{Z}_2$-SPT before gauging is paramerized by an element in $\alpha \in H^3(\mathbb{Z}_2, U(1)) \cong \mathbb{Z}_2$. We get the following theories under gauging [57]

$$\eta, \alpha \text{ trivial} \quad \to \quad \text{Ising} \boxtimes \text{ Toric Code} , \tag{5.11}$$

$$\eta \text{ trivial } \alpha \text{ non-trivial} \quad \to \quad \text{Ising} \boxtimes \text{ Double-Semion} , \tag{5.12}$$

$$\eta \text{ non-trivial } \alpha \text{ trivial} \quad \to \quad \text{Ising}^{(15)} \boxtimes A_4 , \tag{5.13}$$

$$\eta \text{ non-trivial } \alpha \text{ non-trivial} \quad \to \quad \text{Ising}^{(3)} \boxtimes B_4 . \tag{5.14}$$

Since both $\eta$ and $\alpha$ are valued in $\pm 1$, a Galois action on the $\mathbb{Z}_2$-crossed braided structure can only affect the modular subcategory Ising. That is, let $\text{Ising}(\nu, \alpha)$ denote the $\mathbb{Z}_2$-crossed braided theory specified by $\eta$ and $\omega$. A unitarity preserving Galois action on this gives $\text{Ising}^{(q)}(\eta, \alpha)$, where $q$ is specified by the Galois action. Therefore, the electric theories obtained above should also transform in this way.

Since the Toric code and Double-Semion model are invariant under Galois action, we find that the Galois action on (5.11) and (5.12) acts precisely as the $\mathbb{Z}_2$-crossed braided magnetic theory transforms.

Now let us focus on the electric theory, $\text{Ising}^{(15)} \boxtimes A_4$. The data of this theory belongs to the cyclotomic field $\mathbb{Q}(\xi_{16})$. The unitarity preserving Galois actions correspond to $q = 1, 7, 9, 15$. Under these Galois action we get

$$q = 7 : \quad \text{Ising}^{(15)} \boxtimes A_4 \to \text{Ising}^{(9)} \boxtimes B_4 , \tag{5.15}$$

$$q = 9 : \quad \text{Ising}^{(15)} \boxtimes A_4 \to \text{Ising}^{(7)} \boxtimes A_4 , \tag{5.16}$$

$$q = 15 : \quad \text{Ising}^{(15)} \boxtimes A_4 \to \text{Ising}^{(1)} \boxtimes B_4 . \tag{5.17}$$

Recall that $\text{Ising}^{(15)} \boxtimes A_4$ is obtained from gauging $\text{Ising}^{(1)}(\eta = -1, \alpha = +1)$. The three Galois conjugates above are obtained from the $\mathbb{Z}_2$-crossed braided categories $\text{Ising}^{(7)}(\eta = -1, \alpha = +1)$, $\text{Ising}^{(9)}(\eta = -1, \alpha = +1)$ and $\text{Ising}^{(15)}(\eta = -1, \alpha = +1)$, respectively. These are all Galois conjugates of $\text{Ising}^{(1)}(\eta = -1, \alpha = +1)$, as expected.

Similarly, we can check that the Galois conjugates of the electric theory $\text{Ising}^{(3)} \boxtimes B_4$ corresponds to Galois conjugates of the $\mathbb{Z}_2$-crossed braided theory $\text{Ising}^{(1)}(\eta = -1, \alpha = -1)$.

*5.2.2. 3-fermion model with $\mathbb{Z}_2 \times \mathbb{Z}_2$ symmetry*

Consider the prime abelian theory $F_8$. The 64 anyons are labelled by $(m,n)$ where $m, n \in \mathbb{Z}_8$. The bosons $(0,0), (0,4), (4,0), (4,4)$ form a $\text{Rep}(\mathbb{Z}_2 \times \mathbb{Z}_2)$ subcategory which can be condensed. The magnetic theory can be obtained by identifying the anyons which braid trivially with all anyons in $\text{Rep}(\mathbb{Z}_2 \times \mathbb{Z}_2)$. These fall into the following 4 equivalence classes of anyons under fusion with the anyons in $\text{Rep}(\mathbb{Z}_2 \times \mathbb{Z}_2)$

$$(0,0) \ , \ (0,2) \ , \ (2,0) \ , \ (2,2) \ . \tag{5.18}$$

The twists of these anyons are $1, -1, -1, -1$, respectively. Therefore, the magnetic theory is the 3-fermion model. Hence, the $F_8$ prime abelian anyons model can be obtained from $F_2$ by gauging a $\mathbb{Z}_2 \times \mathbb{Z}_2$ natural isomorphism symmetry. Both the magnetic and electric theory are invariant under Galois conjugation.

For the $F_2$ abelian model with $\mathbb{Z}_2 \times \mathbb{Z}_2$ symmetry, the Postnikov class vanishes and the fractionalization class belongs to the group $H^2(\mathbb{Z}_2 \times \mathbb{Z}_2, \mathbb{Z}_2 \times \mathbb{Z}_2) \cong \mathbb{Z}_2^6$. Group cohomology allows for a defectification obstruction since $H^4(\mathbb{Z}_2 \times \mathbb{Z}_2, U(1)) \cong \mathbb{Z}_2 \times \mathbb{Z}_2$. For a given choice of the fractionalization class, if this obstruction vanishes, then the freedom to stack a $\mathbb{Z}_2 \times \mathbb{Z}_2$-SPT before gauging is parametrized by $H^3(\mathbb{Z}_2 \times \mathbb{Z}_2, U(1)) \cong \mathbb{Z}_2^3$. Therefore, we have several possible electric theories in this case based on the choice of fractionalization class and SPT stacking.

*5.3. Non-trivial magnetic theory with non-trivial symmetry*

*5.3.1. Toric code with $\mathbb{Z}_2$ electric-magnetic symmetry*

Let us consider the Toric code with a non-trivial $\mathbb{Z}_2$ symmetry which permutes the two bosons. It is known that the Postnikov class vanishes for this symmetry. We have $H^2_{[\rho]}(\mathbb{Z}_2, \mathbb{Z}_2 \times \mathbb{Z}_2) \cong \mathbb{Z}_1$, $H^4(\mathbb{Z}_2, U(1)) \cong \mathbb{Z}_1$ and $H^3(\mathbb{Z}_2, U(1)) = \mathbb{Z}_2$. Therefore, there is a unique fractionalization class for which group cohomology guarantees that the defectification obstruction vanishes. We have the freedom to stack a $\mathbb{Z}_2$-SPT corresponding to $\alpha \in H^3(\mathbb{Z}_2, U(1)) \cong \mathbb{Z}_2$ before gauging. We get the following theories under gauging

$$\alpha \text{ trivial} \quad \to \quad \text{Ising}^{(1)} \boxtimes \text{Ising}^{(15)} \ , \tag{5.19}$$

$$\alpha \text{ non-trivial} \quad \to \quad \text{Ising}^{(3)} \boxtimes \text{Ising}^{(13)} \ . \tag{5.20}$$

Since the Toric code is Galois invariant, and since the $\mathbb{Z}_2$ crossed braided theory is completely rigid except for the choice of $\alpha$ (which is valued in $\pm 1$), the $\mathbb{Z}_2$-crossed braided theory is invariant under all unitarity-preserving Galois actions. Therefore, the electric theories obtained above should also be invariant under all such Galois actions. Indeed, the MTCs Ising$^{(1)}\boxtimes$ Ising$^{(15)}$ and Ising$^{(3)}\boxtimes$ Ising$^{(13)}$ are both invariant under all unitarity preserving Galois actions.

### 5.3.2. 3-fermion model with $\mathbb{Z}_2$ symmetry

Let us consider the 3-fermion model with a non-trivial $\mathbb{Z}_2$ symmetry which permutes any two of the three fermions in the theory. It is known that the Postnikov class vanishes for this symmetry. We have $H^2_{[\rho]}(\mathbb{Z}_2, \mathbb{Z}_2 \times \mathbb{Z}_2) \cong \mathbb{Z}_1$, $H^4(\mathbb{Z}_2, U(1)) \cong \mathbb{Z}_1$ and $H^3(\mathbb{Z}_2, U(1)) = \mathbb{Z}_2$. Therefore, there is a unique fractionalization class for which group cohomology guarantees that the defectification obstruction vanishes. We have the freedom to stack a $\mathbb{Z}_2$-SPT corresponding to $\alpha \in H^3(\mathbb{Z}_2, U(1)) \cong \mathbb{Z}_2$ before gauging. We get the following theories under gauging

$$\alpha \text{ trivial} \quad \to \quad \text{Ising}^{(1)} \boxtimes \text{Ising}^{(7)} \ , \tag{5.21}$$

$$\alpha \text{ non-trivial} \quad \to \quad \text{Ising}^{(3)} \boxtimes \text{Ising}^{(5)} \ . \tag{5.22}$$

Since the magnetic theory is Galois invariant, and since the $\mathbb{Z}_2$ crossed braided theory is completely rigid except for the choice of $\alpha$ (which is valued in $\pm 1$), the $\mathbb{Z}_2$-crossed braided theory is invariant under all unitarity preserving Galois actions. Therefore, the electric theories obtained above should also be invariant under all such Galois actions. Indeed, the MTCs Ising$^{(1)}\boxtimes$ Ising$^{(7)}$ and Ising$^{(3)}\boxtimes$ Ising$^{(5)}$ are both invariant under all unitarity preserving Galois actions (Galois action leads to permutations of the three anyons with $\sqrt{2}$ quantum dimensions).

### 5.3.3. 3-fermion model with $\mathbb{Z}_3$ symmetry

Let us consider the 3-fermion model with a non-trivial $\mathbb{Z}_3$ symmetry which cyclically permutes the three fermions in the theory. It is known that the Postnikov class vanishes for this symmetry. We have $H^2_{[\rho]}(\mathbb{Z}_3, \mathbb{Z}_2 \times \mathbb{Z}_2) \cong \mathbb{Z}_1$, $H^4(\mathbb{Z}_3, U(1)) \cong \mathbb{Z}_1$ and $H^3(\mathbb{Z}_3, U(1)) = \mathbb{Z}_3$. Therefore, there is a unique fractionalization class for which group cohomology guarantees that the defectification obstruction vanishes. We have the freedom to stack a $\mathbb{Z}_3$-SPT corresponding to $\alpha \in H^3(\mathbb{Z}_3, U(1)) \cong \mathbb{Z}_3$ before gauging. It is known that for non-trivial $\alpha$ and its inverse we get the MTC $SU(3)_3$ and its complex conjugate under gauging [57].

$SU(3)_3$ has only one non-trivial Galois conjugate [74], which is the complex conjugate of $SU(3)_3$. Therefore, Galois conjugation of the electric theory corresponds to changing the $\mathbb{Z}_3$-SPT being stacked before gauging ($\alpha \to \bar{\alpha}$).

For trivial $\alpha$ the resulting TQFT is integral and has different fusion rules than that of $SU(3)_3$ (The explicit fusion rules are given in [57]). Since the magnetic theory is Galois invariant and since the $\mathbb{Z}_3$-crossed braided theory with $\alpha$ trivial is invariant under Galois action, the electric theory is invariant under unitarity preserving Galois actions. Moreover, since the electric theory is integral and unitary, all Galois actions preserve unitarity (using Theorem 2.7). Therefore, the electric theory obtained for trivial $\alpha$ is in fact completely Galois invariant. Indeed, one can check that the modular data for this theory given in [74] is invariant under Galois conjugation (up to permutation of the anyons). This is an example of a Galois invariant non-abelian TQFT which is not a discrete gauge theory.

## 6. Conclusion

We explored several aspects of Galois actions on TQFTs and gave a sufficient condition for producing unitary Galois orbits. We also discussed how Galois conjugation of a bulk TQFT changes its gapped boundary. Using the fact that certain TQFTs are uniquely determined by their gapped boundaries, we studied how the Galois action on gapped boundaries affects the bulk TQFTs. By determining the relationship between Galois action on theories related by gauging, we showed that (assuming a conjecture in the literature) arbitrary Galois-invariant TQFTs are closely related to simple abelian Galois-invariant TQFTs.

These results, along with our earlier work [11], show that, while Galois conjugation usually results in distinct TQFTs, the TQFTs in a Galois orbit are closely related to each other. They have the same symmetry structure (modulo mild assumptions in the defining number field of the $G$-crossed braided theory), and their gapped boundaries are related to each other. This situation is unlike other operations, such as gauging or condensation, which can drastically change the anyon content and symmetry structure of the theory.

Finally, we constructed the defining number field $K_C$ of an MTC using the $F$ and $R$ symbols, and Galois conjugation of the TQFT acted directly on the $F$ and $R$ data. In general, the total quantum dimension, $\mathcal{D}$, is not an element of $K_C$. Moreover, we defined Galois action on the TQFT such that it doesn't change the sign of $\mathcal{D}$ (we can consider taking $\mathcal{D} \to -\mathcal{D}$ as a second step, supplementing our Galois conjugation, when exploring particular orbits).[46] Explicitly including a $\mathcal{D} \to -\mathcal{D}$ transformation leads to certain simple extensions of our results.

We conclude with some comments:

---

[46]Recall that the sign of $\mathcal{D}$ is important for TQFT unitarity.

- Galois conjugation has played a major role in finding counter examples to the conjecture that the modular data determines a topological phase of matter [7]. A general strategy to use Galois conjugation to find modular isotopes is as follows. Let $K_M$ be the cyclotomic field containing the components of the $S$ and $T$ matrices of an MTC $C$. Let $L$ be another link invariant and let $K_L$ be the Galois field containing the component of $L$. If $K_L$ is not the same field extension as $K_M$, then there exists some element $q \in \text{Gal}(K_L)$ such that the action of $q$ on $S$ and $T$ is trivial, while $q(L) \neq L$. If $q(L)$ and $L$ are not related by a permutation of the anyon labels, then the MTCs $C$ and $q(C)$ are modular isotopes. It would be interesting to explore this direction further.

- Another interesting operation which takes us between TQFTs is Zesting [74]. Like Galois conjugation, zesting can be used to find modular isotopes [75]. The $SU(3)_3$ Chern-Simons theory and its time reversal are related by a Galois conjugation. These two theories are also related by zesting. It would be interesting to explore the relationship between Galois action and zesting, and understand when zesting produces Galois conjugate TQFTs.

- Galois invariant TQFTs are very special, and Theorem 4.12 relates them to discrete gauge theories, the 3-fermion model and $A_p \boxtimes A_p$. However, gauging an arbitrary symmetry of these theories can give us a Galois non-invariant TQFT due to a kind of Galois conjugation-0-form symmetry mixed anomaly. It would be interesting to fully define the Galois conjugation-0-form symmetry anomaly (and the Galois conjugation-anyon condensation anomaly) and give sufficient and necessary criteria for its vanishing.

- We saw that in order to argue that certain symmetries were preserved under Galois conjugation, we needed to make some mild assumptions on the underlying number fields. It would be interesting to understand if these assumptions are ever violated. If so, it would be intriguing to understand if one can think of these situations as representing certain number-theoretical anomalies.

- In $2+1$D, discrete gauge theories and quantum groups form two important classes of TQFTs. In contrast, $3+1$D TQFTs are mostly governed by discrete gauge theories. For example, $3+1$D TQFTs with bosonic line operators are known to be classified by $3+1$D discrete gauge theories [76]. These are Drinfeld centers of fusion 2-categories [42], and they have many parallels with $2+1$D discrete gauge theories. This begs the question of how our results generalizes to these higher dimensional TQFTs.

- Along with entanglement entropy, complexity and magic are important quantities which characterize link states [77]. It will be interesting to analyze the behavior of these quantities under Galois action.

- Finally, recall that the Witt group of TQFTs [78] may play an important role in the classification of MTCs and related structures. In this construction, two MTCs, $\mathcal{C}_1$ and $\mathcal{C}_2$, are Witt equivalent if they satisfy $\mathcal{C}_1 \boxtimes \mathcal{Z}(A_1) \simeq \mathcal{C}_2 \boxtimes \mathcal{Z}(A_2)$ (where $\mathcal{Z}(\cdots)$ is the Drinfeld center of the enclosed fusion category). It would be interesting to define and explore a notion of "Galois equivalence" of MTCs $\mathcal{C}_{1,2}$. Here we could define $\mathcal{C}_1$ and $\mathcal{C}_2$ to be Galois equivalent if $\mathcal{C}_1 \boxtimes \mathcal{C}'_1 = \mathcal{C}_2 \boxtimes \mathcal{C}'_2$ where $\mathcal{C}'_{1,2}$ are Galois invariant.

## Acknowledgments

M. B. is funded by the Royal Society grant, "Relations, Transformations, and Emergence in Quantum Field Theory." M. B. and R. R. are funded by the Royal Society grant, "New Aspects of Conformal and Topological Field Theories Across Dimensions." The STFC also partially supported our work under the grant, "String Theory, Gauge Theory and Duality."

## Appendix A. Galois Theory

Let us recall some definitions and results from Galois theory which we use in our arguments. For a standard reference, see [79]. Consider the field of rational numbers $\mathbb{Q}$. Let $\mathbb{Q}(S)$ be a finite field extension of $\mathbb{Q}$, where $S$ is a minimal set of generators of the field extension. For example, $S = e^{\frac{2\pi i}{N}}$ gives the cyclotomic field of $N^{\text{th}}$ roots of unity, and $S = \sqrt{N}$ is a quadratic extension of the rationals. We will also use the notation $K/\mathbb{Q}$ to denote a finite field extension $K$ over $\mathbb{Q}$.

We will only deal with finite field extensions of $\mathbb{Q}$. Such an extension is always separable, though it may not be normal. Given a non-normal extension, $\mathbb{Q}(S)$, we can construct the normal closure of $\mathbb{Q}(S)$, $N(\mathbb{Q}(S))$, as the unqiue and minimal field extension of $\mathbb{Q}$ containing $\mathbb{Q}(S)$ such that $N(\mathbb{Q}(S))$ is normal. $N(\mathbb{Q}(S))$ is a finite field extension, and therefore it is also separable. Thus, $N(\mathbb{Q}(S))$ is a Galois extension.

We will denote the group of automorphisms of a field extension by $\text{Aut}(\mathbb{Q}(S))$, and the automorphisms of a Galois extension by the Galois group $\text{Gal}(\mathbb{Q}(S))$. Consider a non-normal field extension $\mathbb{Q}(S)$. For every element $\sigma \in \text{Aut}(\mathbb{Q}(S))$, we have some $\sigma' \in \text{Gal}(N(\mathbb{Q}(S)))$ such that $\sigma'|_{\mathbb{Q}(S)} = \sigma$. However, every element of $\text{Gal}(N(\mathbb{Q}(S)))$ does not act as an automorphism of $\mathbb{Q}(S)$ since we have automorphisms which take elements of $\mathbb{Q}(S)$ and map them to elements of the field $N(\mathbb{Q}(S))$ outside $\mathbb{Q}(S)$.

For a tower of field extensions $K/M/\mathbb{Q}$, where $K$ and $M$ are normal extensions over $\mathbb{Q}$, we have a map $\text{Aut}(K/\mathbb{Q}) \to \text{Aut}(M/\mathbb{Q})$ given by the restriction

$$\sigma|_M \ , \tag{A.1}$$

for $\sigma \in \text{Gal}(K/\mathbb{Q})$. This map is surjective with kernel $\text{Aut}(K/M)$. Therefore, this map is injective if and only if $K = M$.

### A.1. Composite Extensions

Since we can always take the normal closure of a finite field extension to make it Galois, we will only discuss Galois extensions in the following. Consider the Galois extensions $K$ and $M$ over $\mathbb{Q}$. The composite extension $KM/\mathbb{Q}$ is the minimal extension of $\mathbb{Q}$ containing $K$ and $M$. If $k_1, \cdots, k_n$ and $m_1, \cdots, m_l$ are a set of basis vectors of $K$ and $M$, respectively, as a vector space over $\mathbb{Q}$ then $KM/\mathbb{Q}$ is generated by the vectors $k_1, \cdots, k_n, m_1, \cdots, m_l$. In other words, if $K = \mathbb{Q}(S_1)$ and $M = \mathbb{Q}(S_2)$, then $KM/\mathbb{Q} = \mathbb{Q}(S_1 \cup S_2)$.

If $K/\mathbb{Q}$ and $M/\mathbb{Q}$ are Galois extensions, then so is $KM/\mathbb{Q}$. The Galois group of the

composite extension is

$$\text{Gal}(KM/\mathbb{Q}) = \{(\sigma, \tau) \in \text{Gal}(K/\mathbb{Q}) \times \text{Gal}(M/\mathbb{Q}) : \sigma_{K \cap M} = \tau|_{K \cap M}\} \ . \qquad (A.2)$$

Note that every non-trivial element of $\text{Gal}(KM/\mathbb{Q})$ acts non-trivially on $K$ or $M$. It is clear that

$$\text{Gal}(KM/\mathbb{Q}) = \text{Gal}(K/\mathbb{Q}) \times \text{Gal}(M/\mathbb{Q}) \text{ iff } K \cap M = \phi \ . \qquad (A.3)$$

Given Galois fields $K$ and $M$ over $\mathbb{Q}$, we are interested in going from a Galois action on $K$ to a Galois action on $M$ using the composite extension as follows

$$\begin{array}{ccc}
 & \text{Gal}(KM/\mathbb{Q}) & \\
\sigma|^{KM/\mathbb{Q}} \nearrow & & \searrow (\sigma|^{KM/\mathbb{Q}})|_{M/\mathbb{Q}} \\
\text{Gal}(K/\mathbb{Q}) & & \text{Gal}(M/\mathbb{Q}) \ .
\end{array}$$

For example, let $K = \mathbb{Q}(i, \sqrt{2})$ and let $M = \mathbb{Q}(i, \sqrt{3})$. Then we have $\text{Gal}(K/\mathbb{Q}) \cong \mathbb{Z}_2 \times \mathbb{Z}_2 = \{e, \sigma_1, \tau_1, \sigma_1 \tau_1\}$ where $\sigma_1$ acts non-trivially only on $\sqrt{2}$ as $\sigma_1(\sqrt{2}) = -\sqrt{2}$ and $\tau_1$ acts non-trivially only on $i$ as $\tau_1(i) = -i$. Also, we have $\text{Gal}(M/\mathbb{Q}) \cong \mathbb{Z}_2 \times \mathbb{Z}_2 = \{e, \sigma_2, \tau_2, \sigma_2 \tau_2\}$ where $\sigma_2$ acts non-trivially only on $\sqrt{3}$ as $\sigma_2(\sqrt{3}) = -\sqrt{3}$ and $\tau_2$ acts non-trivially only on $i$ as $\tau_2(i) = -i$. The composite field is $KM/\mathbb{Q} = \mathbb{Q}(i, \sqrt{2}, \sqrt{3})$. Since $K \cap M = \mathbb{Q}(i)$, we have the Galois group

$$\text{Gal}(KM/\mathbb{Q}) = \{(e, e), (e, \sigma_2), (\sigma_1, e), (\sigma_1, \sigma_2), (\tau_1, \tau_2), (\tau_1, \sigma_2 \tau_2), (\sigma_1 \tau_1, \tau_2), (\sigma_1 \tau_1, \sigma_2 \tau_2)\} \ . \qquad (A.4)$$

Suppose we have Galois action on $K/\mathbb{Q}$ by $\sigma_1$. Then we can lift it to a Galois action on $KM/\mathbb{Q}$ to get $(\sigma_1, e)$ or $(\sigma_1, \sigma_2)$. This lift is not unique since $K/\mathbb{Q}$ is a proper subfield of $KM/\mathbb{Q}$. Depending on our choice of the lift, we can restrict the group action on $KM/\mathbb{Q}$ to $M/\mathbb{Q}$ to get either $e, \sigma_2$. Note that we have a choice of the lift such that the action on $M/\mathbb{Q}$ can be taken to be trivial. This is not always the case. For example, if we have the action on $K/\mathbb{Q}$ by $\tau_1$, there is no lift to $KM/\mathbb{Q}$ such that its restriction on $M/\mathbb{Q}$ is trivial. This is of course true because $\tau_1$ acts non-trivially on $i$ which is a common element of both $K$ and $M$.

## Appendix B. TQFTs and Modular Tensor Categories

In this apppendix, we will go through the essential aspects of a modular tensor category which we use in our arguments (we presented a similar review in [11], but we include it here for completeness). A modular tensor category is an algebraic structure which captures the operator content and correlation functions of a $2+1$D TQFT. A TQFT does not have any local operators. In $2+1$D, we can have non-trivial line and surface operators. From a general theorem in [3], the absence of local operators imply that the surface operators in $2+1$D TQFT can be constructed from its line operators. Therefore, if we want to capture the minimal data required to define a $2+1$D TQFT, we only need to keep track of the line operators and their correlation functions. In the following, we will assume that the TQFT has a finite number of line operators.

An MTC consists of a finite set of labels, $\{a, b, \cdots\}$. They satisfy the fusion rules

$$a \otimes b = \sum_c N_{ab}^c c \ , \quad N_{ab}^c \in \mathbb{Z}_{\geq 0} \ . \tag{B.1}$$

The labels denote the different line operators in the TQFT and their fusion rules capture the position-independent operator product expansion (OPE) of these operators. Among the labels, there is a distinguished label, **1**, which denotes the trivial line operator (sometimes, in an additive notation for abelian theories, the trivial line is labeled **0**). Since MTCs describe topological phases of matter, we can also interpret the labels as charges of the quasiparticles in the topological phase. In this language, the label **1** denotes the vacuum. The fusion rules describe the ways in which these particles combine to form new ones.

The non-negative integers, $N_{ab}^c$, count the different ways in which $a$ and $b$ combine to form $c$. Note that the fusion $a \otimes b = c$ is allowed if and only if $N_{ab}^c > 0$. In fact, the $N_{ab}^c$ fusion coefficient is the dimension of the $V_{ab}^c$ fusion Hilbert space. This is the fusion space associated with the anyons $a$ and $b$ fusing to give $c$. More generally, the fusion space corresponding to the anyons $a_1, \cdots, a_n$ fusing to give anyon $b$ is written as $V_{a_1 a_2 \cdots a_n}^b$.

Given the fusion rules, we can define the Frobenius-Perron dimension of an anyon $a$, denoted FPdim($a$), as the maximal non-negative eigenvalue of the matrix $N_a$, where $(N_a)_{b,c} := N_{ab}^c$. The Frobenius-Perron dimension of the MTC $C$ is defined as

$$\text{FPdim}(C) := \sum_a \text{FPdim}(a)^2 \ . \tag{B.2}$$

An MTC is called integral if FPdim($a$) $\in \mathbb{Z}$ $\forall a$. An MTC is called weakly integral if FPdim($C$) $\in \mathbb{Z}$.

The fusion of two anyons is commutative. This fact implies the existence of an isomorphism, $V_{ab}^c \cong V_{ba}^c$, and the associated linear map corresponding to this isomorphism is called the $R$ matrix (see Fig. 3). Moreover, the fusion of three anyons is associative. This statement implies that the fusion space $V_{abc}^d = \sum_f V_{ab}^f \otimes V_{fc}^d$ can also be decomposed as $V_{abc}^d = \sum_e V_{bc}^e \otimes V_{ea}^d$. The $F$ matrix is the linear map associated with the isomorphism $\sum_f V_{ab}^f \otimes V_{fc}^d \cong \sum_e V_{bc}^e \otimes V_{ea}^d$ (see Fig. 2).

From this discussion, we see that

$$F_{abc}^d : \sum_f V_{ab}^f \otimes V_{fc}^d \to \sum_e V_{bc}^e \otimes V_{ea}^d \ , \quad R_{ab}^c : V_{ab}^c \to V_{ba}^c \ . \tag{B.3}$$

Next, from the action of $F$ on $V_{abcd}^e$, the "Pentagon" consistency equation follows

$$\left(F_{a,b,k}^e\right)_l^i \left(F_{i,c,d}^e\right)_j^k = \sum_m \left(F_{b,c,d}^l\right)_m^k \left(F_{a,m,d}^e\right)_j^l \left(F_{a,b,c}^j\right)_i^m \ . \tag{B.4}$$

Moreover, the braiding of anyons captured by the $R$ matrix should be consistent with the associativity of the fusion rules. In other words, the action of the $R$ and $F$ matrices on $V_{abc}^d$ should be consistent. This requirement leads to two "Hexagon" equations. The first takes the form

$$R_{a,c}^k \left(F_{b,a,c}^d\right)_i^k R_{a,b}^i = \sum_j \left(F_{b,c,a}^d\right)_j^k R_{a,j}^d \left(F_{a,b,c}^d\right)_i^j \ , \tag{B.5}$$

and the second is

$$R_{c,a}^k \left(\left(F_{b,a,c}^d\right)_i^k\right)^{-1} R_{b,a}^i = \sum_j \left(\left(F_{b,c,a}^d\right)_j^k\right)^{-1} R_{j,a}^d \left(\left(F_{a,b,c}^d\right)_i^j\right)^{-1} \ . \tag{B.6}$$

Suppressing all indices, we will refer to solutions of (B.4), (B.5), and (B.6) simply as $F$ and $R$. Even though once can start with any set of labels and fusion rules, a consistent MTC exists only if (B.4), (B.5), and (B.6) are satisfied [59, 80].

If we wish to calculate $F$ and $R$ explicitly, we have to choose a basis for the fusion spaces, $V_{ab}^c$. The solutions to the Hexagon and Pentagon equations obtained by choosing different sets of basis vectors should be considered equivalent. This equivalence is known as the "gauge freedom" in defining $F$ and $R$. The Pentagon and Hexagon equations have at most a finite number of inequivalent solutions [4, 29]. To summarize, we have captured the line operators and their OPEs via the labels and fusion rules. The commutativity and associativity of the fusion rules lead to the Pentagon and Hexagon equations. At this level of structure, we have defined a braided fusion category.

To add more structure note that, for every anyon $a$, there is a dual anyon, $\bar{a}$, such that $a \otimes \bar{a}$ involves the vacuum. In other words, $\bar{a}$ is the anti-particle of $a$, and $\bar{\bar{a}} = a$. To capture this fact in our algebraic construction, we need to define a ribbon structure on the braided fusion category by defining isomorphisms from $a$ to $\bar{\bar{a}}$. These isomorphisms are captured by phases, $\epsilon_a$, for each label $a$, satisfying the constraint

$$\epsilon_a^{-1} \epsilon_b^{-1} \epsilon_c = (F_{a,b,\bar{c}}^{\mathbf{1}})_c^{\bar{a}} (F_{b,\bar{c},a}^{\mathbf{1}})_a^{\bar{a}} (F_{\bar{c},a,b}^{\mathbf{1}})_b^{\bar{b}} \ . \tag{B.7}$$

We also require there to be a gauge in which $\epsilon_a \in \{\pm 1\}$ $\forall a$. In general, if there is a solution to these constraints, it need not be unique, though the number of distinct solutions is always finite and has been classified [26]. Using these, we can define the quantum dimension of an anyon $a$ as follows

$$d_a := (\epsilon_a (F^a_{a\bar{a}a})^{\mathbf{1}}_{\mathbf{1}})^{-1} \ . \tag{B.8}$$

This expression is valid only in a particular basis as chosen in Lemma 3.4 of [14]. $d_a$ is the $S^3$ link invariant of an unknot labelled by $a$. Note that $d_a$ depends on several choices and it is not, in general, equal to the Frobenius-Perron dimension of an anyon. In fact, FPdim$(a)$ is always positive, while $d_a$ can be negative for certain choices of solutons $\epsilon_a$ to (B.7). In a unitary TQFT, the quantum dimensions are required to be positive, and in this case $d_a$=FPdim$(a)$ $\forall a$. The total quantum dimension of the TQFT is defined as

$$\mathcal{D} := \sqrt{\sum_a d_a^2} \ , \tag{B.9}$$

where he have picked a particular sign that is necessary for the TQFT to be unitary.

At this level of structure we have defined a ribbon fusion category. We want an MTC to describe systems with no transparent anyons. That is, all non-trivial anyons should braid non-trivially with at least one anyon. This condition is captured by the invertibility of the matrix

$$S_{ab} = \frac{1}{\mathcal{D}} \sum_c d_c \text{Tr}(R^c_{ab} R^c_{ba}) = \frac{1}{\mathcal{D}} \tilde{S}_{ab} \ . \tag{B.10}$$

Here, $\tilde{S}_{ab}$ is the invariant of the Hopf link, which captures the creation of two anyon-antianyon pairs, their braiding and their annihilation. In fact, along with

$$T_{aa} = d_a^{-1} \sum_c d_c R^c_{aa} = \theta(a) \ , \tag{B.11}$$

$S_{ab}$ gives rise to a unitary (projective) representation of the modular group,[47] $SL(2,\mathbb{Z})$. Indeed, these quantities obey the following equations

$$(ST)^3 = \Theta C \ , \quad S^2 = C \ , \quad C^2 = I \ , \tag{B.12}$$

where $\Theta = \frac{1}{\sqrt{\sum_c d_c^2}} \sum_a d_a^2 T_{aa}$, and $C$ is the charge conjugation matrix. The fusion coefficients, $N^c_{ab}$, are determined by the $S$ matrix elements via the Verlinde formula

$$N^c_{ab} = \sum_e \frac{S_{ae} S_{be} S_{ec^*}}{S_{0e}} \ . \tag{B.13}$$

---

[47]The unitarity of this representation does not imply unitarity of the TQFT.

The solutions to (B.4) and (B.5) admit a cohomological interpretation, where the relevant coboundaries capture the gauge freedom. For example, in the case of abelian MTCs, $(F, R)$ are valued in so-called abelian group cohomology. Given a collection of labels and fusion rules, a $2 + 1$D TQFT with non-trivial labels/anyons is a cohomologically non-trivial solution to these polynomial equations.[48] We will refer to the collection, $(N_{ab}^c, R, F)$, as the "MTC data", and to the $(S, T)$ pair (or, depending on the context, the $(\tilde{S}, T)$ pair) as the "modular" data.

Finally, note that we can take the total quantum dimension to be

$$\mathcal{D}^{(-)} = -\sqrt{\sum_a d_a^2} \tag{B.14}$$

In this case, the expression for the normalized $S$ matrix changes by a sign. In fact, given the modular data $(S, T)$ of an MTC, there also exists an MTC realizing the modular data $(-S, T)$. Unless otherwise stated, we will use the definition of the total quantum dimension with positive sign.

---

[48]In particular, the space of consistent $2 + 1$D TQFTs satisfying the MTC axioms is discrete [29].